\documentclass[aps,prl,twocolumn,floats,epsfig,pdflatex]{revtex4-2}
\usepackage{amsmath}
\usepackage{amsfonts}
\usepackage{graphicx}
\usepackage{tikz}
\usepackage{braket}
\usetikzlibrary{matrix,fit,positioning}
\usepackage{amssymb}
\usepackage{amsbsy}
\usepackage{comment}
\usepackage{color}
\usepackage{ulem}
\usepackage[colorlinks=true, citecolor=blue, urlcolor=blue, linkcolor=blue, pdfencoding=auto, psdextra]{hyperref}
\usepackage{hyperref}

\newcommand{\beq}{\begin{equation}}
\newcommand{\eeq}{\end{equation}}
\newcommand{\bea}{\begin{eqnarray}}
\newcommand{\eea}{\end{eqnarray}}

\begin{document}

\title{Emergent prethermal Bethe integrability in a periodically driven Rydberg chain}
\author{Saptadip Roy$^1$, Arnab Sen$^1$, Diptiman Sen$^2$, and K. Sengupta$^1$}
\affiliation{$^1$School of Physical Sciences, Indian Association for the
Cultivation of Science, Kolkata 700032, India \\
$^2$Center for High Energy Physics, Indian Institute of Science, Bengaluru 560012, India}
\date{\today}

\begin{abstract}

We study a chain of periodically driven Rydberg atoms and identify a class of drive protocols for which the system exhibits emergent prethermal Bethe integrability at special drive frequencies. We provide a perturbative analytic expression of its Floquet Hamiltonian in the large drive amplitude regime. We demonstrate integrability of the leading term of this Floquet Hamiltonian at special drive frequencies, which we identify, by mapping it to the Hamiltonian of the paradigmatic spin-$1/2$ ${\rm XXZ}$ chain. We support our analytical results by exact diagonalization studies on finite chains. Our numerical results on level statistics, half-chain entanglement entropy, and longitudinal magnetization of the driven chain brings out its emergent integrable nature at the special drive frequencies which persists up to a large prethermal timescale. 

\end{abstract}


\maketitle

{\it Introduction} : Periodically driven closed quantum systems \cite{rev2,rev5,rev6,rev7,rev8,rev9,rev10,rev11,rev12,rev13,rev14,rev15,rev16,rev17} may host a long prethermal regime before their eventual thermalization to an infinite temperature steady state \cite{mori1,da1,da2,vk1,vk2,dc1}. In this experimentally relevant prethermal regime, such systems show several interesting phenomena such as dynamical freezing \cite{dynfr1,dynfr2,dynfr3,dynfr4,dynfr5,dynfr6,dynfr7} and localization \cite{dynloc1,dynloc2,dynloc3,dynloc4,dynloc5,dynloc6,dynloc7}, drive induced topology \cite{topo1,topo2,topo3,topo4,topo5,topo6,topo7,topo8,topo9}, realization of Floquet scars \cite{scar1,scar2,scar3,scar4,scar5,scar6,scar7,scar8}, prethermal Hilbert space fragmentation \cite{hsf1,hsf2,hsf3,hsf4,hsf5,hsf7} and suppression of heating arising from the presence of Floquet flat bands \cite{tb1,kg1}. Some of these phenomena can be understood in terms of emergent, additional, symmetry of the leading term of the Floquet Hamiltonian of these driven systems, which control their dynamics in the prethermal regime. 

These theoretical studies have received experimental support from several platforms \cite{exp1,exp2,rev13,rev14}; a one-dimensional array of ultracold Rydberg atoms in an optical lattice constitutes an important example of such a platform \cite{exp3,exp4,exp5,exp6,exp7}. The physics of such chains is usually described in terms of the ground and a Rydberg excited state of the atoms at each site; their Hamiltonian can therefore be written in terms of the Pauli matrices $\sigma_j^{\alpha}$, where $\alpha=x,y,z$ \cite{exp3,exp5}. In this notation, the Rydberg excitation density $\hat n_j$ on a site $j$ is related to $\sigma_j^z$ by $\hat n_j = (1+ \sigma_j^z)/2$, while $\sigma_j^x$ provides a finite matrix element between the ground and Rydberg states of the atoms. Moreover, the atoms in their Rydberg excited state experience a strong van der Waals repulsion leading to the well-known Rydberg blockade \cite{exp3,exp5}; this interaction makes the Hamiltonian of these chains non-integrable. 

In this work we show that a periodically driven Rydberg atom chain, driven using either a two-tone or an asymmetric single-tone square pulse drive protocol, can exhibit signatures of Bethe integrability. This phenomenon, which constitutes the emergence of a extensive number of conserved charges, occurs in the large drive amplitude regime and at special drive frequencies which we identify. The signature of this integrability persists for an exponentially long (in drive amplitude) prethermal timescale making it experimentally relevant. We provide an analytic expression of the Floquet Hamiltonian in the large drive amplitude regime using Floquet perturbation theory (FPT). We identify special drive frequencies at which the leading term of this Floquet Hamiltonian exhibits Bethe integrability. This integrable nature is explicitly demonstrated by an exact mapping of the Floquet Hamiltonian to the Hamiltonian of the paradigmatic spin-$1/2$ ${\rm XXZ}$ chain. 

\begin{figure}
\rotatebox{0}{\includegraphics*[width=\linewidth]{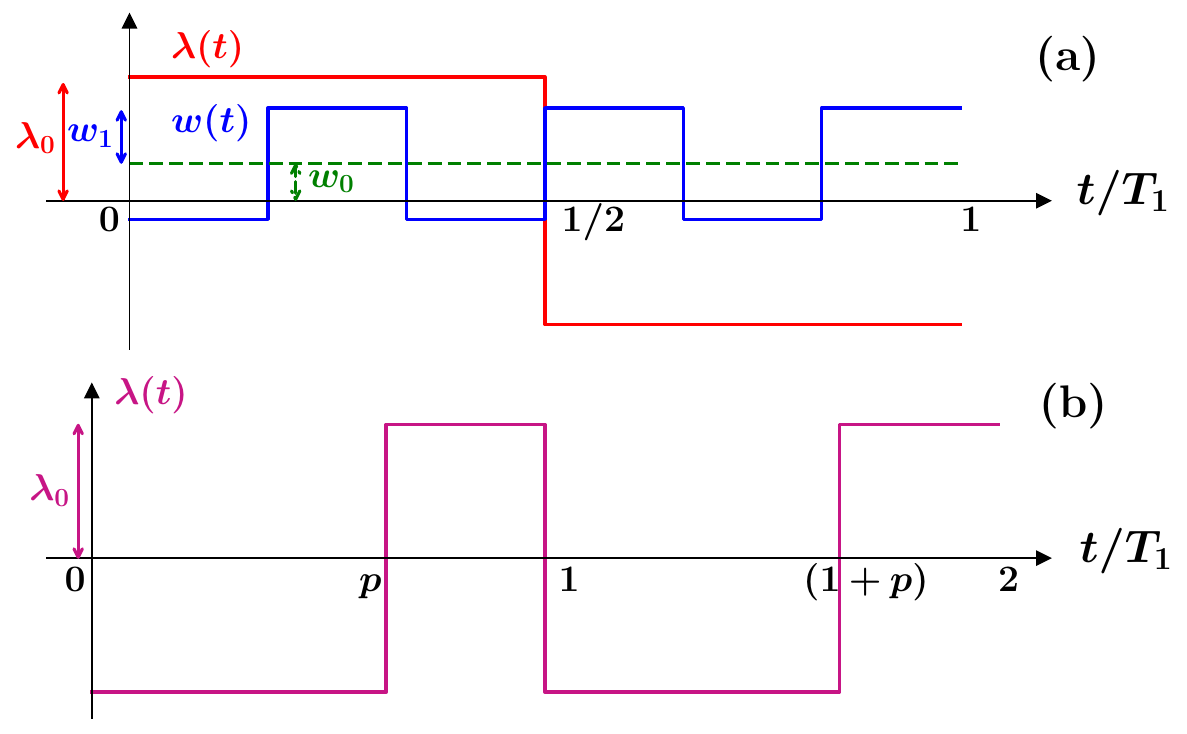}}
\caption{(a) Schematic representation of the two-tone square pulse drive protocol with $\lambda(t)= +(-) \lambda_0$ for $t\le (>) T_1/2$ and $w(t)= w_0 \pm w_1$ with time-period $T_1/3$ 
as shown. (b) Schematic representation of the single-tone asymmetric square pulse ($p\ne 1/2$) drive protocol where $\lambda(t)= -(+)\lambda_0$ for $t\le(>) pT_1$. \label{fig1}}
\end{figure}

Our analytical results are supported by numerical studies of the finite Rydberg chains driven at frequencies $\omega_1= 2\pi/T_1$ and $\omega_2= 3 \omega_1$, where $T_1$ is the time period of the drive (Fig.\ \ref{fig1}(a))~\cite{tb1}. Using exact-diagonalization (ED), we compute the level spacing ratio $r(T_1)$, half-chain entanglement entropy $S_{L/2}$, and the longitudinal magnetization $M^z(nT_1)= \langle S^z(nT_1) \rangle$ after $n$ drive cycles, where 
\begin{eqnarray}
S^z(nT_1)= U^{\dagger}(nT_1,0)\,\left( \sum_j \sigma^z_j /L \right) \,U(nT_1,0). \label{mzop}
\end{eqnarray} 

We find that $r(T_1)$ stays close to a value appropriate for ergodic driven systems for a generic drive frequency\cite{mr1,rev7}. In contrast, for the special drive frequencies identified from an analysis of the perturbative Floquet Hamiltonian, it dips close to $0.39$ consistent with Poisson statistics \cite{rev7}. A study of the level spacing distribution shows a change from Wigner-Dyson to Poisson statistics at these frequencies. The behavior of $S_{L/2}$ of the Floquet eigenstates also changes significantly at these frequencies. Concomitantly, $M^z$ remains constant up to a long time before their 
eventual decay due to thermalization; this timescale $\tau^{\ast}$ increases exponentially with the drive amplitude. This behavior is consistent with the integrable nature of the leading order Floquet Hamiltonian and is to be contrasted with the decay of $M^z$ at generic drive frequencies. Our analysis of the steady state value of $M^z$ allows us to identify a crossover at an intermediate drive amplitude above which the emergent integrability remains stable. 

{\it Model and drive protocol} : The Hamiltonian of a Rydberg chain, in the strong Rydberg interaction limit, can be written as 
\begin{eqnarray} 
H &=& \sum_j (w \tilde \sigma_j^x -\lambda \sigma_j^z), \label{ham1} 
\end{eqnarray} 
where $j$ denotes the lattice sites, $w$ is the coupling between the ground and the Rydberg excited state, and $\lambda$ denotes the detuning of the Rydberg atoms \cite{exp3}. Here $\tilde \sigma_j^{\alpha} = P_{j-1} \sigma_j^{\alpha} P_{j+1}$, where 
$P_j= (1-\sigma_j^z)/2$ projects a local spin to $\downarrow$. They ensure the absence of neighboring up-spins and encodes the effect of the van der Waals interaction between Rydberg excited atoms in the regime where the Rydberg blockade radius extends to neighboring sites \cite{exp3,exp5}. 

In what follows, we drive the Rydberg chain using a two-tone square pulse drive protocol as shown schematically in Fig.\ \ref{fig1}(a). The ratio of the drive frequencies $\omega_2/\omega_1=q$, where $q\in \mathbb{Z^+}$ for the shown protocol; for numerical results, we set $q=3$. An analogous two-tone cosine drive protocol is analyzed in the SM~\cite{si}. For $w_0=0$, the driven chain hosts exact Floquet flat bands for all $\omega_1$, $\lambda_0,$ and $w_1$ \cite{tb1,kg1}. Here we focus on a different regime where $w_0\ne 0$ and $\lambda_0 \gg w_0,w_1$. We note here that another asymmetric single-tone protocol, shown in Fig.\ \ref{fig1}(b), also yields similar emergent Bethe integrability for $p\ne 1/2$ as discussed in the SM~\cite{si}. 

\begin{figure}
\rotatebox{0}{\includegraphics*[width=\linewidth]{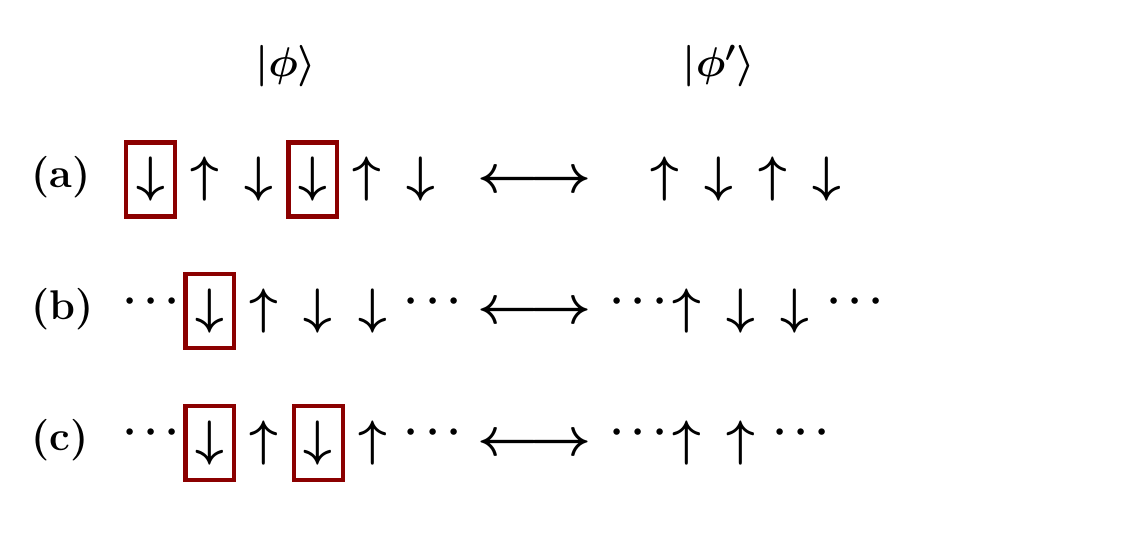}}
\caption{Schematic representation of the mapping between Fock states in the constrained Hilbert space of $H_F^{(2)}$ (denoted by $|\phi\rangle$) and those of $H_{\rm XXZ}$ (denoted by $|\phi'\rangle$). The mapping involves deleting a down-spin ($\downarrow$) marked by red squares on the left of any up-spin ($\uparrow$) of the PXP chain. (a) A state $\ket{\phi}$ with periodicity $R=3$ and a number of up-spins in a block $m_1=1$ is mapped to $\ket{\phi'}$ with $R'=2$ and the same $m_1$. (b) Example of a flippable up-spin in $|\phi\rangle$; this gets mapped to a $\uparrow \downarrow$ pair in $|\phi'\rangle$. (c) Example of a non-flippable up-spin in $|\phi\rangle$ which gets mapped to consecutive up-spins in $|\phi'\rangle$. See text for details. \label{fig2}}
\end{figure} 

For $\lambda_0 \gg w_0,w_1$, one can use FPT to obtain analytic, albeit perturbative, form of the Floquet Hamiltonian. To this end, we construct the zero-th order evolution operator $U_0(t,0) = \exp[i \alpha(t) \sum \sigma_j^z/\hbar] $, where $\alpha(t) = \lambda_0 t$ for $0\le t \le T_1/2$ and $\lambda_0(T_1-t)$ for $T_1/2 < t \le T_1$. We note that at this order $H_F^{(0)}= iT_1 U_0(T_1,0)/\hbar=0$. The higher order corrections to the Floquet Hamiltonian can be computed in a straightforward manner using standard perturbation theory \cite{rev12,tb1}. One obtains $H_F = H_F^{(1)} + H_F^{(2)} + \cdots$, where the ellipsis indicates higher-order terms in perturbation theory, and \cite{si} 
\begin{eqnarray} 
H_F^{(1)} &=& \frac{w_0 \, \sin{\gamma}}{\gamma} \, \sum_j \left( e^{i \gamma} \ \tilde{\sigma}_j^+ \ + \ \text{H.c.} \right), \label{flham1} \\
H_F^{(2)} &=& N(\gamma) \sum_j \left[
P_{j-2} (\sigma_{j-1}^+ \sigma_j^- + \sigma_{j-1}^- 
\sigma_j^+) P_{j+1} + \tilde{\sigma}_j^z \right],\nonumber
\end{eqnarray}
where $\gamma=\lambda_0 T_1/(2\hbar)$, $N(\gamma)= w_0 w_1 A(\gamma)/(3 \lambda_0)$ with $A(\gamma) = (3/(2\gamma)) [2 (\sin (2\gamma/3) - \sin (4\gamma/3) + \sin (2\gamma)] - 1$. We note that $H_F^{(1)}$ vanishes at $\gamma= m\pi$, where $m\in \mathbb{Z^+}$. At these frequencies $\omega_1=\omega_m^{\ast}= \lambda_0/(m \hbar)$, the properties of the driven chain is controlled by $H_F^{(2)}$. In what follows, we show that $H_F^{(2)}$ can be exactly mapped to the Hamiltonian of a spin-$1/2$ ${\rm XXZ}$ chain and therefore leads to 
emergent prethermal integrability \cite{intryd1,intryd2}.

{\it Exact mapping to ${\rm XXZ}$ spin chain} : We now demonstrate the integrable nature of $H_F^{(2)}$ for a chain of length $L$ with periodic boundary condition (PBC) and in the sector with total momentum $K=0$. We note that $H_F^{(2)}$ conserves the total number of up-spins, $N$, for any given Fock state. We then provide a map showing that all matrix elements of $H_F^{(2)}$ in the Fock basis are identical (up to a constant proportional to the identity) to those for a ${\rm XXZ}$ chain of length $L-N$ having a Hamiltonian 
\begin{eqnarray} 
H_{\rm XXZ}= J\sum_{j} [2 (\tau_j^+ \tau_{j+1}^- +{\rm H.c.}) + \Delta \tau_i^z \tau_{j+1}^z], \label{xxzham}
\end{eqnarray}
where $\vec \tau =(\tau^x,\tau^y,\tau^z)$ denote the Pauli matrices, and $\tau_j^{\pm}= (\tau_j^x\pm i \tau_j^y)/2$. The Fock states of the ${\rm XXZ}$ chain are constructed by deleting a spin-$\downarrow$ to the left of each spin-$\uparrow$ from the Fock states of the constrained Hilbert space of the PXP chain, as shown in Fig.\ \ref{fig2} \cite{prlspin}. We note at the outset that $H_{\rm XXZ}$ conserves $N$. 

For the mapping, we first consider the diagonal term of $H_F^{(2)}$. Keeping in mind the constraint of not having two consecutive up-spins, one can write $\sum_j \tilde \sigma_j^z =\sum_j [4(1-P_j)-(1- P_{j-1}) P_j (1- P_{j+1})]$. The first term in the sum is an irrelevant constant since $N$ is fixed. The second term 
contributes $-1$ only for a $\cdots \uparrow \downarrow \uparrow \cdots$ spin pattern, and yields zero otherwise. Since such a pattern gets mapped to consecutive up-spins in the ${\rm XXZ}$ chain (Fig.\ \ref{fig2}(c)), such terms can mapped to ${\mathcal T}_1= -\sum_{\langle ij\rangle} (1+\tau_i^z)(1+\tau_j^z)/4$. Since $H_{\rm XXZ}$ also conserves $N$, the part of ${\mathcal T}_1$ which is linear in $\tau^z$ leads to a constant. Ignoring this constant, we find that the map between two models mandates $\Delta= - N(m \pi)/(4J)$. 

For mapping the off-diagonal terms of $H_F^{(2)}$, it is necessary to examine the structure of Fock states in the $K=0$ sector which has $N_1$ up-spins. A generic Fock state $|\phi_{R_1}\rangle$ can have a periodicity of $R_1$; the cases for $R_1=3$ is shown in Fig.\ \ref{fig2}(a). Thus a normalized Fock state in the $K=0$ sector can be written as 
\begin{eqnarray} 
|\phi_1\rangle &=& \sum_{r=0}^{R_1-1} \frac{1}{\sqrt{R_{1}}} T_r |\phi_{R_{1}}\rangle, \label{statepxp1}
\end{eqnarray} 
where $q_1= L/R_{1}$ is the number of blocks, $R_{1}=L$ for states with no periodicity, and $T_r$ is the translation operator by $r$ sites. We note that if $m_1$ is the number of up-spins within each periodic block, one has $N_1= m_1 q_1$; for the state shown in Fig.\ \ref{fig2}(a), $m_1=1$ and $q_1=L/3$. The matrix element of the off-diagonal terms of $H_F^{(2)} \equiv H_2$ with another state $|\phi_2\rangle$ is then given by
\begin{eqnarray} 
\langle \phi_2|H_2|\phi_1\rangle &=& \frac{N(\gamma) m_0 R_{1}}{\sqrt{R_{1} R_{2}}} = N(\gamma) m_0 \sqrt{q_2/q_1}, \label{matpxp}
\end{eqnarray} 
where $m_0$ is the number of flippable up-spins in each block. Such flippable spins constitute 
a $\uparrow_j \downarrow_{j+1}$ pair at neighboring sites which can be flipped to $\downarrow_j \uparrow_{j+1}$; this requires that there are two $\downarrow$ spins at sites $j-1$ and $j+2$. One such configuration is shown in Fig.\ \ref{fig2}(b). In contrast, the non-fippable up-spins, shown schematically in Fig.\ \ref{fig2}(c), have a 
down-spin between two up-spins: $\uparrow_j \downarrow_{j+1} \uparrow_{j+2}$. 

Next we consider mapping $|\phi_1\rangle$ to the corresponding Fock state of the ${\rm XXZ}$ model. Since this involves deletion of a down-spin to the left of each up-spin, $m_1$ is conserved; the number of sites in the ${\rm XXZ}$ chain is $L-N_1= q_1(R_{1}-m_1)$. The periodicity of each of the blocks is now $R'_{1}= R_{1}-m_1$ 
(Fig.\ \ref{fig2}(a)) due to the deletion of a $\downarrow$ spin to the left of every $\uparrow$ spin within the block. The number of such periodic blocks remains invariant under such a mapping: $q'_1= (L-N_1)/R'_{1}=q_1$. A Fock state for the ${\rm XXZ}$ chain can be written as 
\begin{eqnarray} 
|\phi'_1\rangle &=& \sum_{r=0}^{R'_{1}-1} \frac{1}{\sqrt{R'_{1}}} T_r |\phi_{R'_{1}\rangle}. \label{statexxz1}
\end{eqnarray} 
 
 We note that non-flippable PXP Fock states, discussed above, leads to consecutive $\uparrow$ spins on neighboring sites (Fig.\ \ref{fig2}(c)); such states are annihilated by the $(\tau_j^{+}\tau_j^- +{\rm H.c.})$ term of the ${\rm XXZ}$ model. The flippable up-spins, in contrast, are mapped onto $\uparrow \downarrow$ combinations (Fig.\ \ref{fig2}(b)); the term $H'_2 = 2J \sum_j (\tau_j^{+}\tau_{j+1}^- +{\rm H.c.})$ counts the number of such pairs. The number of such pairs do not change under the mapping; for the state $|\phi'_1\rangle$ it is given by $m_0$. Thus the matrix element of $\langle \phi'_2|H'_2|\phi'_1\rangle$ is given by 
\begin{eqnarray} 
\langle \phi'_2|H'_2|\phi'_1\rangle &=& 2 J m_0 \sqrt{q'_1/q'_2} = 2 J m_0 \sqrt{q_1/q_2}. \label{matxxz}
\end{eqnarray} 
Comparing Eqs.~\eqref{matpxp} and \eqref{matxxz}, we find that choosing $J= N(m\pi)/2$, all the off-diagonal matrix elements of the two models are identical. This choice fixes $\Delta= -1/2$ and establishes the integrability of $H_F^{(2)}$ with PBC in the $K=0$ sector. In the
SM we present an alternative proof of the equivalence of 
the $H_F^{(2)}$ and $XXZ$ models using the Bethe ansatz for 
two particles in the $K=0$ sector~\cite{si}.

A related discussion for chains with open boundary condition (OBC) is presented in the SM~\cite{si}. We also note that whereas it is known that $H_F^{(2)}$ is integrable for PBC in $K\ne 0$ sectors \cite{intryd1,intryd2}, a mapping of it to an unconstrained integrable model such as the ${\rm XXZ}$ chain does not seem to exist. We note here that the net magnetization and the Hamiltonian (the first and the second charges) are easily obtained for both models. In the End Matter, we present explicit construction of the third conserved charge for $H_F^{(2)}$ and demonstrate its conservation for $\gamma=m\pi$.

{\it Numerical Results} : Our analytical results are supplemented by exact numerics using ED for $L\le 28$. For numerical results, we consider a square pulse protocol with $q=3$ for which one can write $H$ at any instant as $H[a,b]= \sum_j [(w_0+b w_1) \tilde \sigma_j^x - a \lambda_0 \sigma_j^z]$. The eigenvalues of $\epsilon_{a b}$ and eigenstates $|m_{ab}\rangle$ can be obtained using ED with PBC in the $K=0$ sector. One can then express $U(T_1,0)$ as a matrix using these eigenvalues and eigenvectors (see Ref.\ \cite{si} for details). Finally, one diagonalizes $U(T_1,0)$ using ED to obtain its eigenvalues $\Lambda_p= e^{i \theta_p(T_1;q)T_1/\hbar}$
and eigenvectors $|p\rangle$; the Floquet quasienergies can be then obtained as $E_F^p =\hbar \arccos({\rm Re}[\Lambda_p])/T_1$. The state of the driven system can also be numerically computed as $|\psi(n T_1)\rangle = \sum_p c_p e^{i n T_1 \theta_p} |p\rangle$, where $c_p=\langle p|\psi(0)\rangle$ and $|\psi(0)\rangle$ is the initial state.

\begin{figure}
\rotatebox{0}{\includegraphics*[width=0.49\linewidth]{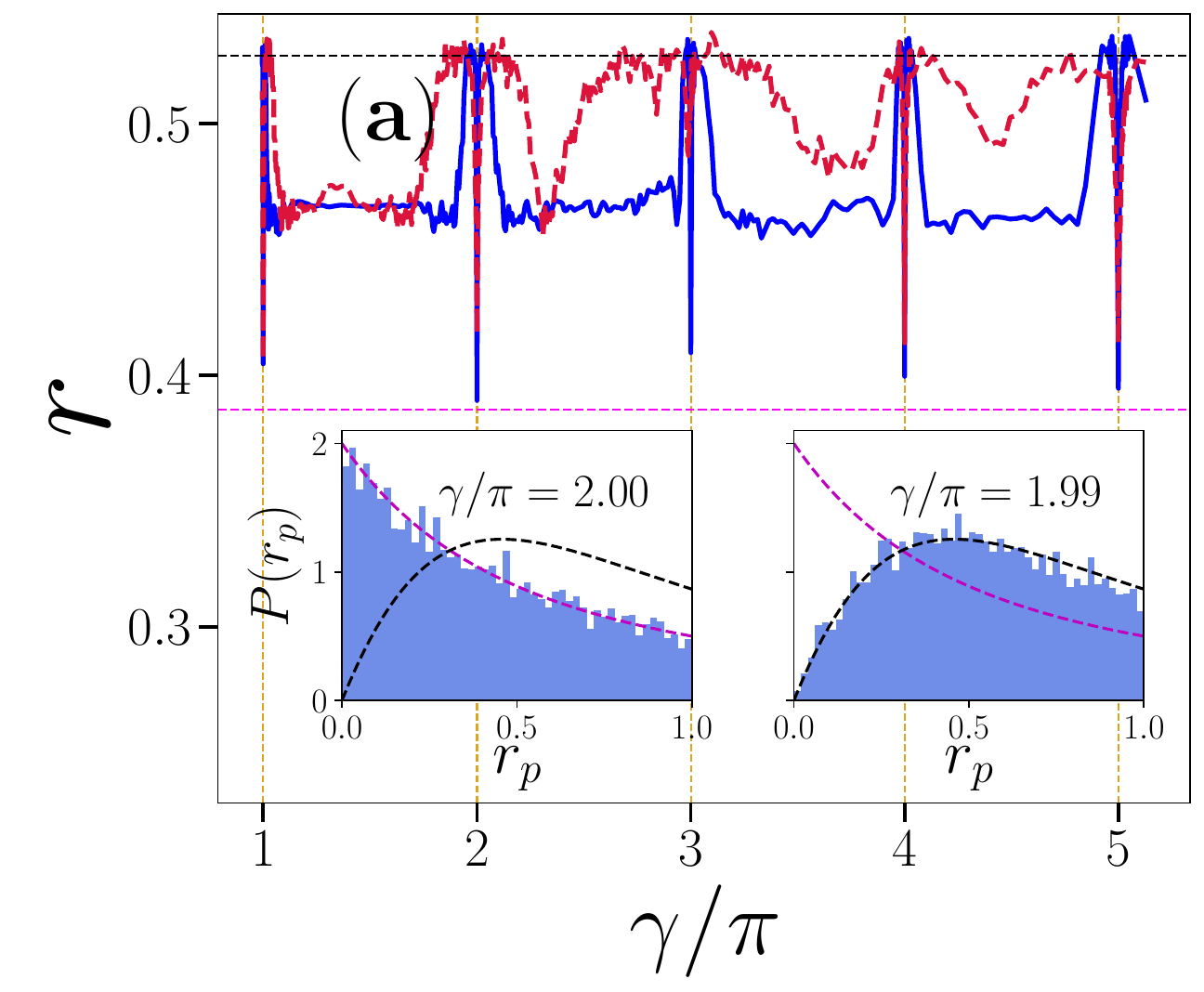}}
\rotatebox{0}{\includegraphics*[width=0.49\linewidth]{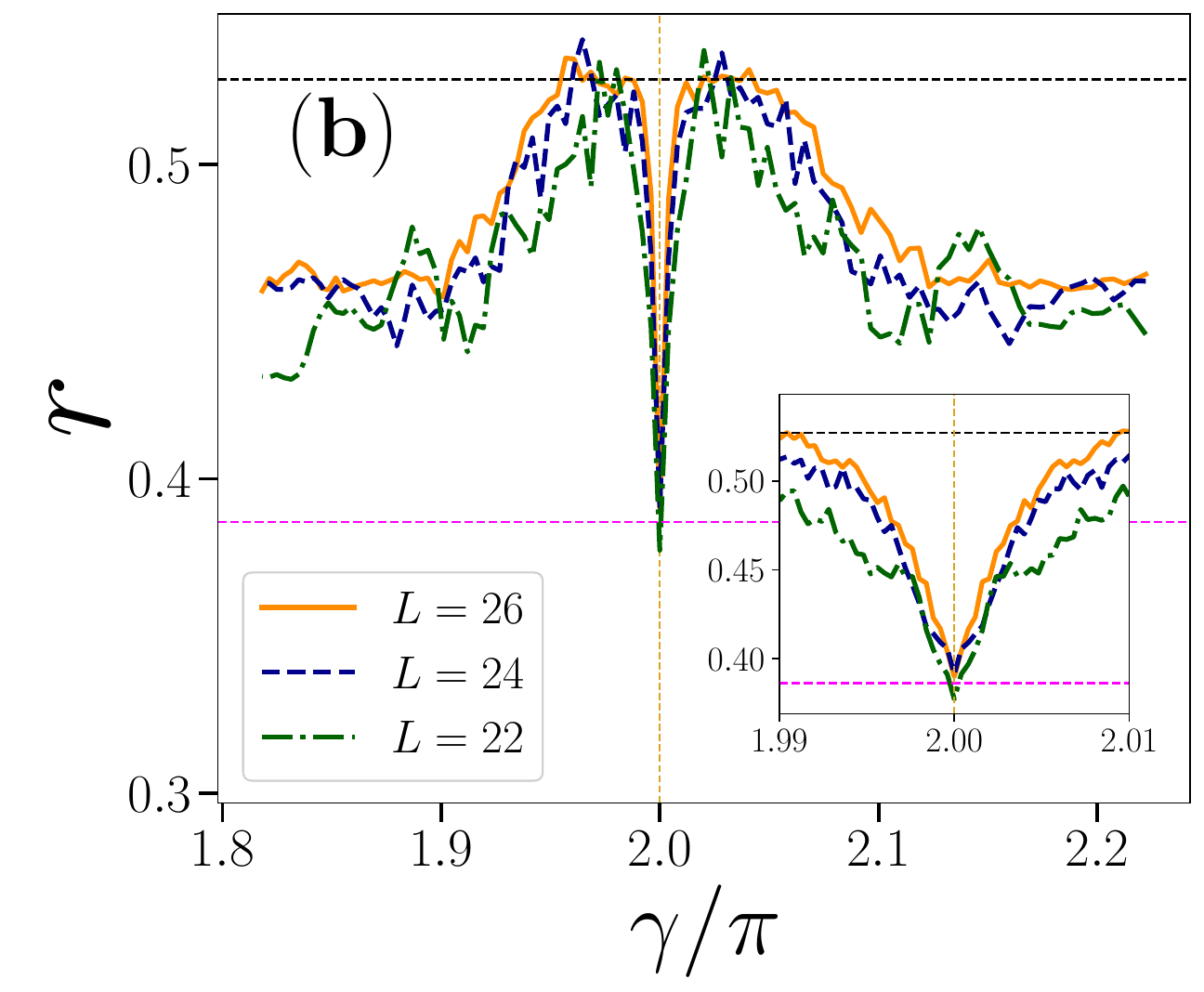}}
\rotatebox{0}{\includegraphics*[width=0.49\linewidth]{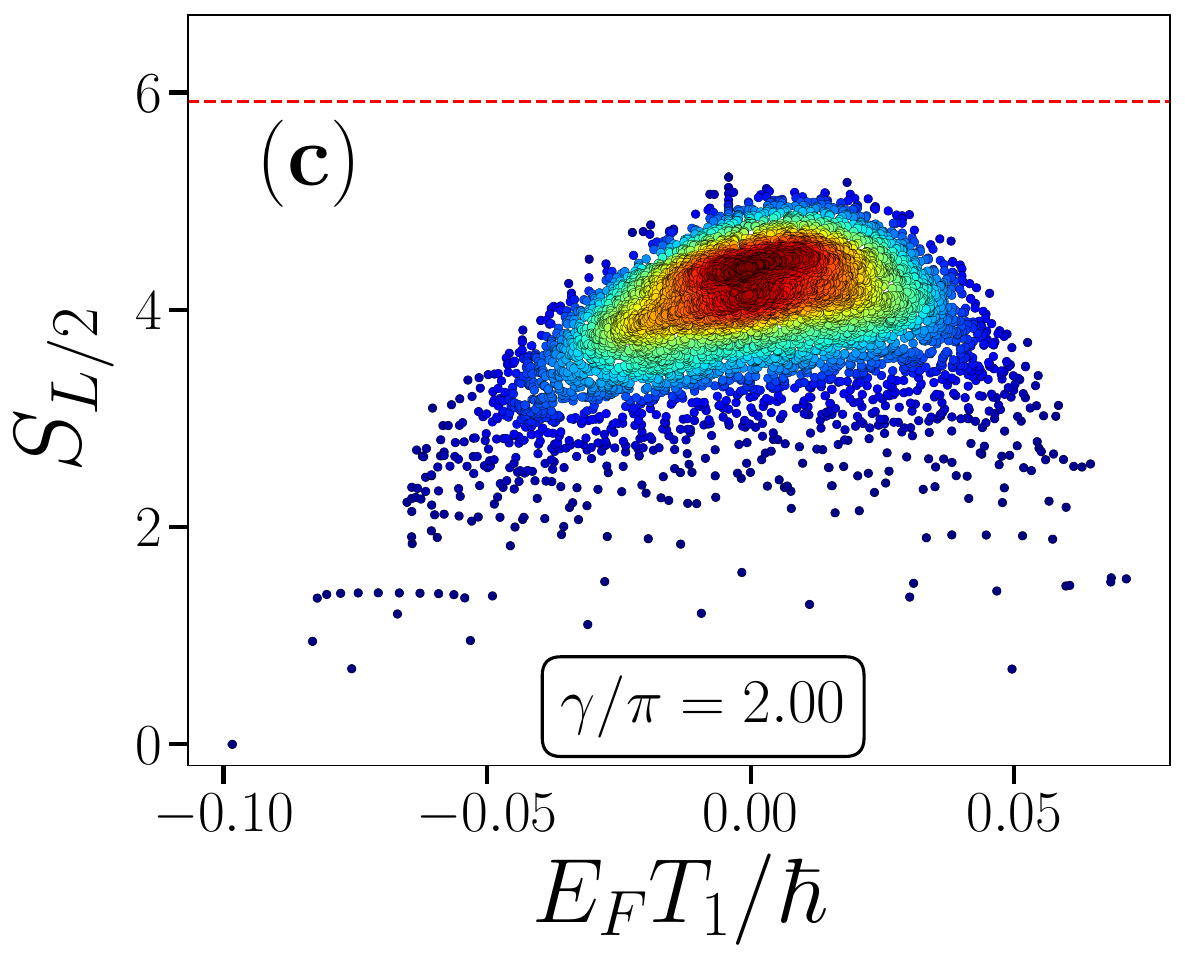}}
\rotatebox{0}{\includegraphics*[width=0.49\linewidth]{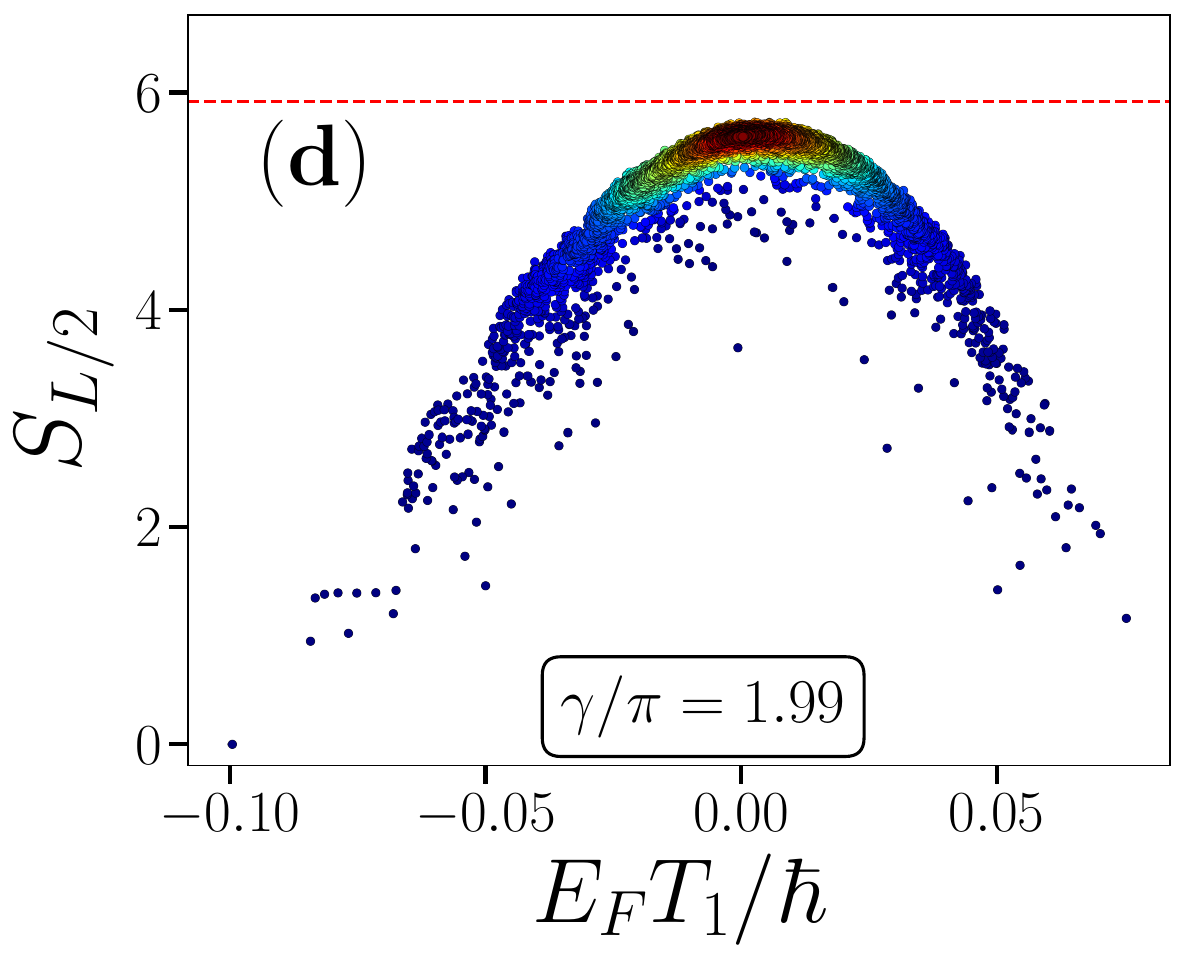}}
\caption{(a) Plot of $r$ as a function of $\gamma/\pi$ for $w_0/\lambda_0=0.05$, $L=26$, and $w_1/\lambda_0=0.05$ (blue) and $0.1$ (red), showing sharp dips at $\gamma/\pi=m$ for integer $m$ ($\omega_1=\omega_m^{\ast}$). The magenta (black) dashed lines $r=0.39 ~(0.527)$ correspond to values of $r$ for the Poisson ensemble (COE). The insets show $P(r_p)$ displaying Poisson (Wigner-Dyson) statistics at $\gamma=2 ~(1.99) \pi$. (b) Same as in (a) with $w_1/\lambda_0=0.05$ but around $\gamma/\pi=2$ and for different $L$. The inset shows a closeup of the sharp rise of $r$ from $0.39$ to $0.527$. (c) Plot of $S_{L/2}$ as a function of $E_F T_1/\hbar$ for Floquet quasienergies for $\gamma/\pi=2$, $L=26$, $w_0/\lambda_0=0.05$ and $w_1/\lambda_0=0.05$, displaying a wide spread. (d) Same as in (c) but for $\gamma/\pi=1.99$ showing a narrow thermal band. For all plots $\omega_2/\omega_1=q=3$. See text for details. \label{fig3}}
\end{figure} 

To test the emergent integrability of the driven system, we first compute the statistics of the Floquet quasienergies. A key marker of integrability is the statistics of ratios of consecutive level spacings defined as $r_p= {\rm Max}[s_p,s_{p+1}]/{\rm Min}[s_p,s_{p+1}]$, where $s_p = (E^F_{p+1}-E^F_p)$. We denote the distribution as $P(r_p)$ and $r$ as the average of $r_p$ \cite{huse1,mr1,gapstat1,rev7,si}. It is well known that $r=0.39 ~(0.53)$ for systems with Poisson ensemble (circular orthogonal ensemble ~(COE)). A plot of $r$ as a function of $\gamma/\pi$ shown in Fig.\ \ref{fig3}(a) 
displays clear dips towards $0.39$ for $\gamma/\pi=m$ where $H_F^{(2)}$ is the leading order term. A small deviation from this frequency shows a sharp rise in $r$ to close to its COE value. The insets of Fig.\ \ref{fig3}(a) show the concomitant change of $P(r_p)$ from Posissonian at $\gamma=2 \pi$ to Wigner-Dyson at $\gamma=1.99 \pi$. At these frequencies, $H_F^{(1)}$ (which has the form of an ergodic PXP Hamiltonian) and $H_F^{(2)}$ have the same magnitude and $H_F$ obeys ETH. Moving further away from the integrable point we find $||H_F^{(1)} || \gg ||H_F^{(2)}||$. Here $H_F \simeq H_F^{(1)}$ to leading order; $H_F^{(1)}$ is known to have strong finite-size effects leading to a lower value of $r$ \cite{abanin1}.

The bottom panels of Fig.\ \ref{fig3} shows the half-chain entanglement $S_{L/2}$ of the Floquet eigenstates (see Ref.\cite{si} for details) for $\gamma/\pi=2$ (panel (c)) and $1.99$ (panel (d)). Fig.\ \ref{fig3}(c) shows a wide spread of $S_{L/2}$ for mid-spectrum states. This is in sharp contrast to the ETH predicted behavior where $S_{L/2}$ for mid-spectrum states lie within a narrow band \cite{rev7,page1}. This behavior is seen in Fig.\ \ref{fig3}(d) even for a small departure from the integrable point; such a feature is a consequence of the sharp increase in the relative amplitudes of $H_F^{(1)}$ and $H_F^{(2)}$
due to such a departure.

\begin{figure}
\rotatebox{0}{\includegraphics*[width=0.49\linewidth]{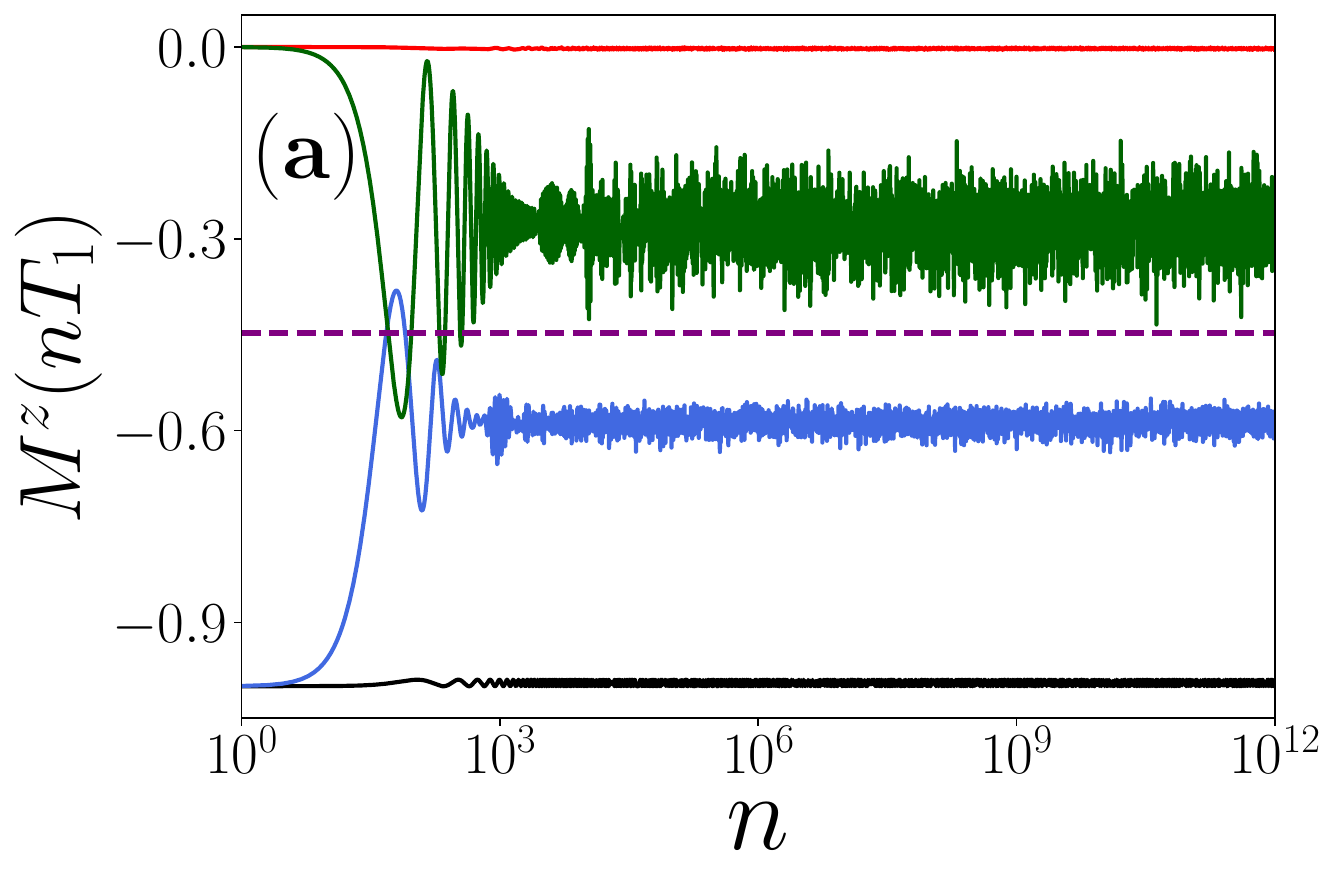}}
\rotatebox{0}{\includegraphics*[width=0.49\linewidth]{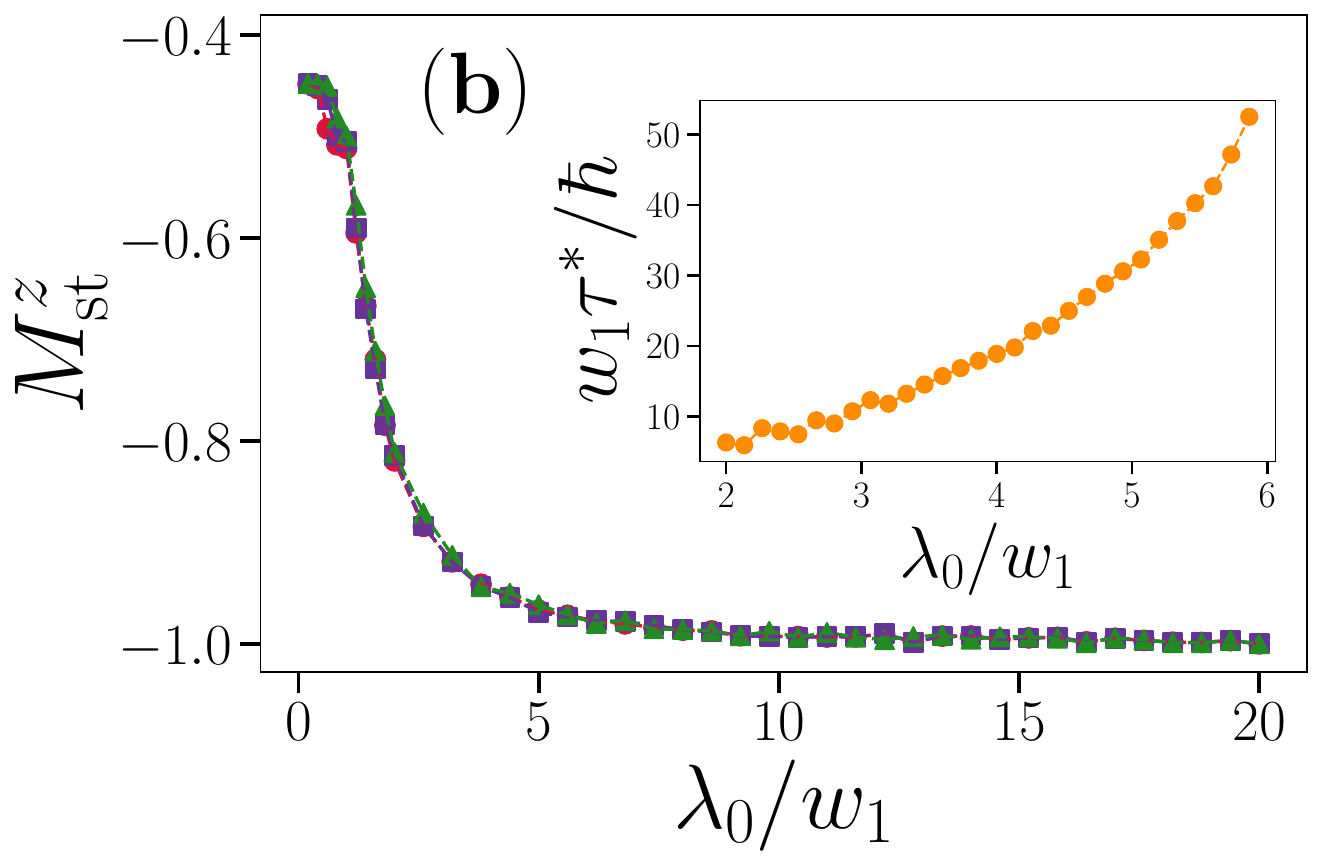}}
\caption{(a) Plot of $M^z(nT_1)$ as a function of $n$ starting from $|{\rm vac}\rangle$ for $\gamma/\pi=2$ (black) and $1.9$ (blue), and similar plots starting from an $|{\rm AFM}\rangle$ for $\gamma/\pi=2$ (red) and $1.9$ (green). The purple dotted line indicates the ITE average value of $M^z$. For all plots $\lambda_0/w_1=20$ and $L=28$. (b) Plot of $M_{\rm st}^z$ starting from $|{\rm vac}\rangle$ as a function of $\lambda_0/w_1$ for $L=24$ (red), $26$ (blue) and $28$ (green) at $\gamma/\pi=1$. The plots show a gradual crossover to the thermal phase where $M_{\rm st}^z \to -0.45$ for $\lambda_0/w_1 \le 1.0$. The inset shows a plot of $w_1 \tau^{\ast}/\hbar$ (where $\tau^{\ast}= n^{\ast}T_1$) as a function of $\lambda_0/w_1$ for a system size $L=26$, indicating exponential growth of $n^{\ast}$, the minimum number of drive cycles after which $M^z$ changes by $5\%$ of $M^z(0)=-1$. For all plots $w_0/w_1=1.0$ and $q=3$. See text for details.}
\label{fig4} \end{figure}

The dynamics of the driven chain also shows signature of such integrability. The simplest, experimentally testable, signature is seen via conservation of the total magnetization $M^z(nT_1)$ as shown in Fig.\ \ref{fig4}(a) starting from two simple initial Fock states $|{\rm vac}\rangle = |\cdots \downarrow \downarrow \downarrow \cdots\rangle$ and $|{\rm AFM}\rangle= (|Z_2\rangle + |\bar Z_2\rangle)/\sqrt{2}$ where $|Z_2\rangle= |\cdots \downarrow \uparrow \downarrow \uparrow \cdots \rangle$ and $|\bar Z_2\rangle$ is the time-reversed partner of $|Z_2\rangle$. We find that $M^z$ remain pinned to its initial value for $n\le 10^{12}$ cycles for $\gamma= m\pi$; a departure from these integrable points leads to saturation of $M^z$ to superthermal ($|{\rm AFM}\rangle$) or subthermal ($|{\rm vac}\rangle$) values. The deviation of these saturation values from their infinite temperature ensemble (ITE) averages have already been noted and explained in Refs.\ \cite{scar2,scar5}. 

The dependence of the steady state value of the magnetization $M^z_{\rm st}= (1/200)\sum_{n=n_0}^{n_0+200} M^z(nT_1)$, computed with $n_0=10^8$ is plotted for $\gamma=\pi$ as a function of $\lambda_0/w_1$. The plot (Fig.~\ref{fig4} (b)) shows a deviation of $M_{\rm st}^z$ at low $\lambda_0/w_1$ indicating destabilization of the prethermal integrability at lower drive amplitudes. The plot also indicates that this phenomenon does not show significant $L$ dependence. The inset of Fig.~\ref{fig4}(b) shows a plot of the time $\tau^{\ast} = n^{\ast} T_1$, after which $M^z$ changes by $5\%$ of its initial value, as a function of $\lambda_0/w_1$. We find that $\tau^{\ast}$ increases exponentially with $\lambda_0$ indicating a large stable prethermal regime at large $\lambda_0/w_1$. 

{\it Conclusion} : Our analysis of a driven Rydberg chain using a two-tone protocol demonstrates the presence of emergent Bethe integrability at special drive frequencies and in the large drive amplitude regime. We provide an exact mapping of the leading term of the Floquet Hamiltonian, obtained using FPT, to a spin-$1/2$ ${\rm XXZ}$ chain at these frequencies. We note that this phenomenon constitutes the emergence of an $O(L)$ number of approximately conserved charges and is therefore distinct from other prethermal phenomena which usually feature a single conserved charge \cite{rev17,dynloc3,dynfr1,dynfr6,dc2}. We provide an explicit construction of the third conserved charge and show its conservation in End Matter which demonstrates this point. The emergent integrability shows distinct signature in both the statistics of the Floquet quasienergies and in the magnetization dynamics of the driven chain. The conservation of magnetization at special drive frequencies can be experimentally verified in standard Rydberg atom platforms \cite{exp3,exp5} and can therefore serve as a test of the emergent Bethe integrability. 

{\it Acknowledgment} : KS thanks DST, India for support through the project JCB/2021/000030. D.S. thanks SERB, India for support through the project JBR/2920/000043. S.R. thanks Tista Banerjee for insightful discussions and acknowledges the support from UGC under the JRF Scheme.

\bibliography{cit}

@article{rev2,
  title = {Colloquium: Nonequilibrium dynamics of closed interacting quantum systems},
  author = {Polkovnikov, Anatoli and Sengupta, Krishnendu and Silva, Alessandro and Vengalattore, Mukund},
  journal = {Rev. Mod. Phys.},
  volume = {83},
  issue = {3},
  pages = {863--883},
  numpages = {0},
  year = {2011},
  month = {Aug},
  publisher = {American Physical Society},
  doi = {10.1103/RevModPhys.83.863},
  url = {https://link.aps.org/doi/10.1103/RevModPhys.83.863}
}

@article{rev5,
author = {Marin Bukov and Luca D'Alessio and Anatoli Polkovnikov},
title = {Universal high-frequency behavior of periodically driven systems: from dynamical stabilization to {F}loquet engineering},
journal = {Advances in Physics},
volume = {64},
number = {2},
pages = {139--226},
year = {2015},
publisher = {Taylor \& Francis},
doi = {10.1080/00018732.2015.1055918},
URL = {https://doi.org/10.1080/00018732.2015.1055918},
eprint = {https://doi.org/10.1080/00018732.2015.1055918}

}

@article{rev6,
title = {Many-body energy localization transition in periodically driven systems},
journal = {Annals of Physics},
volume = {333},
pages = {19-33},
year = {2013},
issn = {0003-4916},
doi = {https://doi.org/10.1016/j.aop.2013.02.011},
url = {https://www.sciencedirect.com/science/article/pii/S0003491613000389},
author = {Luca D’Alessio and Anatoli Polkovnikov}
}

@article{rev7,
author = {Luca D'Alessio and Yariv Kafri and Anatoli Polkovnikov and Marcos Rigol},
title = {From quantum chaos and eigenstate thermalization to statistical mechanics and thermodynamics},
journal = {Advances in Physics},
volume = {65},
number = {3},
pages = {239--362},
year = {2016},
publisher = {Taylor \& Francis},
doi = {10.1080/00018732.2016.1198134},
URL = {https://doi.org/10.1080/00018732.2016.1198134},
eprint = {https://doi.org/10.1080/00018732.2016.1198134}

}

@article{rev8,
  title = {Landau–{Z}ener–{S}t\"{u}ckelberg interferometry},
  volume = {492},
  ISSN = {0370-1573},
  url = {http://dx.doi.org/10.1016/j.physrep.2010.03.002},
  DOI = {10.1016/j.physrep.2010.03.002},
  number = {1},
  journal = {Physics Reports},
  publisher = {Elsevier BV},
  author = {Shevchenko,  S.N. and Ashhab,  S. and Nori,  Franco},
  year = {2010},
  month = jul,
  pages = {1-30}
}

@article{rev9,
  title = {Floquet Engineering of Quantum Materials},
  volume = {10},
  ISSN = {1947-5462},
  url = {http://dx.doi.org/10.1146/annurev-conmatphys-031218-013423},
  DOI = {10.1146/annurev-conmatphys-031218-013423},
  number = {1},
  journal = {Annual Review of Condensed Matter Physics},
  publisher = {Annual Reviews},
  author = {Oka,  Takashi and Kitamura,  Sota},
  year = {2019},
  month = mar,
  pages = {387–408}
}

@article{rev10,
  title = {The {M}agnus expansion and some of its applications},
  volume = {470},
  ISSN = {0370-1573},
  url = {http://dx.doi.org/10.1016/j.physrep.2008.11.001},
  DOI = {10.1016/j.physrep.2008.11.001},
  number = {5–6},
  journal = {Physics Reports},
  publisher = {Elsevier BV},
  author = {Blanes,  S. and Casas,  F. and Oteo,  J.A. and Ros,  J.},
  year = {2009},
  month = jan,
  pages = {151–238}
}

@article{rev11,
  title = {Colloquium: Atomic quantum gases in periodically driven optical lattices},
  author = {Eckardt, Andr\'e},
  journal = {Rev. Mod. Phys.},
  volume = {89},
  issue = {1},
  pages = {011004},
  numpages = {30},
  year = {2017},
  month = {Mar},
  publisher = {American Physical Society},
  doi = {10.1103/RevModPhys.89.011004},
  url = {https://link.aps.org/doi/10.1103/RevModPhys.89.011004}
}

@article{rev12,
doi = {10.1088/1361-648X/ac1b61},
url = {https://dx.doi.org/10.1088/1361-648X/ac1b61},
year = {2021},
month = {aug},
publisher = {IOP Publishing},
volume = {33},
number = {44},
pages = {443003},
author = {Sen, Arnab and Sen, Diptiman and Sengupta, K},
title = {Analytic approaches to periodically driven closed quantum systems: methods and applications},
journal = {Journal of Physics: Condensed Matter}
}

@article{rev13,
  title = {Many-body physics with ultracold gases},
  author = {Bloch, Immanuel and Dalibard, Jean and Zwerger, Wilhelm},
  journal = {Rev. Mod. Phys.},
  volume = {80},
  issue = {3},
  pages = {885--964},
  numpages = {0},
  year = {2008},
  month = {Jul},
  publisher = {American Physical Society},
  doi = {10.1103/RevModPhys.80.885},
  url = {https://link.aps.org/doi/10.1103/RevModPhys.80.885}
}

@article{rev14,
  title = {Quantum simulation of the {H}ubbard model with ultracold fermions in optical lattices},
  volume = {19},
  ISSN = {1878-1535},
  url = {http://dx.doi.org/10.1016/j.crhy.2018.10.013},
  DOI = {10.1016/j.crhy.2018.10.013},
  number = {6},
  journal = {Comptes Rendus. Physique},
  publisher = {Cellule MathDoc/Centre Mersenne},
  author = {Tarruell,  Leticia and Sanchez-Palencia,  Laurent},
  year = {2018},
  month = sep,
  pages = {365–393}
}

@article{rev15,
  title = {Quantum and classical Floquet prethermalization},
  volume = {454},
  ISSN = {0003-4916},
  url = {http://dx.doi.org/10.1016/j.aop.2023.169297},
  DOI = {10.1016/j.aop.2023.169297},
  journal = {Annals of Physics},
  publisher = {Elsevier BV},
  author = {Ho,  Wen Wei and Mori,  Takashi and Abanin,  Dmitry A. and Dalla Torre,  Emanuele G.},
  year = {2023},
  month = jul,
  pages = {169297}
}

@article{rev16,
  title = {Thermalization and prethermalization in isolated quantum systems: a theoretical overview},
  volume = {51},
  ISSN = {1361-6455},
  url = {http://dx.doi.org/10.1088/1361-6455/aabcdf},
  DOI = {10.1088/1361-6455/aabcdf},
  number = {11},
  journal = {Journal of Physics B},
  publisher = {IOP Publishing},
  author = {Mori,  Takashi and Ikeda,  Tatsuhiko N and Kaminishi,  Eriko and Ueda,  Masahito},
  year = {2018},
  month = may,
  pages = {112001}
}

@article{rev17,
  title = {Emergent symmetries in prethermal phases of periodically driven quantum systems},
  volume = {37},
  ISSN = {1361-648X},
  url = {http://dx.doi.org/10.1088/1361-648X/ada860},
  DOI = {10.1088/1361-648x/ada860},
  number = {13},
  journal = {Journal of Physics: Condensed Matter},
  publisher = {IOP Publishing},
  author = {Banerjee,  Tista and Sengupta,  K},
  year = {2025},
  month = feb,
  pages = {133002}
}

@article{exp1,
  title = {A quantum gas microscope for detecting single atoms in a {H}ubbard-regime optical lattice},
  volume = {462},
  ISSN = {1476-4687},
  url = {http://dx.doi.org/10.1038/nature08482},
  DOI = {10.1038/nature08482},
  number = {7269},
  journal = {Nature},
  publisher = {Springer Science and Business Media LLC},
  author = {Bakr,  Waseem S. and Gillen,  Jonathon I. and Peng,  Amy and F\"{o}lling,  Simon and Greiner,  Markus},
  year = {2009},
  month = nov,
  pages = {74}
}

@article{exp2,
  title = {Probing the Superfluid–to–{M}ott Insulator Transition at the Single-Atom Level},
  volume = {329},
  ISSN = {1095-9203},
  url = {http://dx.doi.org/10.1126/science.1192368},
  DOI = {10.1126/science.1192368},
  number = {5991},
  journal = {Science},
  publisher = {American Association for the Advancement of Science (AAAS)},
  author = {Bakr,  W. S. and Peng,  A. and Tai,  M. E. and Ma,  R. and Simon,  J. and Gillen,  J. I. and F\"{o}lling,  S. and Pollet,  L. and Greiner,  M.},
  year = {2010},
  month = jul,
  pages = {547}
}

@article{exp3,
  title = {Probing many-body dynamics on a 51-atom quantum simulator},
  volume = {551},
  ISSN = {1476-4687},
  url = {http://dx.doi.org/10.1038/nature24622},
  DOI = {10.1038/nature24622},
  number = {7682},
  journal = {Nature},
  publisher = {Springer Science and Business Media LLC},
  author = {Bernien,  Hannes and Schwartz,  Sylvain and Keesling,  Alexander and Levine,  Harry and Omran,  Ahmed and Pichler,  Hannes and Choi,  Soonwon and Zibrov,  Alexander S. and Endres,  Manuel and Greiner,  Markus and Vuletić,  Vladan and Lukin,  Mikhail D.},
  year = {2017},
  month = nov,
  pages = {579}
}

@article{exp4,
  title = {High-Fidelity Control and Entanglement of {R}ydberg-Atom Qubits},
  author = {Levine, Harry and Keesling, Alexander and Omran, Ahmed and Bernien, Hannes and Schwartz, Sylvain and Zibrov, Alexander S. and Endres, Manuel and Greiner, Markus and Vuleti\ifmmode \acute{c}\else \'{c}\fi{}, Vladan and Lukin, Mikhail D.},
  journal = {Phys. Rev. Lett.},
  volume = {121},
  issue = {12},
  pages = {123603},
  numpages = {6},
  year = {2018},
  month = {Sep},
  publisher = {American Physical Society},
  doi = {10.1103/PhysRevLett.121.123603},
  url = {https://link.aps.org/doi/10.1103/PhysRevLett.121.123603}
}

@article{exp5,
  title = {Controlling quantum many-body dynamics in driven {R}ydberg atom arrays},
  author = {D. Bluvstein  and A. Omran  and H. Levine  and A. Keesling  and G. Semeghini  and S. Ebadi  and T. T. Wang  and A. A. Michailidis  and N. Maskara  and W. W. Ho  and S. Choi  and M. Serbyn  and M. Greiner  and V. Vuletić  and M. D. Lukin },
  journal = {Science},
  volume = {371},
  number = {6536},
  pages = {1355-1359},
  year = {2021},
  doi = {10.1126/science.abg2530},
  URL = {https://www.science.org/doi/abs/10.1126/science.abg2530}
}

@article{exp6,
  title={Quantum coarsening and collective dynamics on a programmable simulator},
  author={Manovitz, Tom and Li, Sophie H and Ebadi, Sepehr and Samajdar, Rhine and Geim, Alexandra A and Evered, Simon J and Bluvstein, Dolev and Zhou, Hengyun and Koyluoglu, Nazli Ugur and Feldmeier, Johannes and Dolgirev, Pavel E and Maskara, Nishad and Kalinowski, Marcin and  Sachdev, Subir and Huse, David A and Greiner, Markus and Vuletić, Vladan and Lukin, Mikhail D},
  journal={Nature},
  volume={638},
  number={8049},
  pages={86--92},
  year={2025},
  publisher={Nature Publishing Group UK London},
   doi = {10.1103/w1cp-l5vq},
  url = {https://www.nature.com/articles/s41586-024-08353-5}
}

@article{exp7,
  title = {Observation of Anomalous Information Scrambling in a {R}ydberg Atom Array},
  author = {Liang, Xinhui and Yue, Zongpei and Chao, Yu-Xin and Hua, Zhen-Xing and Lin, Yige and Tey, Meng Khoon and You, Li},
  journal = {Phys. Rev. Lett.},
  volume = {135},
  issue = {5},
  pages = {050201},
  numpages = {7},
  year = {2025},
  month = {Jul},
  publisher = {American Physical Society},
  doi = {10.1103/w1cp-l5vq},
}

@article{dynloc1,
  title = {From dynamical localization to bunching in interacting Floquet systems},
  volume = {5},
  ISSN = {2542-4653},
  url = {http://dx.doi.org/10.21468/SciPostPhys.5.2.017},
  DOI = {10.21468/scipostphys.5.2.017},
  number = {2},
  journal = {SciPost Physics},
  publisher = {Stichting SciPost},
  author = {Baum,  Yuval and van Nieuwenburg,  Everard and Refael,  Gil},
  year = {2018},
  month = aug 
}

@article{dynloc2,
  title = {Absence of dynamical localization in interacting driven systems},
  volume = {3},
  ISSN = {2542-4653},
  url = {http://dx.doi.org/10.21468/SciPostPhys.3.4.029},
  DOI = {10.21468/scipostphys.3.4.029},
  number = {4},
  journal = {SciPost Physics},
  publisher = {Stichting SciPost},
  author = {Luitz,  David J. and Bar Lev,  Yevgeny and Lazarides,  Achilleas},
  year = {2017},
  month = oct 
}

@article{dynloc3,
  title = {Effects of interactions on periodically driven dynamically localized systems},
  volume = {95},
  ISSN = {2469-9969},
  url = {http://dx.doi.org/10.1103/PhysRevB.95.014305},
  DOI = {10.1103/physrevb.95.014305},
  number = {1},
  pages={014305},
  journal = {Phys. Rev. B},
  publisher = {American Physical Society (APS)},
  author = {Agarwala,  Adhip and Sen,  Diptiman},
  year = {2017},
  month = jan 
}

@Article{dynloc4,
	title={{Dynamical localization and slow thermalization in a class of disorder-free periodically driven one-dimensional interacting systems}},
	author={Sreemayee Aditya and Diptiman Sen},
	journal={SciPost Phys. Core},
	volume={6},
	pages={083},
	year={2023},
	publisher={SciPost},
	doi={10.21468/SciPostPhysCore.6.4.083},
	url={https://scipost.org/10.21468/SciPostPhysCore.6.4.083},
}

@article{dynloc5,
  title = {Dynamical localization in a chain of hard core bosons under periodic driving},
  author = {Nag, Tanay and Roy, Sthitadhi and Dutta, Amit and Sen, Diptiman},
  journal = {Phys. Rev. B},
  volume = {89},
  issue = {16},
  pages = {165425},
  numpages = {5},
  year = {2014},
  month = {Apr},
  publisher = {American Physical Society},
  doi = {10.1103/PhysRevB.89.165425},
  url = {https://link.aps.org/doi/10.1103/PhysRevB.89.165425}
}

@article{dynloc6,
  title = {Floquet engineering of low-energy dispersions and dynamical localization in a periodically kicked three-band system},
  volume = {104},
  ISSN = {2469-9969},
  url = {http://dx.doi.org/10.1103/PhysRevB.104.174308},
  DOI = {10.1103/physrevb.104.174308},
  number = {17},
  pages={174308},
  journal = {Phys. Rev. B},
  publisher = {American Physical Society (APS)},
  author = {Tamang,  Lakpa and Nag,  Tanay and Biswas,  Tutul},
  year = {2021},
  month = nov 
}

@article{dynloc7,
  title = {Many-body dynamical localization in the kicked {B}ose-{H}ubbard chain},
  volume = {101},
  ISSN = {2469-9969},
  url = {http://dx.doi.org/10.1103/PhysRevB.101.064302},
  DOI = {10.1103/physrevb.101.064302},
  pages={064302},
  number = {6},
  journal = {Phys. Rev. B},
  publisher = {American Physical Society (APS)},
  author = {Fava,  Michele and Fazio,  Rosario and Russomanno,  Angelo},
  year = {2020},
  month = feb 
}

@article{dynfr1,
  title = {Exotic freezing of response in a quantum many-body system},
  author = {Das, Arnab},
  journal = {Phys. Rev. B},
  volume = {82},
  issue = {17},
  pages = {172402},
  numpages = {4},
  year = {2010},
  month = {Nov},
  publisher = {American Physical Society},
  doi = {10.1103/PhysRevB.82.172402},
  url = {https://link.aps.org/doi/10.1103/PhysRevB.82.172402}
}

@article{dynfr2,
  title = {Freezing a quantum magnet by repeated quantum interference: An experimental realization},
  author = {Hegde, Swathi S. and Katiyar, Hemant and Mahesh, T. S. and Das, Arnab},
  journal = {Phys. Rev. B},
  volume = {90},
  issue = {17},
  pages = {174407},
  numpages = {7},
  year = {2014},
  month = {Nov},
  publisher = {American Physical Society},
  doi = {10.1103/PhysRevB.90.174407},
  url = {https://link.aps.org/doi/10.1103/PhysRevB.90.174407}
}

@article{dynfr3,
  title = {Dynamics-induced freezing of strongly correlated ultracold bosons},
  volume = {100},
  ISSN = {1286-4854},
  url = {http://dx.doi.org/10.1209/0295-5075/100/60007},
  DOI = {10.1209/0295-5075/100/60007},
  number = {6},
  journal = {EPL (Europhysics Letters)},
  publisher = {IOP Publishing},
  author = {Mondal,  S. and Pekker,  D. and Sengupta,  K.},
  year = {2012},
  month = dec,
  pages = {60007}
}

@article{dynfr4,
  title = {Dynamical Freezing in Exactly Solvable Models of Driven Chaotic Quantum Dots},
  volume = {134},
  pages={226501},
  ISSN = {1079-7114},
  url = {http://dx.doi.org/10.1103/ggk3-6cf8},
  DOI = {10.1103/ggk3-6cf8},
  number = {22},
  journal = {Phys. Rev. Lett.},
  publisher = {American Physical Society (APS)},
  author = {Guo,  Haoyu and Mukherjee,  Rohit and Chowdhury,  Debanjan},
  year = {2025},
  month = jun 
}

@article{dynfr5,
  title = {Dynamic freezing and defect suppression in the tilted one-dimensional {B}ose-{H}ubbard model},
  volume = {90},
  ISSN = {1550-235X},
  url = {http://dx.doi.org/10.1103/PhysRevB.90.184303},
  pages={184303},
  DOI = {10.1103/physrevb.90.184303},
  number = {18},
  journal = {Phys. Rev. B},
  publisher = {American Physical Society (APS)},
  author = {Divakaran,  U. and Sengupta,  K.},
  year = {2014},
  month = nov 
}

@article{dynfr6,
  title = {Dynamical Freezing and Scar Points in Strongly Driven {F}loquet Matter: Resonance vs Emergent Conservation Laws},
  author = {Haldar, Asmi and Sen, Diptiman and Moessner, Roderich and Das, Arnab},
  journal = {Phys. Rev. X},
  volume = {11},
  issue = {2},
  pages = {021008},
  numpages = {25},
  year = {2021},
  month = {Apr},
  publisher = {American Physical Society},
  doi = {10.1103/PhysRevX.11.021008},
  url = {https://link.aps.org/doi/10.1103/PhysRevX.11.021008}
}

@article{dynfr7,
  title = {Counterdiabatic Route to Entanglement Steering and Dynamical Freezing in the {F}loquet {L}ipkin-{M}eshkov-{G}lick Model},
  author = {Gangopadhay, Nakshatra and Choudhury, Sayan},
  journal = {Phys. Rev. Lett.},
  volume = {135},
  issue = {2},
  pages = {020407},
  numpages = {7},
  year = {2025},
  month = {Jul},
  publisher = {American Physical Society},
  doi = {10.1103/bzcf-gm89},
  url = {https://link.aps.org/doi/10.1103/bzcf-gm89}
}

@article{topo1,
  title = {Topological characterization of periodically driven quantum systems},
  author = {Kitagawa, Takuya and Berg, Erez and Rudner, Mark and Demler, Eugene},
  journal = {Phys. Rev. B},
  volume = {82},
  issue = {23},
  pages = {235114},
  numpages = {12},
  year = {2010},
  month = {Dec},
  publisher = {American Physical Society},
  doi = {10.1103/PhysRevB.82.235114},
  url = {https://link.aps.org/doi/10.1103/PhysRevB.82.235114}
}

@article{topo2,
  title = {{F}loquet topological insulator in semiconductor quantum wells},
  volume = {7},
  ISSN = {1745-2481},
  url = {http://dx.doi.org/10.1038/nphys1926},
  DOI = {10.1038/nphys1926},
  number = {6},
  journal = {Nature Physics},
  publisher = {Springer Science and Business Media LLC},
  author = {Lindner,  Netanel H. and Refael,  Gil and Galitski,  Victor},
  year = {2011},
  month = mar,
  pages = {490}
}

@article{topo3,
  title = {Transport properties of nonequilibrium systems under the application of light: Photoinduced quantum Hall insulators without Landau levels},
  author = {Kitagawa, Takuya and Oka, Takashi and Brataas, Arne and Fu, Liang and Demler, Eugene},
  journal = {Phys. Rev. B},
  volume = {84},
  issue = {23},
  pages = {235108},
  numpages = {13},
  year = {2011},
  month = {Dec},
  publisher = {American Physical Society},
  doi = {10.1103/PhysRevB.84.235108},
  url = {https://link.aps.org/doi/10.1103/PhysRevB.84.235108}
}

@article{topo4,
  title = {{F}loquet generation of {M}ajorana end modes and topological invariants},
  author = {Thakurathi, Manisha and Patel, Aavishkar A. and Sen, Diptiman and Dutta, Amit},
  journal = {Phys. Rev. B},
  volume = {88},
  issue = {15},
  pages = {155133},
  numpages = {13},
  year = {2013},
  month = {Oct},
  publisher = {American Physical Society},
  doi = {10.1103/PhysRevB.88.155133},
  url = {https://link.aps.org/doi/10.1103/PhysRevB.88.155133}
}

@article{topo5,
  title = {Effective Theory of {F}loquet Topological Transitions},
  author = {Kundu, Arijit and Fertig, H. A. and Seradjeh, Babak},
  journal = {Phys. Rev. Lett.},
  volume = {113},
  issue = {23},
  pages = {236803},
  numpages = {5},
  year = {2014},
  month = {Dec},
  publisher = {American Physical Society},
  doi = {10.1103/PhysRevLett.113.236803},
  url = {https://link.aps.org/doi/10.1103/PhysRevLett.113.236803}
}

@article{topo6,
doi = {10.1088/1367-2630/17/12/125014},
url = {https://dx.doi.org/10.1088/1367-2630/17/12/125014},
year = {2015},
month = {dec},
publisher = {IOP Publishing},
volume = {17},
number = {12},
pages = {125014},
author = {Nathan, Frederik and Rudner, Mark S},
title = {Topological singularities and the general classification of {F}loquet–{B}loch systems},
journal = {New Journal of Physics}
}

@article{topo7,
  title = {Signatures and conditions for phase band crossings in periodically driven integrable systems},
  author = {Mukherjee, Bhaskar and Sen, Arnab and Sen, Diptiman and Sengupta, K.},
  journal = {Phys. Rev. B},
  volume = {94},
  issue = {15},
  pages = {155122},
  numpages = {12},
  year = {2016},
  month = {Oct},
  publisher = {American Physical Society},
  doi = {10.1103/PhysRevB.94.155122},
  url = {https://link.aps.org/doi/10.1103/PhysRevB.94.155122}
}

@article{topo8,
  title = {Low-frequency phase diagram of irradiated graphene and a periodically driven spin-$\frac{1}{2}$ {XY} chain},
  author = {Mukherjee, Bhaskar and Mohan, Priyanka and Sen, Diptiman and Sengupta, K.},
  journal = {Phys. Rev. B},
  volume = {97},
  issue = {20},
  pages = {205415},
  numpages = {23},
  year = {2018},
  month = {May},
  publisher = {American Physical Society},
  doi = {10.1103/PhysRevB.97.205415},
  url = {https://link.aps.org/doi/10.1103/PhysRevB.97.205415}
}

@article{topo9,
  title = {{F}loquet topological transition by unpolarized light},
  volume = {98},
  ISSN = {2469-9969},
  url = {http://dx.doi.org/10.1103/PhysRevB.98.235112},
  DOI = {10.1103/physrevb.98.235112},
  number = {23},
  pages={235112},
  journal = {Phys. Rev. B},
  publisher = {American Physical Society (APS)},
  author = {Mukherjee,  Bhaskar},
  year = {2018},
  month = dec 
}

@article{scar1,
  title = {Dynamical Scar States in Driven Fracton Systems},
  author = {Pai, Shriya and Pretko, Michael},
  journal = {Phys. Rev. Lett.},
  volume = {123},
  issue = {13},
  pages = {136401},
  numpages = {5},
  year = {2019},
  month = {Sep},
  publisher = {American Physical Society},
  doi = {10.1103/PhysRevLett.123.136401},
  url = {https://link.aps.org/doi/10.1103/PhysRevLett.123.136401}
}

@article{scar2,
  title = {Collapse and revival of quantum many-body scars via {F}loquet engineering},
  author = {Mukherjee, Bhaskar and Nandy, Sourav and Sen, Arnab and Sen, Diptiman and Sengupta, K.},
  journal = {Phys. Rev. B},
  volume = {101},
  issue = {24},
  pages = {245107},
  numpages = {12},
  year = {2020},
  month = {Jun},
  publisher = {American Physical Society},
  doi = {10.1103/PhysRevB.101.245107},
  url = {https://link.aps.org/doi/10.1103/PhysRevB.101.245107}
}

@article{scar3,
  title = {Exact {F}loquet quantum many-body scars under {R}ydberg blockade},
  author = {Mizuta, Kaoru and Takasan, Kazuaki and Kawakami, Norio},
  journal = {Phys. Rev. Res.},
  volume = {2},
  issue = {3},
  pages = {033284},
  numpages = {13},
  year = {2020},
  month = {Aug},
  publisher = {American Physical Society},
  doi = {10.1103/PhysRevResearch.2.033284},
  url = {https://link.aps.org/doi/10.1103/PhysRevResearch.2.033284}
}

@article{scar4,
  title = {Many-body scar state intrinsic to periodically driven system},
  author = {Sugiura, Sho and Kuwahara, Tomotaka and Saito, Keiji},
  journal = {Phys. Rev. Res.},
  volume = {3},
  issue = {1},
  pages = {L012010},
  numpages = {6},
  year = {2021},
  month = {Feb},
  publisher = {American Physical Society},
  doi = {10.1103/PhysRevResearch.3.L012010},
  url = {https://link.aps.org/doi/10.1103/PhysRevResearch.3.L012010}
}

@article{scar5,
  title = {Dynamics of the vacuum state in a periodically driven {R}ydberg chain},
  author = {Mukherjee, Bhaskar and Sen, Arnab and Sen, Diptiman and Sengupta, K.},
  journal = {Phys. Rev. B},
  volume = {102},
  issue = {7},
  pages = {075123},
  numpages = {16},
  year = {2020},
  month = {Aug},
  publisher = {American Physical Society},
  doi = {10.1103/PhysRevB.102.075123},
  url = {https://link.aps.org/doi/10.1103/PhysRevB.102.075123}
}

@article{scar6,
  title = {Discrete Time-Crystalline Order Enabled by Quantum Many-Body Scars: Entanglement Steering via Periodic Driving},
  author = {Maskara, N. and Michailidis, A. A. and Ho, W. W. and Bluvstein, D. and Choi, S. and Lukin, M. D. and Serbyn, M.},
  journal = {Phys. Rev. Lett.},
  volume = {127},
  issue = {9},
  pages = {090602},
  numpages = {7},
  year = {2021},
  month = {Aug},
  publisher = {American Physical Society},
  doi = {10.1103/PhysRevLett.127.090602},
  url = {https://link.aps.org/doi/10.1103/PhysRevLett.127.090602}
}

@article{scar7,
  title = {Driving quantum many-body scars in the {PXP} model},
  volume = {106},
  ISSN = {2469-9969},
  url = {http://dx.doi.org/10.1103/PhysRevB.106.104302},
  DOI = {10.1103/physrevb.106.104302},
  pages={104302},
  number = {10},
  journal = {Phys. Rev. B},
  publisher = {American Physical Society (APS)},
  author = {Hudomal,  Ana and Desaules,  Jean-Yves and Mukherjee,  Bhaskar and Su,  Guo-Xian and Halimeh,  Jad C. and Papić,  Zlatko},
  year = {2022},
  month = sep 
}

@article{scar8,
  title = {Discrete Time Crystals Enforced by {F}loquet-{B}loch Scars},
  author = {Huang, Biao and Leung, Tsz-Him and Stamper-Kurn, Dan M. and Liu, W. Vincent},
  journal = {Phys. Rev. Lett.},
  volume = {129},
  issue = {13},
  pages = {133001},
  numpages = {7},
  year = {2022},
  month = {Sep},
  publisher = {American Physical Society},
  doi = {10.1103/PhysRevLett.129.133001},
  url = {https://link.aps.org/doi/10.1103/PhysRevLett.129.133001}
}

@article{hsf1,
  title = {Prethermal Fragmentation in a Periodically Driven Fermionic Chain},
  author = {Ghosh, Somsubhra and Paul, Indranil and Sengupta, K.},
  journal = {Phys. Rev. Lett.},
  volume = {130},
  issue = {12},
  pages = {120401},
  numpages = {6},
  year = {2023},
  month = {Mar},
  publisher = {American Physical Society},
  doi = {10.1103/PhysRevLett.130.120401},
  url = {https://link.aps.org/doi/10.1103/PhysRevLett.130.120401}
}

@article{hsf2,
  title = {Signatures of fragmentation for periodically driven fermions},
  author = {Ghosh, Somsubhra and Paul, Indranil and Sengupta, K.},
  journal = {Phys. Rev. B},
  volume = {109},
  issue = {21},
  pages = {214304},
  numpages = {14},
  year = {2024},
  month = {Jun},
  publisher = {American Physical Society},
  doi = {10.1103/PhysRevB.109.214304},
  url = {https://link.aps.org/doi/10.1103/PhysRevB.109.214304}
}

@article{hsf3,
  title = {Hilbert space fragmentation and exact scars of generalized {F}redkin spin chains},
  volume = {103},
  ISSN = {2469-9969},
  url = {http://dx.doi.org/10.1103/PhysRevB.103.L220304},
  DOI = {10.1103/physrevb.103.l220304},
  number = {22},
  pages={L220304},
  journal = {Phys. Rev. B},
  publisher = {American Physical Society (APS)},
  author = {Langlett,  Christopher M. and Xu,  Shenglong},
  year = {2021},
  month = jun 
}

@article{hsf4,
  title = {Floquet engineering of {H}ilbert space fragmentation in {S}tark lattices},
  volume = {109},
  ISSN = {2469-9969},
  url = {http://dx.doi.org/10.1103/PhysRevB.109.184313},
  DOI = {10.1103/physrevb.109.184313},
  number = {18},
  pages={184313},
  journal = {Phys. Rev. B},
  publisher = {American Physical Society (APS)},
  author = {Zhang,  Li and Ke,  Yongguan and Lin,  Ling and Lee,  Chaohong},
  year = {2024},
  month = may 
}

@misc{hsf5,
      title={Destructive Interference induced constraints in {F}loquet systems}, 
      author={Somsubhra Ghosh and Indranil Paul and K. Sengupta and Lev Vidmar},
      year={2025},
      eprint={2508.18368},
      archivePrefix={arXiv},
      primaryClass={cond-mat.str-el},
      url={https://arxiv.org/abs/2508.18368}, 
}

@misc{hsf7,
      title={Floquet realization of prethermal 
      {M}eissner phase in a two-leg flux ladder}, 
      author={Biswajit Paul and Tapan Mishra and K. Sengupta},
      year={2025},
      eprint={2504.11017},
      archivePrefix={arXiv},
      primaryClass={cond-mat.quant-gas},
      url={https://arxiv.org/abs/2504.11017}, 
}

@article{tb1,
doi = {10.1088/1367-2630/adfd07},
url = {https://doi.org/10.1088/1367-2630/adfd07},
year = {2025},
month = {aug},
publisher = {IOP Publishing},
volume = {27},
number = {8},
pages = {084506},
author = {Banerjee, Tista and Choudhury, Sayan and Sengupta, K},
title = {Exact {F}loquet flat band and heating suppression via two-tone drive protocols},
journal = {New Journal of Physics}
}

@article{kg1,
  title = {Heating suppression via two-rate random and quasiperiodic drive protocols},
  author = {Ghosh, Krishanu and Choudhury, Sayan and Sen, Diptiman and Sengupta, K.},
  journal = {Phys. Rev. B},
  volume = {112},
  issue = {15},
  pages = {155142},
  numpages = {17},
  year = {2025},
  month = {Oct},
  publisher = {American Physical Society},
  doi = {10.1103/3slf-yvtq},
  url = {https://link.aps.org/doi/10.1103/3slf-yvtq}
}

@article{page1,
  title = {Average entropy of a subsystem},
  author = {Page, Don N.},
  journal = {Phys. Rev. Lett.},
  volume = {71},
  issue = {9},
  pages = {1291--1294},
  numpages = {0},
  year = {1993},
  month = {Aug},
  publisher = {American Physical Society},
  doi = {10.1103/PhysRevLett.71.1291},
  url = {https://link.aps.org/doi/10.1103/PhysRevLett.71.1291}
}

@article{mori1,
  title = {Rigorous Bound on Energy Absorption and Generic Relaxation in Periodically Driven Quantum Systems},
  author = {Mori, Takashi and Kuwahara, Tomotaka and Saito, Keiji},
  journal = {Phys. Rev. Lett.},
  volume = {116},
  issue = {12},
  pages = {120401},
  numpages = {5},
  year = {2016},
  month = {Mar},
  publisher = {American Physical Society},
  doi = {10.1103/PhysRevLett.116.120401},
  url = {https://link.aps.org/doi/10.1103/PhysRevLett.116.120401}
}

@Article{da1,
author={Abanin, Dmitry
and De Roeck, Wojciech
and Ho, Wen Wei
and Huveneers, Fran{\c{c}}ois},
title={A Rigorous Theory of Many-Body Prethermalization for Periodically Driven and Closed Quantum Systems},
journal={Communications in Mathematical Physics},
year={2017},
month={Sep},
day={01},
volume={354},
number={3},
pages={809-827},
issn={1432-0916},
doi={10.1007/s00220-017-2930-x},
url={https://doi.org/10.1007/s00220-017-2930-x}
}

@article{da2,
  title = {Effective {H}amiltonians, prethermalization, and slow energy absorption in periodically driven many-body systems},
  author = {Abanin, Dmitry A. and De Roeck, Wojciech and Ho, Wen Wei and Huveneers, F.},
  journal = {Phys. Rev. B},
  volume = {95},
  issue = {1},
  pages = {014112},
  numpages = {8},
  year = {2017},
  month = {Jan},
  publisher = {American Physical Society},
  doi = {10.1103/PhysRevB.95.014112},
  url = {https://link.aps.org/doi/10.1103/PhysRevB.95.014112}
}

@article{dc1,
  title = {Floquet Thermalization via Instantons near Dynamical Freezing},
  author = {Mukherjee, Rohit and Guo, Haoyu and Chowdhury, Debanjan},
  journal = {Phys. Rev. X},
  volume = {16},
  issue = {1},
  pages = {011041},
  numpages = {26},
  year = {2026},
  month = {Feb},
  publisher = {American Physical Society},
  doi = {10.1103/4w5w-57my},
  url = {https://link.aps.org/doi/10.1103/4w5w-57my}
}

@article{vk1,
  title = {Prethermal Stability of Eigenstates under High Frequency Floquet Driving},
  author = {O'Dea, Nicholas and Burnell, Fiona and Chandran, Anushya and Khemani, Vedika},
  journal = {Phys. Rev. Lett.},
  volume = {132},
  issue = {10},
  pages = {100401},
  numpages = {6},
  year = {2024},
  month = {Mar},
  publisher = {American Physical Society},
  doi = {10.1103/PhysRevLett.132.100401},
  url = {https://link.aps.org/doi/10.1103/PhysRevLett.132.100401}
}

@article{vk2,
  title = {Phenomenology of the Prethermal Many-Body Localized Regime},
  author = {Long, David M. and Crowley, Philip J. D. and Khemani, Vedika and Chandran, Anushya},
  journal = {Phys. Rev. Lett.},
  volume = {131},
  issue = {10},
  pages = {106301},
  numpages = {7},
  year = {2023},
  month = {Sep},
  publisher = {American Physical Society},
  doi = {10.1103/PhysRevLett.131.106301},
  url = {https://link.aps.org/doi/10.1103/PhysRevLett.131.106301}
}

@article{mr1,
  title = {Long-time Behavior of Isolated Periodically Driven Interacting Lattice Systems},
  author = {D'Alessio, Luca and Rigol, Marcos},
  journal = {Phys. Rev. X},
  volume = {4},
  issue = {4},
  pages = {041048},
  numpages = {12},
  year = {2014},
  month = {Dec},
  publisher = {American Physical Society},
  doi = {10.1103/PhysRevX.4.041048},
  url = {https://link.aps.org/doi/10.1103/PhysRevX.4.041048}
}

@article{prlspin,
  title = {Constraint-Induced Delocalization},
  author = {Sierant, Piotr and Lazo, Eduardo Gonzalez and Dalmonte, Marcello and Scardicchio, Antonello and Zakrzewski, Jakub},
  journal = {Phys. Rev. Lett.},
  volume = {127},
  issue = {12},
  pages = {126603},
  numpages = {7},
  year = {2021},
  month = {Sep},
  publisher = {American Physical Society},
  doi = {10.1103/PhysRevLett.127.126603},
  url = {https://link.aps.org/doi/10.1103/PhysRevLett.127.126603}
}

@article{huse1,
  title = {Localization of interacting fermions at high temperature},
  author = {Oganesyan, Vadim and Huse, David A.},
  journal = {Phys. Rev. B},
  volume = {75},
  issue = {15},
  pages = {155111},
  numpages = {5},
  year = {2007},
  month = {Apr},
  publisher = {American Physical Society},
  doi = {10.1103/PhysRevB.75.155111},
  url = {https://link.aps.org/doi/10.1103/PhysRevB.75.155111}
}

@article{gapstat1,
  title = {Distribution of the Ratio of Consecutive Level Spacings in Random Matrix Ensembles},
  author = {Atas, Y. Y. and Bogomolny, E. and Giraud, O. and Roux, G.},
  journal = {Phys. Rev. Lett.},
  volume = {110},
  issue = {8},
  pages = {084101},
  numpages = {5},
  year = {2013},
  month = {Feb},
  publisher = {American Physical Society},
  doi = {10.1103/PhysRevLett.110.084101},
  url = {https://link.aps.org/doi/10.1103/PhysRevLett.110.084101}
}

@article{abanin1,
  title = {Quantum scarred eigenstates in a 
  {R}ydberg atom chain: Entanglement, breakdown of thermalization, and stability to perturbations},
  author = {Turner, C. J. and Michailidis, A. A. and Abanin, D. A. and Serbyn, M. and Papi\ifmmode \acute{c}\else \'{c}\fi{}, Z.},
  journal = {Phys. Rev. B},
  volume = {98},
  issue = {15},
  pages = {155134},
  numpages = {23},
  year = {2018},
  month = {Oct},
  publisher = {American Physical Society},
  doi = {10.1103/PhysRevB.98.155134},
  url = {https://link.aps.org/doi/10.1103/PhysRevB.98.155134}
}

@article{dc2,
title = {Frozonium: Freezing anharmonicity in 
{F}loquet superconducting circuits},
journal = {Newton},
pages = {100434},
year = {2026},
issn = {2950-6360},
doi = {https://doi.org/10.1016/j.newton.2026.100434},
url = {https://www.sciencedirect.com/science/article/pii/S2950636026000368},
author = {Keiran Lewellen and Rohit Mukherjee and Haoyu Guo and Saswata Roy and Valla Fatemi and Debanjan Chowdhury}
}

@misc{si,
  author = { },
  title = {See {S}upplemental {M}aterial for details},
  year = { },
  url = {},
  note = {}
}

@Article{intryd1,
	title={{Integrable models on Rydberg atom chains}},
	author={Luke Corcoran and Marius de Leeuw and Balázs Pozsgay},
	journal={SciPost Phys.},
	volume={18},
	pages={139},
	year={2025},
	publisher={SciPost},
	doi={10.21468/SciPostPhys.18.4.139},
	url={https://scipost.org/10.21468/SciPostPhys.18.4.139},
}

@Article{intryd2,
  title = {Integrable hard-rod deformation of the {H}eisenberg spin chains},
  author = {Pozsgay, Bal\'azs and Gombor, Tam\'as and Hutsalyuk, Arthur},
  journal = {Phys. Rev. E},
  volume = {104},
  issue = {6},
  pages = {064124},
  numpages = {13},
  year = {2021},
  month = {Dec},
  publisher = {American Physical Society},
  doi = {10.1103/PhysRevE.104.064124},
  url = {https://link.aps.org/doi/10.1103/PhysRevE.104.064124}
}

@article{con1,
title = {Structure of the Conservation Laws in Quantum Integrable Spin Chains with Short Range Interactions},
journal = {Annals of Physics},
volume = {243},
number = {2},
pages = {299-371},
year = {1995},
issn = {0003-4916},
doi = {https://doi.org/10.1006/aphy.1995.1101},
url = {https://www.sciencedirect.com/science/article/pii/S0003491685711013},
author = {M.P. Grabowski and P. Mathieu}
}

@article{sff1,
  title = {Measuring the Spectral Form Factor in Many-Body Chaotic and Localized Phases of Quantum Processors},
  author = {Dong, Hang and Zhang, Pengfei and Da\ifmmode \breve{g}\else \u{g}\fi{}, Ceren B. and Gao, Yu and Wang, Ning and Deng, Jinfeng and Zhang, Xu and Chen, Jiachen and Xu, Shibo and Wang, Ke and Wu, Yaozu and Zhang, Chuanyu and Jin, Feitong and Zhu, Xuhao and Zhang, Aosai and Zou, Yiren and Tan, Ziqi and Cui, Zhengyi and Zhu, Zitian and Shen, Fanhao and Li, Tingting and Zhong, Jiarun and Bao, Zehang and Li, Hekang and Wang, Zhen and Guo, Qiujiang and Song, Chao and Liu, Fangli and Chan, Amos and Ying, Lei and Wang, H.},
  journal = {Phys. Rev. Lett.},
  volume = {134},
  issue = {1},
  pages = {010402},
  numpages = {7},
  year = {2025},
  month = {Jan},
  publisher = {American Physical Society},
  doi = {10.1103/PhysRevLett.134.010402},
  url = {https://link.aps.org/doi/10.1103/PhysRevLett.134.010402}
}

@article{sff2,
  title = {Spectral Statistics in Spatially Extended Chaotic Quantum Many-Body Systems},
  author = {Chan, Amos and De Luca, Andrea and Chalker, J. T.},
  journal = {Phys. Rev. Lett.},
  volume = {121},
  issue = {6},
  pages = {060601},
  numpages = {5},
  year = {2018},
  month = {Aug},
  publisher = {American Physical Society},
  doi = {10.1103/PhysRevLett.121.060601},
  url = {https://link.aps.org/doi/10.1103/PhysRevLett.121.060601}
}

@article{sff3,
  title = {Many-Body Quantum Chaos: Analytic Connection to Random Matrix Theory},
  author = {Kos, Pavel and Ljubotina, Marko and Prosen, Tomaz},
  journal = {Phys. Rev. X},
  volume = {8},
  issue = {2},
  pages = {021062},
  numpages = {11},
  year = {2018},
  month = {Jun},
  publisher = {American Physical Society},
  doi = {10.1103/PhysRevX.8.021062},
  url = {https://link.aps.org/doi/10.1103/PhysRevX.8.021062}
}

@article{sff4,
  title = {Spectral statistics in constrained many-body quantum chaotic systems},
  author = {Moudgalya, Sanjay and Prem, Abhinav and Huse, David A. and Chan, Amos},
  journal = {Phys. Rev. Res.},
  volume = {3},
  issue = {2},
  pages = {023176},
  numpages = {27},
  year = {2021},
  month = {Jun},
  publisher = {American Physical Society},
  doi = {10.1103/PhysRevResearch.3.023176},
  url = {https://link.aps.org/doi/10.1103/PhysRevResearch.3.023176}
}

\appendix*

\section{End Matter}

{\it Third conserved charge} : The integrable spin-$1/2$ ${\rm XXZ}$ model has several conserved charges. The first two of these are the net magnetization and the Hamiltonian which has been discussed in the main text in context of $H_F^{(2)}$. Here we focus on the third conserved charge. Following Ref.\ \cite{con1}, we define the operators $\hat \tau_j^{\alpha} =\sqrt{J_{\alpha}} \tau_j^{\alpha}$ for $\alpha=x,y,z$, where $J_{x,y}= J$ and $J_z= -J\Delta$. Also, we define $\tau_j^{' \alpha}= (J_x J_y J_z/J_{\alpha})^{1/2} \tau_j^{\alpha}$. In terms of these variables, the third conserved charge $C_3$ is given by
\begin{eqnarray}
C_3 &=& (\vec {\hat \tau}_j \times \vec {\tau}'_{j+1})\cdot \vec {\hat \tau}_{j+2}, \label{thirdcharge1}
\end{eqnarray} 
which can be expanded to obtain $C_3= (T_1+T_2+T_3)$, where 
\begin{eqnarray}
 T_1 &=& \left(2iJ\Delta\;\tau^{+}_{j}\,\tau^{-}_{j+1}\,\tau^{z}_{j+2} + \mathrm{H.c.}\right), \nonumber\\
T_2 &=& \left(2iJ\;\tau^{-}_{j}\,\tau^{z}_{j+1}\,\tau^{+}_{j+2} + \mathrm{H.c.}\right), \nonumber\\
T_3 &=& \left(2iJ\Delta\;\tau^{z}_{j}\,\tau^{+}_{j+1}\,\tau^{-}_{j+2} + \mathrm{H.c.}\right). \label{thirdcharge2}
\end{eqnarray}

To construct an equivalent operators in the PXP language, we consider the effect of each of these operators on the Fock states of the ${\rm XXZ}$ model. For example, let us consider the first term of $T_1$ in Eq.\ \eqref{thirdcharge2}. We find $ T_1 |1\rangle_{\rm XXZ} = 2i J\Delta |2\rangle_{\rm XXZ}$, where 
\begin{eqnarray} 
|1\rangle_{\rm XXZ} &=& |\cdots \downarrow_j, \uparrow_{j+1}, \alpha_{j+2}, \cdots\rangle \nonumber\\
|2\rangle_{\rm XXZ} &=& {\rm sgn}(\alpha) |\cdots \uparrow_j, \downarrow_{j+1}, \alpha_{j+2}, \cdots\rangle \label{xxzfock1}
\end{eqnarray} 
where $\alpha=\uparrow$ or $\downarrow$ and ${\rm sgn}\alpha = 1(-1)$ for $\alpha=\uparrow(\downarrow)$.

Next, we construct the PXP Focks states which can be mapped to $|1\rangle_{\rm XXZ}$ and $|2\rangle_{\rm XXZ}$ by the mapping considered in the main text. These states, for $\alpha_{j+2}=\downarrow_{j+2}$, are given by 
\begin{eqnarray} 
|1\rangle_{\rm PXP} &=& |\cdots \downarrow_j, \downarrow_{j+1}, \uparrow_{j+2}, \downarrow_{j+3}, \cdots\rangle \nonumber\\
|2\rangle_{\rm PXP} &=& |\cdots \downarrow_{j-1}, \uparrow_j, \downarrow_{j+1}, \downarrow_{j+2}, \cdots\rangle \label{xxzfock2}
\end{eqnarray} 
We now try to find an operator which connects between these two states. In doing so, we note that since we are in the $K=0$ sector, it is enough to find an operator which connect $|1\rangle_{\rm PXP}$ to a translated version of $|2\rangle_{\rm PXP}$ given by $T_{1}|2\rangle_{\rm PXP} = |\cdots \downarrow_{j}, \uparrow_{j+1}, \downarrow_{j+2}, \downarrow_{j+3}, \cdots\rangle$. It is easy to see by inspection that this operator is given by
\begin{eqnarray} 
\hat{\mathcal{O}}_{1a} = -\sum_j P^{\downarrow}_{j-1}\,\sigma^{+}_j\,\sigma^{-}_{j+1}\,P^{\downarrow}_{j+2}, \label{oplist1} 
\end{eqnarray} 
where $P_j^{\downarrow (\uparrow)}= (1-(+)\sigma_j^z)/2$ are projection operators. A similar construction for $\alpha_{j+2} = \uparrow$ leads to the operator 
\begin{eqnarray} 
\hat{\mathcal{O}}_{1b} = \sum_j P^{\downarrow}_{j-1}\,\sigma^{+}_j\,\sigma^{-}_{j+1}\,P^{\downarrow}_{j+2}\,P^{\uparrow}_{j+3} \,P^{\downarrow}_{j+4}. \label{oplist2} 
\end{eqnarray} 
Together these two operators yields $T_1^{\rm pxp}= 2i J \Delta ({\mathcal O}_{1a}+{\mathcal O}_{1b}) +{\rm H.c.}$. Following an exact same procedure, we find that $T_2^{\rm pxp}= 2i J^2 ({\mathcal O}_{2a}+{\mathcal O}_{2b})+{\rm H.c.}$ and $T_3^{\rm pxp}= 2i J \Delta ({\mathcal O}_{3a}+{\mathcal O}_{3b})+{\rm H.c.} $, where 
\begin{eqnarray} 
\hat{\mathcal{O}}_{2a} &=& -\sum_j P^{\downarrow}_{j-1}\,\sigma^{-}_j\,P^{\downarrow}_{j+1}\,\sigma^{+}_{j+2} P^{\downarrow}_{j+3},\nonumber\\
\hat{\mathcal{O}}_{2b} &=& \sum_j P^{\downarrow}_{j-1}\,\sigma^{-}_j\,\sigma^{+}_{j+1}\,\sigma^{-}_{j+2}\,\sigma^{+}_{j+3} P^{\downarrow}_{j+4}, \nonumber\\
\hat{\mathcal{O}}_{3a} &=& -\sum_j P^{\downarrow}_{j-1} P^{\downarrow}_j\,\sigma^{+}_{j+1}\,\sigma^{-}_{j+2} P^{\downarrow}_{j+3},\nonumber\\
\hat{\mathcal{O}}_{3b} &=& \sum_j P^{\downarrow}_{j-1}\,P^{\uparrow}_j\,P^{\downarrow}_{j+1}\,\sigma^{+}_{j+2}\,\sigma^{-}_{j+3} P^{\downarrow}_{j+4}.\label{oplist3} 
\end{eqnarray} 
We note that the span of these operators are larger than their ${\rm XXZ}$ counterparts. Also they conserve magnetization and hence commute with the first charge $S^z= \sum_j \sigma_j^z$. 

To construct $C_3$, we need to find the correct superposition of these operators which commutes with the second charge, namely, $H_F^{(2)}$. We note that, unlike the ${\rm XXZ}$ model, a simple linear combination of $T_i^{\rm PXP}$ operators do not yield $C_3$. To determine this, we note that $C_3$ must odd under parity. A straightforward analysis shows that ${\mathcal P} {\mathcal O}_{2a[2b]} {\mathcal P} ^{-1} = {\mathcal O}_{2a[2b]}^{\dagger}$, where ${\mathcal P} $ denotes the parity operator which satisfies ${\mathcal P} A_j {\mathcal P}= A_{L-j+1}$ for any local operator $A_j$. Thus we find that ${\mathcal P} T_2^{\rm PXP} {\mathcal P} ^{-1} = -T_2^{\rm PXP}$. This mandates that 
\begin{eqnarray} 
{\mathcal P} (T_1^{\rm PXP} +T_3^{\rm PXP}) {\mathcal P} ^{-1} = -(T_1^{\rm PXP} + T_3^{\rm PXP}). \label{pareq1}
\end{eqnarray} 
Furthermore, a simple calculation shows that 
\begin{eqnarray} 
{\mathcal P} {\mathcal O}_{1b(3b)} {\mathcal P} = {\mathcal O}_{3b(1b)}^{\dagger}, \quad {\mathcal P} {\mathcal O}_{1a} {\mathcal P} = {\mathcal O}_{1a}^{\dagger}. \label{parop1}
\end{eqnarray} 

\begin{figure}
\includegraphics[width=\linewidth]{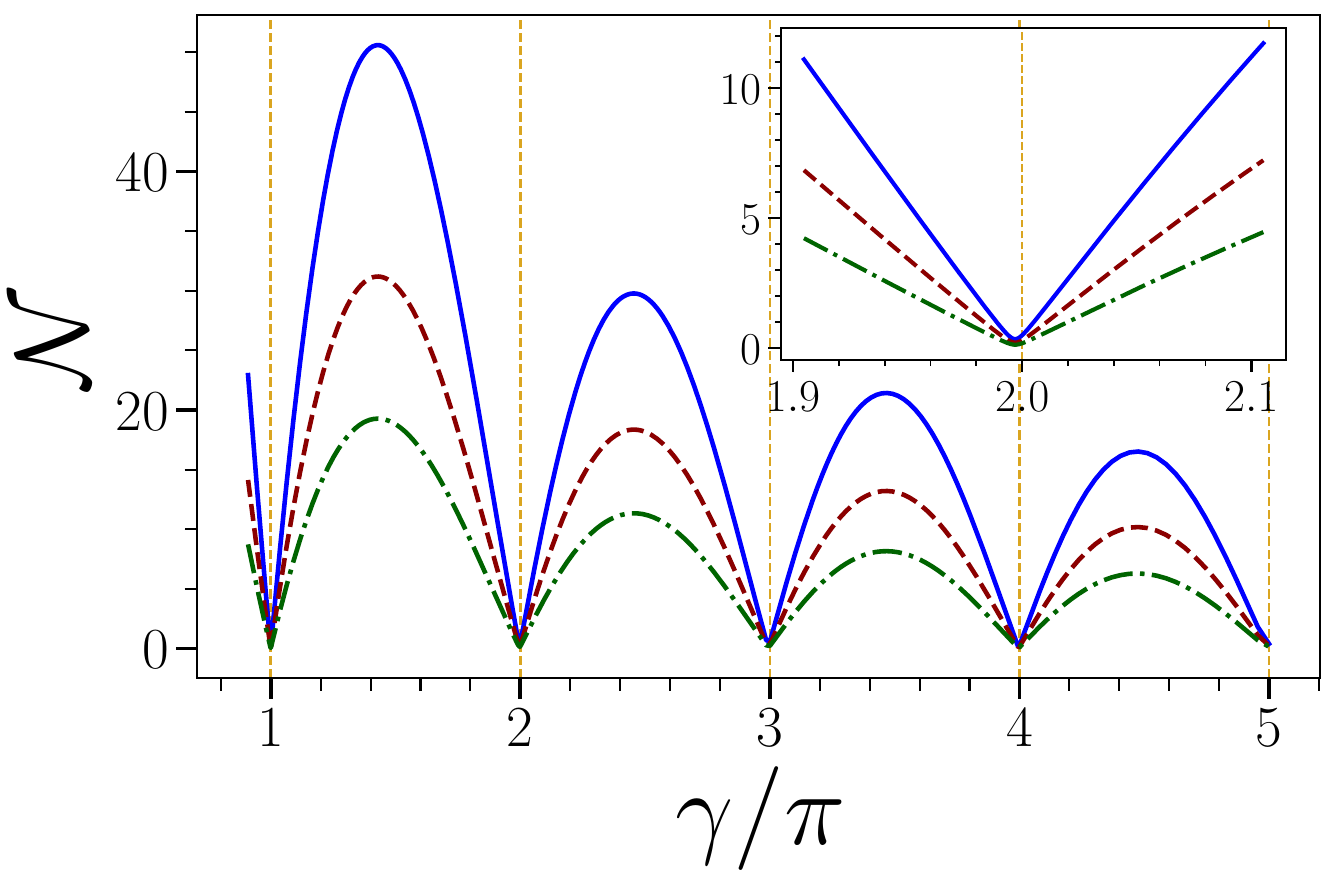}
\caption{Plot of Frobenius norm $\mathcal{N}$ of commutator $[H_F, C_3^{\mathrm{PXP}}]$ as a function of $\gamma/\pi$, where the Floquet Hamiltonian $H_F$ is obtained numerically taking matrix logarithm of time evolution operator $U(T_1,0)$, in $K=0$ sector of system sizes $L=20$ (green), $L=22$ (red) and $L=24$ (blue). The inset shows behavior of ${\mathcal N}$ around $\gamma=2 \pi$. For all plots, $w_0/w_1=1$, $q=3$ and $\lambda_0/w_1=20$. See text for details.} \label{figem1}
\end{figure} 

To satisfy Eq.\ \eqref{pareq1}, we therefore need to find the parity transformation of ${\mathcal O}_{3a}$. A straightforward but cumbersome calculation shows that 
\begin{eqnarray}
 {\mathcal P} ( {\mathcal O}_{3a}-{\mathcal O}_{3a}^{\dagger}) {\mathcal P}^{-1} - ( {\mathcal O}_{3a}-{\mathcal O}_{3a}^{\dagger}) &=& ({\mathcal O}_{3b}+ {\mathcal O}_{1b}^{\dagger} -{\rm H.c.}). \nonumber\\ \label{parop2}
\end{eqnarray}
Using Eqs.\ \eqref{pareq1}, \eqref{parop1}, and \eqref{parop2}, we find that ${\mathcal P} C_3^{\rm PXP} {\mathcal P} ^{-1} = -C_3^{\rm PXP}$ if one uses 
\begin{eqnarray} 
C_3^{\rm PXP} &=& 2i J [ \Delta( {\mathcal O}_{1a} + 2{\mathcal O}_{1b} + {\mathcal O}_{3a} + {\mathcal O}_{3b}) \nonumber\\
&& +( {\mathcal O}_{2a} +{\mathcal O}_{2b})] + {\rm H.c.}.
\end{eqnarray} 
To verify this construction, we numerically compute the Frobenius norm of the commutator ${\mathcal N}= ||[H_F, C_3^{\rm PXP}]||$ in the $K=0$ sector, and plot it as a function of $\gamma/\pi$ as shown in Fig.\ \ref{figem1}. The plot shows clear dip of ${\mathcal N}$ to near zero value for $\gamma=m \pi, \, m \in \mathbb{Z^+}$; the deviation of ${\mathcal N}$ from zero is small but finite due to higher order terms in $H_F$. This deviation increases with decreasing drive frequency (increasing $\gamma$ for fixed $\lambda_0$). The inset shows the behavior of ${\mathcal N}$ around $\gamma/\pi=2$ showing a sharp linear increase in ${\mathcal N}$ for small deviation of $\gamma/\pi$ from its integer value. These features indicate emergent approximate conservation of $C_3^{\rm PXP}$ at special drive frequencies in the prethermal regime. We note that such a conservation provides a clear signature of the emergent integrability discussed in the main text.

\end{document}


\title{Supplemental Material for Emergent prethermal Bethe integrability in a periodically driven Rydberg chain}
\author{Saptadip Roy$^1$, Arnab Sen$^1$, Diptiman Sen$^2$, and K. Sengupta$^1$}
\affiliation{$^1$School of Physical Sciences, Indian Association for the
Cultivation of Science, Kolkata 700032, India \\
$^2$Center for High Energy Physics, Indian Institute of Science, Bengaluru 560012, India}
\date{\today}

\maketitle
\tableofcontents 

\section{Details of Floquet perturbation for the two-tone protocol}
\label{fptdetails}

In this section, we provide the details of FPT for the two-rate protocol. The square pulse protocol is discussed in Sec.\ \ref{sqp1} while the results for the cosine protocol are charted out in Sec.\ \ref{cqp1}. 

\subsection {Square pulse protocol}
\label{sqp1} 

We begin with the Hamiltonian of the driven Rydberg chain given by $H(t) = H_0(t) \ + \ V(t)$ where 
\begin{eqnarray}
H_0 (t) = \lambda(t) \, \sum_j \sigma_j ^z \ \ , \quad \quad V(t) = \left( w_0 + w_1(t)\right) \, \sum_j \tilde{\sigma}_j^x. \label{eq1}
\end{eqnarray}
Here we shall consider the case where $\lambda(t)$ is driven with a period $T_1$ while $w_1(t)$ is driven with a period $T_2= T_1/3$. The piecewise functions $\lambda(t)$ and $w_1(t)$ are defined as 
\begin{align}
\lambda(t) =
\begin{cases}
+ \lambda_0 \ , & 0 \leq t < T_1/2 \label{eq2} \\
- \lambda_0 \ , & T_1/2 \leq t < T_1,
\end{cases} 
\end{align}
and 
\begin{align} 
w_1(t) =
\begin{cases}
- w_1 \ , & 0 \leq t < T_1/6, \ \ T_1/3 \leq t < T_1/2, \ \ 2T_1/3 \leq t < 5T_1/6 \\
+ w_1 \ , & T_1/6 \leq t < T_1/3, \ \ T_1/2 \leq t < 2T_1/3, \ \ 5T_1/6 \leq t < T_1.
\end{cases} 
\end{align}
We consider the regime $\lambda_0 \gg w_0,w_1$; in this regime, one can define the zeroth order evolution operator $U_0(t,0)$ as 
\begin{eqnarray}
U_0(t,0) &=& \exp{\left(-i \, \int_0^t dt' \, H_0(t')\right)} = \prod_j \, \exp{\left(-i \, \theta(t) \, \sigma_j^z\right)}, \quad \theta(t) = \int_0^t dt' \, \lambda(t'). \label{eq3}
\end{eqnarray}
For the present protocol, $\theta(t)= \lambda_0 t$ for $t \le T_1/2$ and $\lambda_0(T_1-t)$ for $T_1/2 <t\le T_1$. Note that $U_0(T_1,0)= \mathbb {I}$ in the present case; this implies $H_F^{(0)}=0$. 

To compute the effect of $V(t)$, we use the interaction picture and define $V_I(t) = U_0^+(t,0) \, V(t) \, U_0(t,0)$. A simple calculation yields
\begin{align}
V_I(t) = \left(w_0 + w_1(t) \right) \, \sum_j \left( \cos{(2\theta(t))}\, \tilde{\sigma}_j^x \ - \ \sin{(2\theta(t))}\, \tilde{\sigma}_j^y \right). \label{eq5}
\end{align}
In what follows, we calculate terms in the Dyson series for $U$ and use it to compute the effective Floquet Hamiltonian $H_F$ at each order in $w_0,w_1$.

The time evolution of $U(T_1,0)$ in the interaction picture is given in terms of the Dyson series $U_I(T_1,0) = \sum_{n=0}^{\infty} U_n (T_1,0)$, where 
\begin{eqnarray}
U_n(T_1,0) &=& \sum_{n=0}^{\infty} (-i)^n \int_0^{T_1} dt_1 \int_0^{t_1} dt_2 \cdots \int_0^{t_{n-1}} dt_n \, V_I(t_1) V_I(t_2) \cdots V_I(t_n). \label{uneq} 
\end{eqnarray}
The propagator is given by $U(T_1,0) = U_0(T_1,0) \, U_I(T_1,0)$. In the present case $U_0(T_1,0)= \mathbb{I}$; this allows us to write 

\begin{eqnarray} 
U(T_1,0) &=& e^{-i T_1 H_F/\hbar} = \mathbb{I} + \, \sum_{n=1}^{\infty} \alpha^n U_n(T_1,0), \label{uperteq} 
\end{eqnarray}
where the parameter $\alpha$ is introduced to keep track of the orders in the perturbation series and will be set to unity at the end of the calculation. The Floquet Hamiltonian is then expressed as 
\begin{eqnarray}
H_F = \sum_{n=0}^{\infty} \alpha^n H_F^{(n)} = \frac{i\hbar}{T_1} \, \ln{(\mathbb{I} + \, \sum_{n=1}^{\infty} \alpha^n U_n(T_1,0))}. \label{flh1}
\end{eqnarray}
Then one equates the powers of $\alpha$ from both sides to obtain (up to the second order) 
\begin{eqnarray} 
H_F^{(1)} &=& \frac{i}{T_1} \, U_1, \quad H_F^{(2)} = \frac{i}{T_1} \Big( U_2 - \frac{1}{2} U_1^2 \Big), \nonumber\\
U_1 &=& - \frac{i}{\hbar} \int_0^{T_1} dt_1 \, V_I(t_1), \quad
U_2 = \left(- \frac{i}{\hbar} \right)^2 \int_0^{T_1} dt_1 \int_0^{t_1} dt_2 \, V_I(t_1) V_I(t_2), \label{flh2} 
\end{eqnarray}

Writing $U(T_1,0) = \exp{(\Omega)}$, we see that $\Omega_1 = U_1$ so that the first-order Floquet Hamiltonian is given by the average Hamiltonian
\begin{align}
H_F^{(1)} = \frac{i\hbar}{T_1} \, \Omega_1 = \frac{1}{T_1} \, \int_0^{T_1} dt \, V_I(t). \label{eq7}
\end{align}
The second-order term can be obtained from $\Omega_2= U_2(T_1,0)-U^2_1(T_1,0)/2$. Using the expressions of $U_2$ and $U_1$ in Eq.\ \eqref{flh2}, we find, after some algebra 
\begin{align*}
\Omega_2 &= - \frac{1}{2\hbar^2} \, \int_0^{T_1} dt_1 \int_0^{t_1} dt_2 [V_I(t_1), V_I(t_2)],
\end{align*}
which yields 
\begin{align}
H_F^{(2)} = \frac{i\hbar}{T_1} \, \Omega_2 = - \frac{i}{2 T_1 \hbar} \, \int_0^{T_1} dt_1 \, \int_0^{t_1} dt_2 \, [V_I(t_1), V_I(t_2)]. \label{eq8}
\end{align}
This brings out the commutator structure of the Floquet Hamiltonian. We note that such a structure is key to ensuring the locality of the perturbative Floquet Hamiltonian. 

Next, we compute explicit expressions for $H_F^{(n)}$ for $n=1,2$. To this end, we consider the following integrals
\begin{eqnarray}
 I_{1a} &=& \int_0^{T_1} (w_0 +w_1(t)) \cos 2\theta(t) dt = \frac{w_0}{\lambda_0} \sin (\lambda_0 T_1/\hbar), \nonumber\\
 I_{1b} &=& \int_0^{T_1} (w_0 +w_1(t)) \cos 2\theta(t) dt =\frac{2w_0}{\lambda_0} \sin^2 (\lambda_0 T_1/(2\hbar)), \nonumber\\
 I_2 &=& \int_0^{T_1} dt_1 \int_0^{t_1} dt_2 \,
\bigl(w_0 + w_1(t_1)\bigr)\bigl(w_0 + w_1(t_2)\bigr)
\sin\!\, \bigl(2\theta(t_1) - 2\theta(t_2)\bigr) \nonumber\\
&=& - \frac{w_0 \, w_1}{3 \, \lambda_0^2} \
\Bigl[
\lambda_0 T_1
- 6 \sin\!\left(\frac{\lambda_0 T_1}{3}\right)
+ 6 \sin\!\left(\frac{2\lambda_0 T_1}{3}\right)
- 3 \sin\!\left(\lambda_0 T_1\right)
\Bigr]. \label{integ1}
\end{eqnarray}

For the first-order term, we find, after some simple calculation, $\int_0^{T_1} dt \, V_I(t)= \sum_j \left( I_{1a} \,\tilde{\sigma}_j^x \ - \ I_{1b} \, \tilde{\sigma}_j^y \right)$. This leads to the final form of the first-order Floquet Hamiltonian 
\begin{align}
H_F^{(1)} = \frac{w_0 \, \sin{\gamma}}{\gamma} \, \sum_j \left( e^{i \gamma} \ \tilde{\sigma}_j^+ \ + \ \text{H.c.} \right),\label{eq10}
\end{align}
where $\gamma= \lambda_0 T_1/(2\hbar)$. This is Eq. (3) of the main text. 

Next, we compute the second-order term of the Floquet Hamiltonian. We begin with noting down operator identities involving the Pauli operators $\sigma_j^{\alpha}$
on site $j$ and the projection operators $P_j^{\downarrow} \equiv P_j = (1-\sigma_j^z)/2$ and $P_j^{\uparrow}= (1-P_j)$. These are given by 
\begin{eqnarray}
P_k \, \sigma_k^x &=& \sigma_k^x \, P_k^\uparrow = \sigma_k^- \,,\quad 
\sigma_k^x \, P_k = P_k^\uparrow \, \sigma_k^x = \sigma_k^+ \,,\nonumber\\
P_k \, \sigma_k^y &=&\sigma_k^y \, P_k^\uparrow = i\,\sigma_k^- \,,\quad 
\sigma_k^y \, P_k = P_k^\uparrow \, \sigma_k^y = -\,i\,\sigma_k^+. \label{rel1} 
\end{eqnarray}

To compute $H_F^{(2)}$, we first note that the commutator in Eq.\ \eqref{flh2} can be written as 
\begin{eqnarray}
[V_I(t_1) , V_I(t_2)] &=&\sum_{j \, , \, j'} \, \left(w_0 + w_1(t_1) \right) \ \left(w_0 + w_1(t_2) \right)\ \left[ A(1;j), A(2;j')\,\right], \nonumber\\
A(a;j) &=& \left( \cos{\theta_a} \ \tilde{\sigma}_j^x \ - \ \sin{\theta_a} \ \tilde{\sigma}_j^y \right), \quad \theta_a = \theta(t_a), \label{comsecond}
\end{eqnarray}
We then note that these commutators can be non-zero only if $j= j$ or $j'= j\pm 1$ ensuring the local structure of $H_F^{(2)}$. Using Eqs.\ \eqref{rel1}, it is possible to group the non-zero terms as 
\begin{eqnarray}
\sum_j \left[ A(1;j), A(2;j\pm 1)\,\right]
&=& \mp \cos(\theta_1 - \theta_2) \, \sum_j [P_{j-2} \,\left(\sigma_{j-1}^+ \, \sigma_j^- \ - \ \sigma_{j-1}^- \, \sigma_j^+ \right) \, P_{j+1} ] \nonumber\\
&& + i \, \sin(\theta_1 - \theta_2) \, \sum_j [P_{j-2} \,\left(\sigma_{j-1}^+ \, \sigma_j^- \ + \ \sigma_{j-1}^- \, \sigma_j^+ \right) \, P_{j+1}], \label{terms1}
\end{eqnarray}
where we have used the fact that $j \to j-1$ leaves the sum invariant. Similarly, for the $j=j'$ terms, we find the following
\begin{eqnarray}
\sum_j \left[ A(1;j), A(2;j)\,\right]\, &=& i \, \sin(\theta_1 - \theta_2) \, \sum_j \tilde{\sigma}_j^z. \label{terms2}
\end{eqnarray} 
Together, Eqs.\ \eqref{terms1} and \eqref{terms2} yield 
\begin{eqnarray} 
[V_I(t_1), V_I(t_2)] 
&=& 2i \, \bigl(w_0 + w_1(t_1)\bigr)\bigl(w_0 + w_1(t_2)\bigr)\sin(\theta_1 - \theta_2) \sum_j \left[
P_{j-2}\bigl(\sigma_{j-1}^+ \sigma_j^- + \sigma_{j-1}^- \sigma_j^+\bigr) P_{j+1}
+ \,\tilde{\sigma}_j^z \right].
\end{eqnarray}
The integral in Eq.\ \eqref{eq8} can now be easily evaluated using Eq.\ \eqref{integ1} and finally one obtains 
\begin{eqnarray}
H_F^{(2)} &=& -\frac{i}{2T_1} \int_0^{T_1} dt_1 \int_0^{t_1} dt_2 \, [V_I(t_1), V_I(t_2)]= \frac{w_0 \, w_1}{3 \, \lambda_0} \ A(\alpha) \ \sum_j \left[
P_{j-2}\bigl(\sigma_{j-1}^+ \sigma_j^- + \sigma_{j-1}^- \sigma_j^+\bigr) P_{j+1}
+ \,\tilde{\sigma}_j^z
\right], \nonumber\\
A(\alpha) &=& \frac{6}{\alpha} \, \Big[2 \sin{\left( \frac{\alpha}{6} \right)} - 2 \sin{\left( \frac{\alpha}{3} \right)} - \sin{\left( \frac{\alpha}{2} \right)}\Big] \ - \ 1,
\end{eqnarray}
where $\alpha =2 \, \lambda_0 T_1/\hbar =4 \gamma$. This leads to Eq. (3) of the main text. 

\subsection{ Cosine protocol} 
\label{cqp1}
The perturbative Floquet Hamiltonian can be computed for cosine protocol using a similar method as the square pulse one. Here we use the protocol
\begin{eqnarray} 
\lambda(t) &=& \lambda_0 \cos \omega_1 t, \quad w(t)= w_0 + w_1 \cos[(2p+1)\omega_1 t], 
\label{cprot}
\end{eqnarray} 
where $p$ is an integer. For $\lambda_0 \gg w_0,w_1$, we denote $H_0(t) = \lambda(t) \sum_j \sigma_j^z$ and obtain 
\begin{eqnarray} 
U_0(t,0) &=& e^{-i \int^t dt' H_0(t')/\hbar} = \exp\left[-i\lambda_0 t \frac{\sin y_1 }{\hbar y_1} \sum_j \sigma_j^z\right], \label{uoeq}
\end{eqnarray} 
where $y_1 =\omega_1 t$. This allows us to identify $\theta (t) = (\lambda_0 t/(\hbar y_1)) \sin y_1$. We note that $U_0(T_1,0)=\mathbb {I}$ for this protocol, so that $H_F^{(0)}=0$. 

Following the computation outlined for the square pulse protocol in Sec.\ \ref{sqp1}, we then find 
\begin{eqnarray} 
V_I(t) &=& (w_0 + w_1 \cos [(2p+1)\omega_1 t] \sum_j (\cos (2\theta(t)) \tilde \sigma_j^x -\sin(2\theta(t)) \tilde \sigma_j^y). \label{veq1}
\end{eqnarray} 
The first-order Floquet Hamiltonian can then be obtained from $U_1(T,0) = (-i/\hbar) \int_0^T V_I(t) dt$ using $H_F^{(1)}= i \hbar U_1(T,0)/T_1$. To carry out the integral, we use the identity 
$\exp[i z \sin x]= \sum_{n=-\infty}^{\infty} J_n(z) \exp[ inx]$, to obtain
\begin{eqnarray} 
I_c &=& w_0 \int_0^{T_1} \cos(\theta(t)) dt = J_0(z_1) w_0 T_1, \quad I_s = w_0 \int_0^{T_1} \sin(\theta(t)) dt=0, 
\end{eqnarray} 
where $z_1= 2 \lambda_0/(\hbar \omega_1)$. This leads to the 
first-order Floquet Hamiltonian 
\begin{eqnarray}
H_F^{(1)} &=& w_0 J_0 (z_1) \sum_j \tilde \sigma_j^x. 
\end{eqnarray}
The special frequencies, at which $H_F^{(1)}$ vanishes, are therefore given by $2\lambda_0/(\hbar \omega_n^{\ast}) =\eta_n$ where $\eta_n$ denotes the $n^{\rm th}$ zero of the Bessel function $J_0$.

To compute the second-order Floquet Hamiltonian given by 
\begin{eqnarray} 
H_F^{(2)}(T_1,0) = - \frac{i\hbar }{2 T_1} \int_0^{T_1} dt_1 \int_0^{t_1} dt_2 [V_I(t_1),V_I(t_2)],
\end{eqnarray} 
we first evaluate the commutator. A straightforward algebra leads to 
\begin{eqnarray} 
[V_I(t_1),V_I(t_2)] &=& \frac{2i}{\hbar^2} w(t_1)w(t_2) \sin [2\theta(t_1) -2 \theta (t_2)] \sum_j \left[P_{j-1} (\sigma_{j}^+ \sigma_{j+1}^- + {\rm H.c.}) P_{j+2} +\tilde \sigma_j^z \right], \label{comeq1}
\end{eqnarray} 
where $w(t) = w_0 + w_1 \cos[(2p+1)\omega_1 t]$. Thus the integral we need to evaluate is 
\begin{eqnarray}
 I_2(T_1) &=& \int_0^{T_1} dt_1 \int_0^{t_1} dt_2 w(t_1)w(t_2) \sin [2\theta(t_1) -2 \theta (t_2)]. \label{comeq2} 
\end{eqnarray}
The evaluation of this integral can be carried out by using standard Bessel function identities. After some lengthy calculation one obtains 
\begin{eqnarray} 
I_2(T_1) &=&-\frac{4w_0^2 T_1}{\omega_1}\, J_0 \sum_{n=0}^{\infty} \frac{J_{2n+1}}{(2n+1)}
+ \frac{2w_0 w_1 T_1}{\omega_1}\, J_0
\left[
\frac{J_{2p+1}}{(2p+1)}
- \sum_{n=0}^{\infty} \frac{J_{2n+1}}{2(n+p+1)}
- \sum_{\substack{n=0\\(n\neq p)}}^{\infty} \frac{J_{2n+1}}{2(n-p)}
\right] \label{comeq3} \\
&& + \frac{2w_0 w_1 T_1}{\omega_1}
\left[
\sum_{n=0}^{\infty}
\left(\frac{1}{(2n+1)} + \frac{1}{2(n+p+1)}\right) J_{2n+1}\, J_{2n+2p+2}
+ \sum_{\substack{n=0\\(n\neq p)}}^{\infty}
\left(\frac{1}{(2n+1)} + \frac{1}{2(n-p)}\right) J_{2n+1}\, J_{2n-2p}
\right], \nonumber
\end{eqnarray} 
where $J_n \equiv J_n(z_1)$ is the Bessel function of order $n$. Using Eqs.\ \eqref{comeq1}, \eqref{comeq2}, and \eqref{comeq3}, we finally obtain 
\begin{eqnarray}
H_F^{(2)} &=& \frac{\hbar }{T_1} I_2(T_1) \sum_j \left[P_{j-1} (\sigma_{j}^+ \sigma_{j+1}^- + {\rm H.c.}) P_{j+2} +\tilde \sigma_j^z \right].
\end{eqnarray}
This completes our derivation of the Floquet Hamiltonian for the cosine drive protocol. We find that emergent Bethe integrability also occurs for such continuous drive protocols, albeit with different values of the special frequencies.

\section{Single-tone Asymmetric drive}
\label{singletone}

In this section, we note that emergent integrability obtained using a two-tone protocol can also occur for a asymmetric single-tone drive, pictorially represented in Fig.\,1(b) of the main text, is given by 
\begin{eqnarray} 
\lambda(t) &=&-\lambda_0 \quad {\rm for}\, \,0\le t < pT_1 \nonumber\\
&=& \lambda_0, \quad {\rm for}\, \,pT_1\le t < T_1, \label{protas1} 
\end{eqnarray}
where $T_1=2\pi/\omega_1$ is the drive period, $p$ is a tunable fraction, and $p=1/2$ corresponds to the symmetric drive studied earlier \cite{scar2,scar5}. We note that for $p=1/2$, the Floquet Hamiltonian satisfies $\{H_F, {\mathcal C}\}=0$, where ${\mathcal C}=\prod_j \sigma_j^z$ is the chirality operator. Due to this property, all even order terms in the usual perturbative expansion of $H_F$ vanishes. However, for $p\ne 1/2$ or for the two-tone protocol discussed in the main text, this anti-commutation relation is not satisfied, leading to a non-zero $H_F^{(2)}$.

The Hamiltonian of the driven chain is given in Eq. (2) of the main text. To derive the evolution operator, we consider the regime $\lambda_0 \gg w$ and use FPT. A straightforward computation shows that 
\begin{eqnarray}
U_0(t,0) &=& e^{i t\lambda_0 \sum_j \sigma_j^z/\hbar} \quad {\rm for}\,\, t\le pT_1 \nonumber\\
&=& e^{i (2pT_1-t)\lambda_0 \sum_j \sigma_j^z/\hbar}\quad {\rm for}\,\, pT_1 < t \le T_1. \label{uoexp} 
\end{eqnarray}
We note that for $p\ne 1/2$, $U_0(T_1,0)$ remain finite and lead to the leading order Floquet Hamiltonian given by $H_F^{(0)} = (i\hbar/T_1)\ln U_0(T_1,0)$. We note that since $\sum_j \sigma_j^z$ acting on any Fock state only yields even integer eigenvalues (for chain length $L$ being even), $U_0(T_1,0) \to I$ for $\lambda_0 T_1(2p-1)/\hbar= m_1\pi$, where $m_1\in Z$. 

To compute $U_1$, we use Floquet perturbation theory (FPT) which yields \cite{rev12}
\begin{eqnarray} 
U_1(T_1,0) &=& \frac{-i}{\hbar} \int_0^{T_1} U_0^{\dagger}(t,0) H_1 U_0(t,0),
\end{eqnarray} 
where $H_1= w_0 \sum_j \tilde \sigma_j^x$. A straightforward computation similar to that carried out in Refs.\ \onlinecite{scar2,scar5} yields 
\begin{eqnarray} 
U_1(T_1,0) &=& \frac{i w_0}{2 \hbar} \sum_{s=\pm} \sum_{j=1}^{L} {C_s^{(1)} \tilde{\sigma_j}^{s}}, \nonumber\\
C_s^{(1)} &=& \frac{2is}{\lambda_0} \left( 1 - 2 e^{2i \lambda_0 s p T_1/\hbar} + e^{2 i \lambda_0 s (2p - 1) T_1/\hbar} \right). \label{u1exp}
\end{eqnarray} 
We note that $C_s^{(1)}$ vanishes if one can simultaneously satisfy $\lambda_0 T_1 p/\hbar= m_2 \pi $ and $\lambda_0 T_1(2p-1)/\hbar=m_1 \pi$, where $m_1,m_2 \in Z$. At these frequencies, both $U_0(T_1,0)$ and $U_1(T_1,0)$ becomes identity matrix lading to $H_F^{(0)}=H_F^{(1)}=0$. 

The computation of $U_2(T_1,0)$ can be carried out within FPT using a straightforward manner
and is given by
\begin{eqnarray}
 U_2(T_1,0) &=& \left(\frac{-i}{\hbar}\right)^2 \int_0^{T_1} dt_1 U_0^{\dagger}(t_1,0) H_1 U_0(t_1,0) \int_0^{t_1} dt_2 U_0^{\dagger}(t_2,0) H_1 U_0(t_2,0). \label{u2expa} 
 \end{eqnarray}
 The computation of $U_2(T_1,0)$ was charted out for similar protocols in Refs.\ \onlinecite{scar5}. Following a similar procedure, we find
 \begin{eqnarray}
 U_2(T_1,0) &=& \left(\frac{-i w_0}{\hbar}\right)^2 \sum_{s_1,s_2=\pm} C_{s_1,s_2}^{(2)} \sum_{j_1,j_2} 
 \tilde \sigma_{j_1}^{s_1} \tilde \sigma_{j_2}^{s_2}, \nonumber\\
 C_{s,s}^{(2)} &=& -\frac{\hbar^2}{8\lambda_0^2} \left( 1- 2 e^{2 i p s T_1\lambda_0/\hbar} + 2e^{4i p s T_1\lambda_0/\hbar} + e^{4 i (2p-1) s T_1\lambda_0/\hbar} -2 e^{2i (3p-1) s T_1\lambda_0/\hbar} \right), \nonumber\\
 C_{s,-s}^{(2)} &=& -\frac{\hbar^2}{8\lambda_0^2} \left( -2 + e^{2 i (1-p) s T_1\lambda_0/\hbar} + e^{2i p s T_1\lambda_0/\hbar} -2i \lambda_0 T_1(2p-1)/\hbar \right). \label{u2expb}
\end{eqnarray}
We note that for $p=1/2$ where $H_F^{(0)}$ vanishes, our expression reduces to that of Refs.\ \onlinecite{scar2,scar5}. Moreover, at special points where $U_0= U_1= I$, one can compute $H_F^{(2)}$ using Eq.\ \eqref{u2expb}. At these points $C^{(2)}_{s,s}=0$ and $C^{(2)}_{s,-s}= -\pi m_1 \hbar^2/(4\lambda_0^2)$, we find 
\begin{eqnarray}
 H_F^{(2)}= (i \hbar/T_1) \ U_2(T_1,0)= {\mathcal N}_1 \left[\sum_j P_{j-1} (\sigma_j^{+} \sigma_{j+1}^- + {\rm H.c.}) P_{j+2} + \sum_j \tilde \sigma_j^z \right], \label{hf2exp}
\end{eqnarray}
 where ${\mathcal N}_1= -w_0^2 \hbar \pi m_1/(4 T_1\lambda_0^2)$. Thus, at these points, the leading order Floquet Hamiltonian is the same as the one obtained from the two-tone protocol.
 
\begin{figure}[t]
\centering
\includegraphics[width=0.245\linewidth]{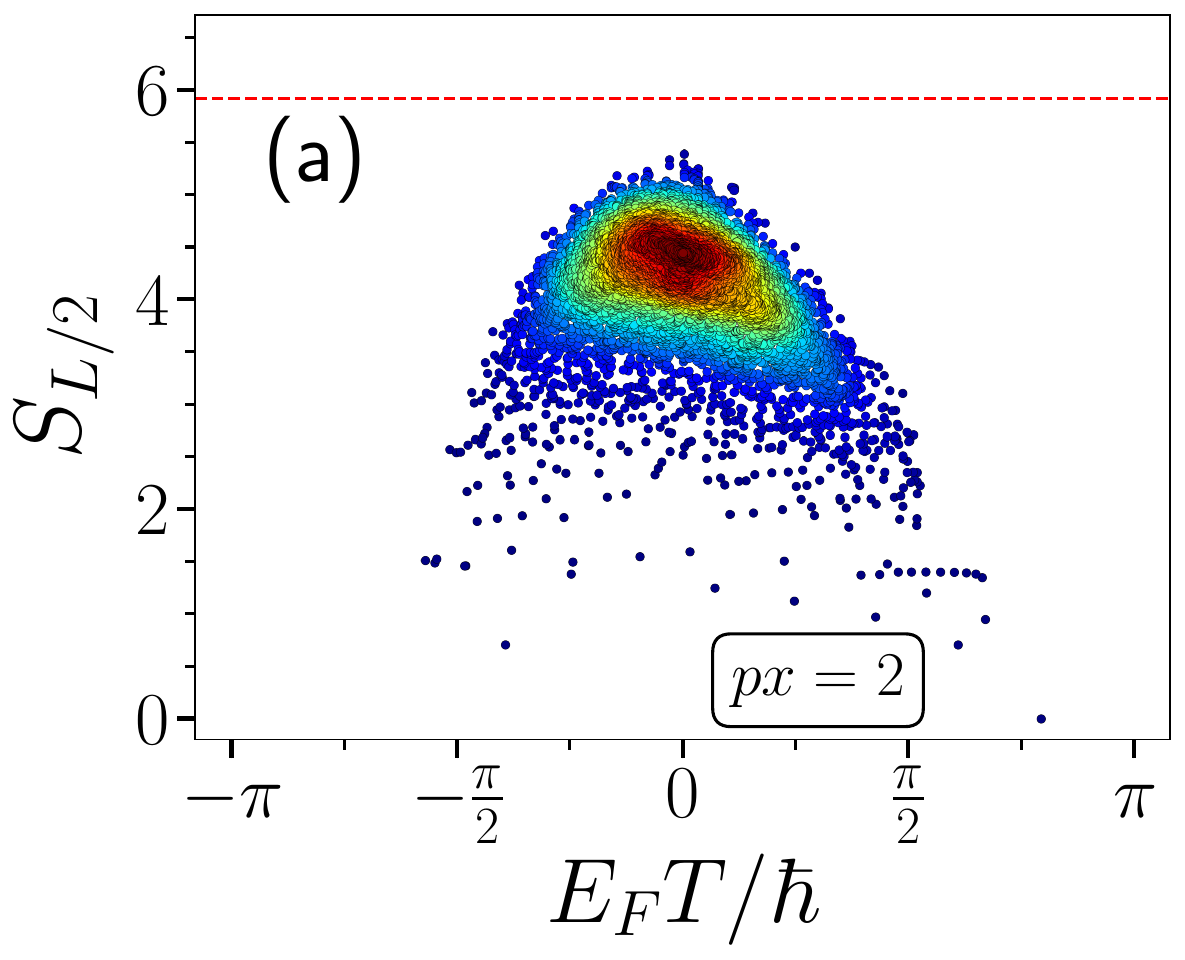}
\includegraphics[width=0.245\linewidth]{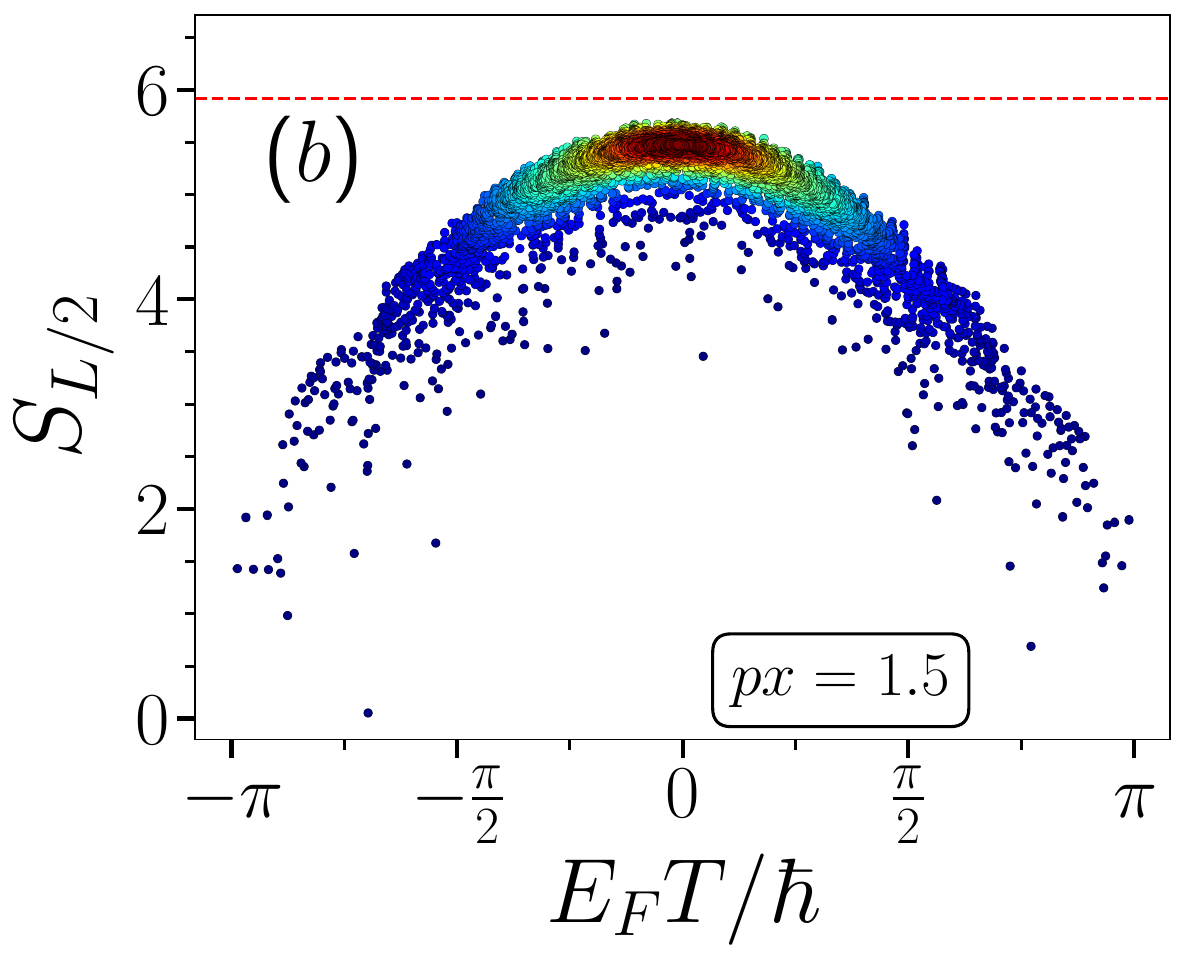}
\includegraphics[width=0.245\linewidth]{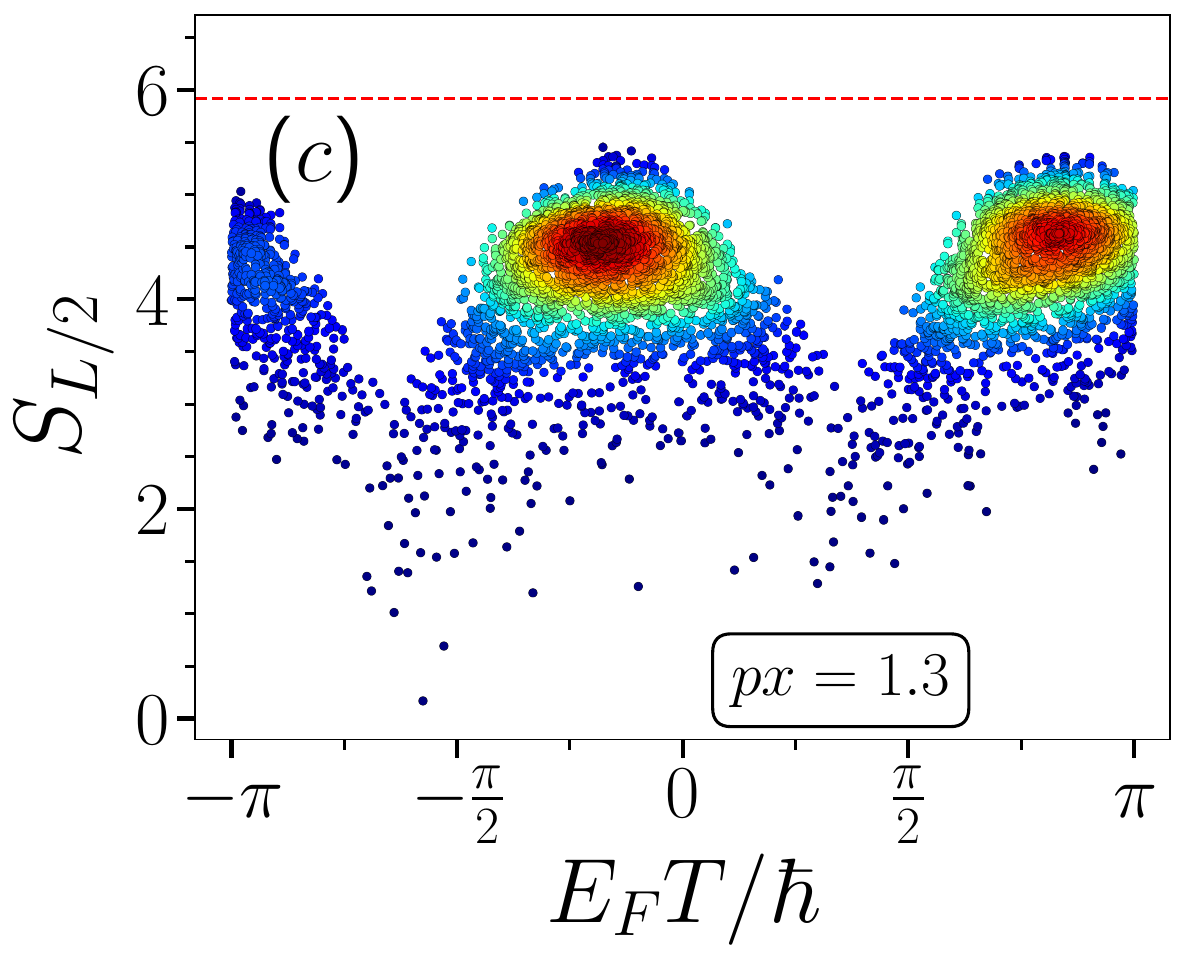}
\includegraphics[width=0.245\linewidth]{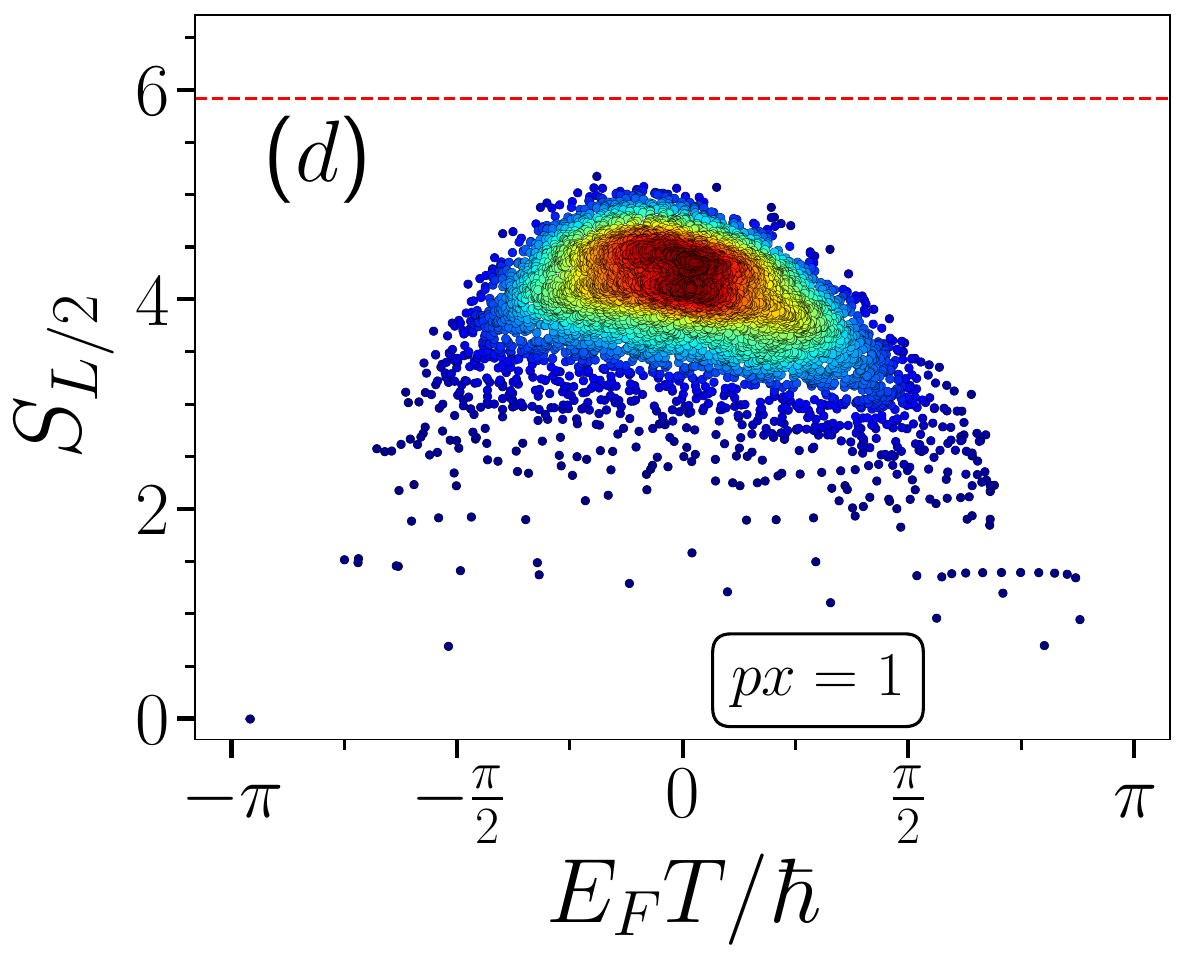}
\centering
\caption{Plot of half-chain entanglement $S_{L/2}$ as a function of $E_FT_1/\hbar$ for Floquet quasienergies for $x=\lambda_0 T_1/(\pi \hbar)=10$, (a) $px=2$, (b) $px=1.5$, (c) $px=1.3$, and (d) $px=1$. For all plots the system size is $L=26$ with PBC, and we have considered the translation symmetric and parity symmetric sector.
\label{fig1_SI}}
\end{figure} 

 Next, we study the dependence of the average consecutive level spacing ratio, $r$ as a function of the drive frequency. To this end, we define $x=\lambda_0 T_1/(\pi \hbar)$ and choose $x=10$ for our numerics. This means that special frequencies occur when $p=0.1 m$, where $m \in Z^+$. The special points, for which $px$ and $(1-p)x$ are integers, correspond then to $p = 0.1,0.2,0.3, \cdots$ and the points where $U_0(T_1,0) = \mathbbm{1}$ are at $p =0.05,0.1,0.15,0.2,0.25,0.3, \cdots$ provided $L$ is even.

 A plot of the half-chain entanglement, $S_{L/2}$ for the Floquet eigenstates is shown in Fig.\ \ref{fig1_SI} for $x= \lambda_0 T_1/(\pi \hbar)=10$ and several representative values of $p$. We find that at special frequencies $px \in Z$, where $U_0=U_1=I$ and the dynamics is controlled by $H_F^{(2)}$, $S_{L/2}$ shows a broad spread as shown in Fig.\ \ref{fig1_SI}(a) and (d) for $px=2$ and $1$ respectively; the entanglement of the Floquet eigenstates does not fall within a thermal band as expected for an ergodic driven model. The latter behavior is seen for $px \in Z+1/2$, where $U_0= I$ and $H_F^{(1)}$ is finite; this is shown in Fig.\ \ref{fig1_SI}(b) for $px=1.5$. Finally, for $px=1.3$, where the dynamics is controlled by $U_0$, we find that the eigenstates are separated into groups labeled by the total $S_z = \sum_j \sigma_j^z$ and, once again, show a wide spread. This behavior, shown in Fig.\ \ref{fig1_SI}(c), is expected since the dominant term in the perturbative Floquet Hamiltonian is trivially integrable in this case.

\begin{figure}[t]
\centering
\rotatebox{0}{\includegraphics*[width=0.49\linewidth]{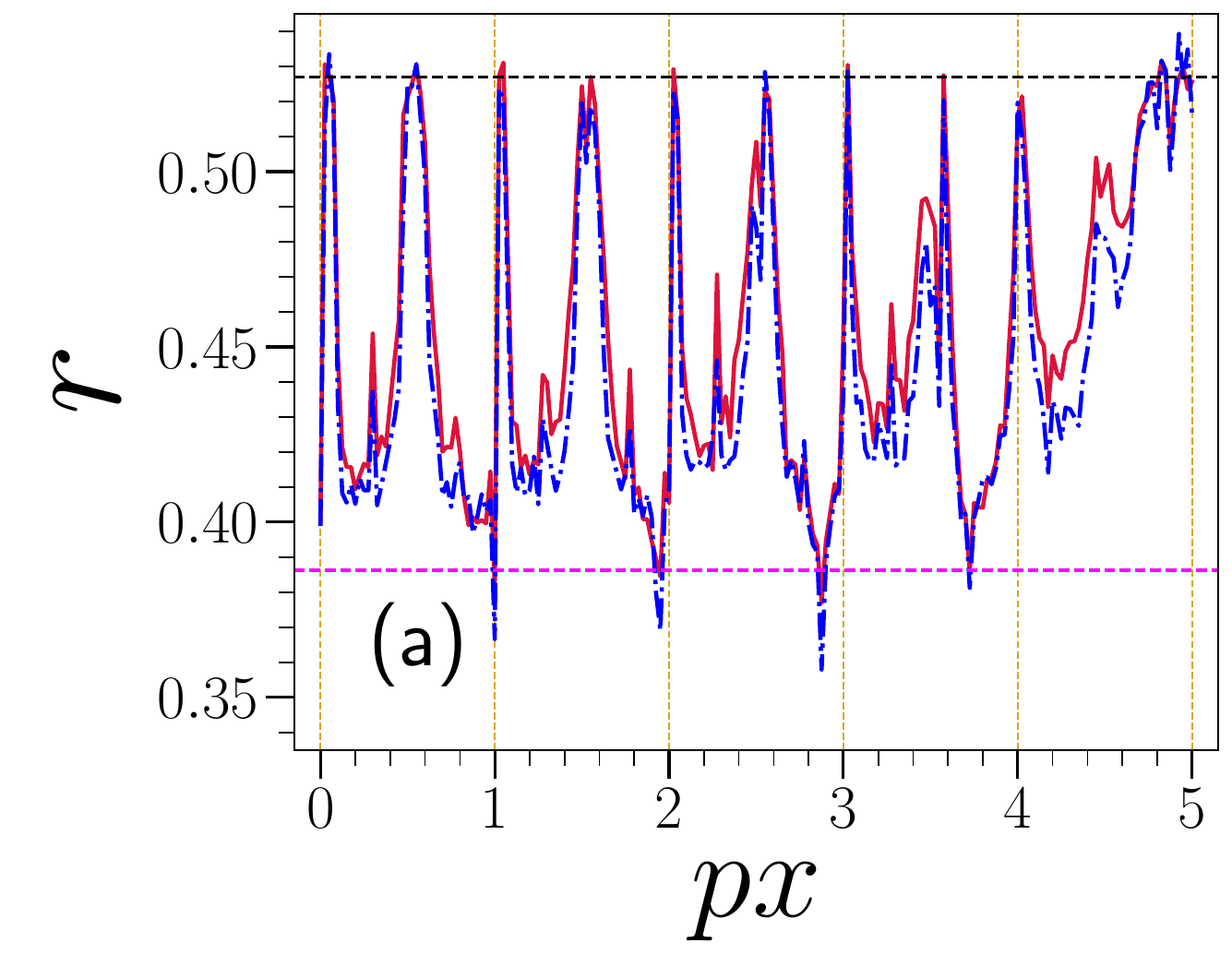}}
\rotatebox{0}{\includegraphics*[width=0.49\linewidth]{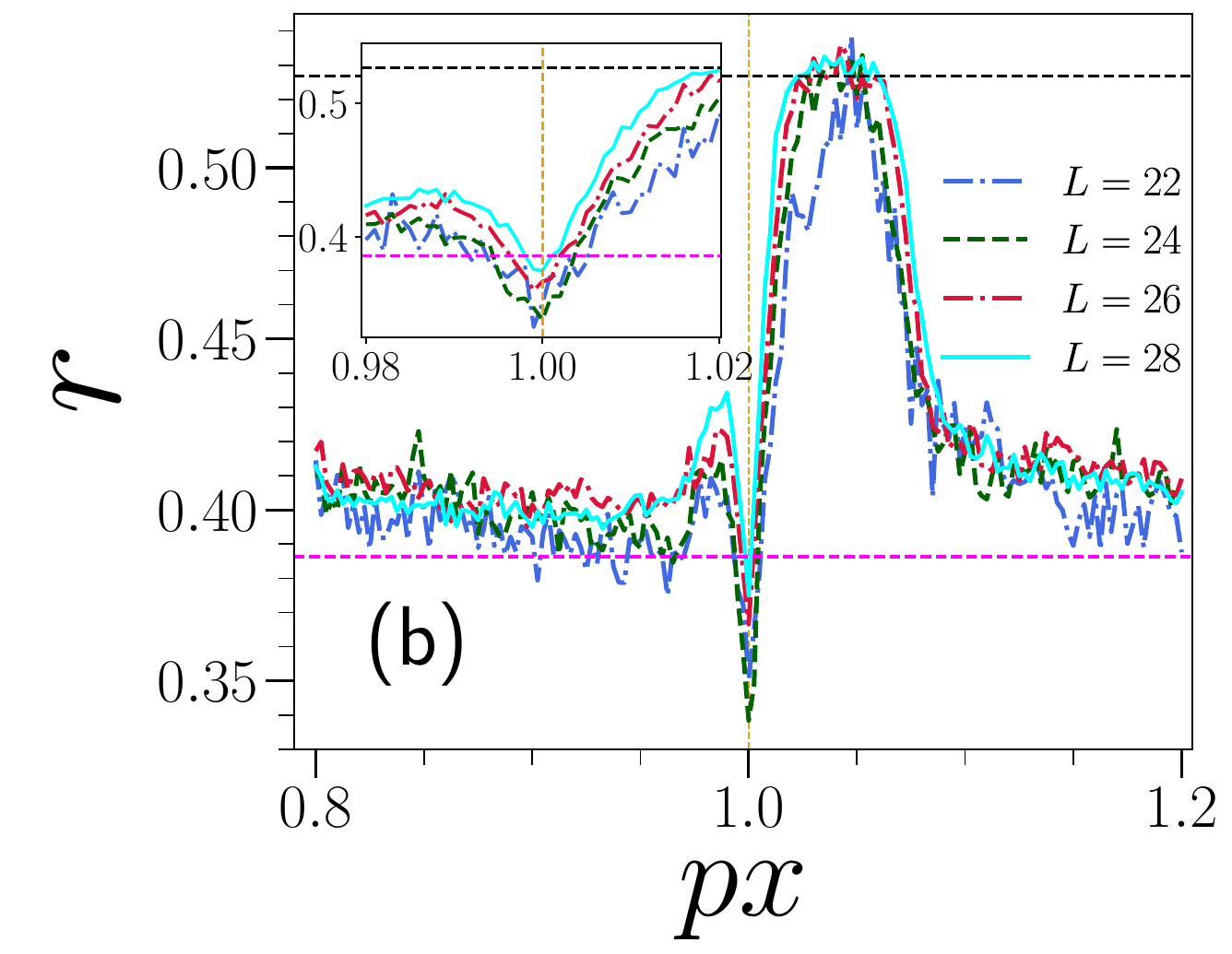}}
\centering
\caption{(a) Plot of $r$ as a function of $px$ where $x=\lambda_0 T_1/(\pi \hbar) = 10$ for $L=20$, $R=+1$ sector with OBC (red) and $L=26$, $K=0$, $P=+1$ sector with PBC showing sharp dips at $px=m$ for integer $m$ which correspond to special points. The magenta (black) dashed lines $r=0.39 ~(0.527)$ correspond to the value of $r$ for the Poisson ensemble (COE). (b) Same as in (a) for $K=0,P=+1$ sector with PBC but around $px=1$, for different chain lengths $L$. The inset shows a closeup of a sharp rise of $r$ from $0.39$ to $0.527$.
\label{fig2_SI}}
\end{figure} 

A plot of the average consecutive level spacing ratio, $r$, of the Floquet eigenstates is shown in Fig.\ \ref{fig2_SI}. Fig.\ \ref{fig2_SI}(a) shows $r$ as a function of $px$; for low $px$, one finds dips for $px \in Z$ leading to $r \sim 0.38$ which is consistent with the emergent integrability. At higher values of $px >2$, these dips deviate from their integer values and disappear from $px>4$. The shift of these dips can be understood to be renormalization of the special drive frequencies due to higher order terms in the Floquet Hamiltonian. In contrast, the absence of dips indicates that the second-order Floquet Hamiltonian does not control the dynamics at lower frequency; in this regime, the emergent integrability disappears. A close-up of the dip in $r$ around $px=1$ is shown in Fig.\ \ref{fig2_SI}(b) for several $L$. The behavior of $r$ stays close to $0.4$ for $px<1$ whereas it rises sharply to the COE value of 0.53 for $px>1$. The behavior of $r$ for $px>1$ mimics that of the two-tone drive. However, for $px<1$, this is not the case; this asymmetric response constitutes a key difference between the two-tone and the single-tone drive protocols. We have checked that the behavior of $r$ around the special points becomes symmetric for much larger drive amplitudes ($\lambda_0/w \sim 100$). This indicates that such an asymmetry arises from higher order terms in the Floquet Hamiltonian; a detailed analysis of this feature is left for future study.

Before ending this section, we note that if an asymmetry in drive amplitude is also incorporated, such as 
\begin{eqnarray} 
\lambda(t) &=&-\lambda_1 \quad {\rm for}\, \,0\le t < pT_1 \nonumber\\
&=& \lambda_2, \quad {\rm for}\, \,pT_1\le t < T_1, \label{new_da} 
\end{eqnarray}
with the ratio adjusted for a given $p$
\begin{equation}
    \chi = \frac{\lambda_2}{\lambda_1} = \frac{p}{1-p} \ , \label{cond_amp}
\end{equation}
one gets $U_0(T,0)=\mathbbm{1}$. However in such a set up $\{H_F,\mathcal{C}\}\neq0$ in general and an analogous calculation leads to the following terms in FPT.
\begin{align} 
H_F^{(1)} &= \frac{w_0 \, \sin{\gamma}}{\gamma} \, \sum_j \left( e^{-i \gamma} \ \tilde{\sigma}_j^+ \ + \ \text{H.c.} \right) \label{eq:new_HF1}\\
H_F^{(2)} &= (2p-1) w_0^2 \, T_1\  \left( \frac{\sin (2 \gamma) - 2 \gamma}{4 \gamma^2} \right) \sum_j \left[
P_{j-2}\bigl(\sigma_{j-1}^+ \sigma_j^- + \sigma_{j-1}^- \sigma_j^+\bigr) P_{j+1}
+ \,\tilde{\sigma}_j^z \right] \label{eq:new_HF2}
\end{align}
with $\gamma:= \lambda_1p\, T_1$. Zeros of $H_F^{(1)}$ now corresponds to $\gamma=m \pi\, , \ m\in \mathbb{Z^+}$ where the dominant contribution (in perturbative regime) is given by $H_F^{(2)}$ that contains the same operator structure as \eqref{hf2exp}. Qualitatively this study is similar to the two tone protocol studied in Sec.\ \ref{sqp1}, since control over two tuning parameters is required in either cases.

\section{Mapping to XXZ chain for open boundary condition (OBC)}
\label{mapping}

\begin{table}[b]
\centering
\begin{tabular}{|c | c | c|}
\hline
$(x_1,x_2)$ & State & Mapped state $(y_1,y_2)$ \\
\hline
$(4,6)$ & $\downarrow\downarrow\downarrow\uparrow\downarrow\uparrow$ & $\downarrow\downarrow\downarrow\uparrow\uparrow \, = (4,5)$ \\
$(3,6)$ & $\downarrow\downarrow\uparrow\downarrow\downarrow\uparrow$ & $\downarrow\downarrow\uparrow\downarrow\uparrow \, = (3,5)$ \\
$(3,5)$ & $\downarrow\downarrow\uparrow\downarrow\uparrow\downarrow$ & $\downarrow\downarrow\uparrow\uparrow\downarrow \, = (3,4)$ \\
$(2,6)$ & $\downarrow\uparrow\downarrow\downarrow\downarrow\uparrow$ & $\downarrow\uparrow\downarrow\downarrow\uparrow \, = (2,5)$ \\
$(2,5)$ & $\downarrow\uparrow\downarrow\downarrow\uparrow\downarrow$ & $\downarrow\uparrow\downarrow\uparrow\downarrow \, = (2,4)$ \\
$(2,4)$ & $\downarrow\uparrow\downarrow\uparrow\downarrow\downarrow$ & $\downarrow\uparrow\uparrow\downarrow\downarrow \, = (2,3)$ \\
$(1,6)$ & $\uparrow\downarrow\downarrow\downarrow\downarrow\uparrow$ & $\uparrow\downarrow\downarrow\downarrow\uparrow \, = (1,5)$ \\
$(1,5)$ & $\uparrow\downarrow\downarrow\downarrow\uparrow\downarrow$ & $\uparrow\downarrow\downarrow\uparrow\downarrow \, = (1,4)$ \\
$(1,4)$ & $\uparrow\downarrow\downarrow\uparrow\downarrow\downarrow$ & $\uparrow\downarrow\uparrow\downarrow\downarrow \, = (1,3)$ \\
$(1,3)$ & $\uparrow\downarrow\uparrow\downarrow\downarrow\downarrow$ & $\uparrow\uparrow\downarrow\downarrow\downarrow \, = (1,2)$ \\
\hline
\end{tabular}
\caption{Bijection between hard-core configurations for $L=6$ and $N=2$ and ordinary configurations for $L'=5, ~N=2$ via the
mapping $y_i = x_i - (i-1)$.}
\label{tab1}
\end{table}

In this section, we shall extend the mapping, outlined for periodic boundary condition (PBC) with $K=0$ in the main text, to chains with open boundary condition (OBC). To this end, we first consider $L$ spins on a chain of length $L$ (with lattice spacing set to unity) with OBC such that $N$ of these spins are up (eigenstates of $\sigma_j^z$ on the site $j$ with eigenvalue $1$) and no two up-spins are neighbors.

The computation of all allowed configurations (Fock states) involves choosing $N$ out of $M$ vacant sites on an open chain followed by addition of a particle (up spin) at a chosen site along with an extra empty site (down spin) to its right. The latter operation is done to avoid any configurations with neighboring up-spins. We note that the addition of down spins is needed only for $N-1$ up spins since the $N^{\rm th}$ up spin can never be a neighbor of any of the added up spins by construction. The total number of sites under this construction is therefore $N+M-1= L$ for a spin chain fo length $L$. The number of such configurations (Fock states) is therefore given by $\binom{M}{N} = \binom{L-N+1}{N}$. We note that this process of inserting an up spin followed by a down spin to its right is equivalent to insertion of a hard rod of length $2$ (here the lattice spacing is set to unity). In that language, the mapping reduces to a problem of insertion of $(N-1)$ hard rods and then addition of a particle at $N^{\rm th}$ step. 

Since $\binom{M}{N}$ represents the number of ways of distributing $N$ particles (up spins) on a chain of $M$ vacant sites, the idea of hard rod immediately gives a bijection (one to one, onto, {\it i.e.} invertible) map of the constrained Hilbert space to an unconstrained one once we identify $L$ as $N+M-1$ . This identification can be done by keeping track of the locations of 
the up spins $(x_1,x_2,\cdots,x_N)$ in the original problem as the follows. 

First, we consider an allowed Fock state of length $L$ having $N$ up spins (for example, $\uparrow\downarrow\uparrow\downarrow\downarrow\uparrow\downarrow$ corresponds to $L=7$ and $N=3$) and label it by the locations of the up spins $(x_1,x_2,\cdots,x_N)$ (for example, $\uparrow\downarrow\uparrow\downarrow\downarrow\uparrow\downarrow \equiv (1,3,6)$), where $x_i$ is the location of the $i$-th up spin. Note that we have 
\begin{equation*}
 \begin{aligned}
 1 \leq x_1<x_2< \cdots<x_N \leq L \quad \text{and} \quad x_{i+1} -x_i \geq 2
 \end{aligned}
\end{equation*}
which charts out the constraint. 

Next, we introduce the mapping
\begin{equation} \label{eq2M}
 \begin{aligned}
 y_i = x_i - (i-1) , ~~\ i \in \{1,2,\cdots,N\}
 \end{aligned}
\end{equation}
as the location of the $i$-th up spin in a spin chain with unconstrained $(L-N+1)$ spins on a chain. We note that $N$ of these spins are up by construction since 
\begin{equation*}
 \begin{aligned}
 & 1 \leq y_1<y_2< \cdots<y_N \leq L-N+1 \qquad \\& \text{and} \quad y_{i+1} -y_i = x_{i+1} - x_i -1 \ , \Rightarrow \ y_{i+1} - y_i >=1.
 \end{aligned}
\end{equation*}
The last inequality represents a trivial constraint that no two or more particles can sit on top of each other. The map therefore reduces the distances $(x_2-x_1),(x_3-x_2),\cdots ,(x_N-x_{N-1})$ by an unit. This is equivalent to deleting a down spin to the left of each up spins at $x_2,x_3,\cdots,x_N$ except at $x_1$. An example of this mapping for $L=5$ and $N=2$ is shown in \ref{tab1}. We note that the mapping is a bijection for PBC as well if one eliminates states having up spins at both site $1$ and site $L$ in the constrained Hilbert space; for example the state $(1,6)$ in Table
\ref{tab1} is not allowed and one needs to manually delete it in the constrained space for PBC. 

Next we decouple the Hamiltonian $H_F^{(2)}$ in the main text into two parts $H_{\text{od}}$ and $H_{\text{d}}$ which are given by 
\begin{align}
 H_{\text{od}} &= J \sum_j P_{j-1} \left( \sigma_j^- \sigma_{j+1}^+ \ + \ \sigma_j^+ \sigma_{j+1}^- \right) \, P_{j+2}, \label{eq3M}\\
 H_{\text{d}} &= J \sum_j P_{j-1} \, \sigma_j^z \, P_{j+2}. \label{eq4M}
\end{align}
In what follows, we consider the effect of the mapping described above for each of these terms as done in the main text for PBC. 

\subsection {Off-diagonal term}
\label{odt} 

We note that $H_{{\rm od}}$ involves the flipping of spins on neighboring sites and yields non-zero matrix elements between states with $|\cdots \downarrow \uparrow \downarrow \downarrow \cdots\rangle$ and $|\cdots \downarrow \downarrow \uparrow \downarrow \cdots\rangle$. In the particle language, this corresponds to a hopping
to the right for the constrained model with the positions of the particles being denoted by $x_i$. The hopping $x_i \to x_{i}+1$ is allowed iff $x_{i+1} \geq x_i +3$ , i.e. the tagged up spin at $x_i$ must be separated from the next up spin at its right at $x_{i+1}$ by at least three lattice sites. This is ensured by the projectors in Eq.\ \eqref{eq3M}. 

The condition under the mapping gets translated as follows. First, we note that $y_i \to y_i +1$ hopping will only be allowed
\begin{eqnarray} 
y_{i+1} + i &\geq&  y_i + (i-1) + 3,  \quad  {\it i.e.}\,\, \ y_{i+1} \geq y_i + 2. \label{hopeq} 
\end{eqnarray} 
Eq.\ \eqref{hopeq} indicates that a hop of a particle from site $i$ to site $i+1$ requires that the site $i+1$ initially be empty. Hence,  the next particle must be separated from it by at least two units. $H_{\text{od}}$ thus gets mapped to a `free' hopping Hamiltonian $ J \sum_j \left( \tau_j^- \tau_{j+1}^+ \ + {\rm H.c.}\right)$ under OBC where $\tau^{\alpha}$ are Pauli matrices. This term can be represented by free fermions whose eigenstates are plane waves.

Notice that under PBC this mapping immediately breaks down since the hopping at the boundaries produces states which are out of the constrained Hilbert space. For example, in the mapped space (in table-\ref{tab1}) the Hamiltonian can give a state $(1,5)=\uparrow \downarrow \downarrow \downarrow \uparrow$ acting on $(1,4)=\uparrow \downarrow \downarrow \uparrow \downarrow $ although pre-image of $(1,5)$ violates the constraint 
under PBC. This ambiguity can be resolved if we disallow hopping when one of the sites $1$ or $L$ are involved (hopping between sites $1$ and $L$ is allowed); however, imposition of such a restriction breaks translation invariance of the mapped Hamiltonian. Since we have already provided an exact mapping for PBC in the $K=0$ sector in the main text, we shall not discuss PBC here any further. 

We now show explicitly that mapping of this constrained hopping term to a free-fermionic model under OBC does not require such an extra measure. We write the off-diagonal term (\ref{eq3M}) explicitly under open boundaries
\begin{align} \label{5M}
H_{\text{od}} &= J \sum_{j=2}^{L-2} P_{j-1} \left( \sigma_j^- \sigma_{j+1}^+ + \sigma_j^+ \sigma_{j+1}^- \right) \, P_{j+2} \ + J \left( \sigma_1^- \sigma_2^+ + \sigma_1^+ \sigma_2^- \right) P_3 \ + J \, P_{L-2} \left( \sigma_{L-1}^- \sigma_L^+ + \sigma_{L-1}^+ \sigma_L^-\right).
\end{align}
Using the same argument as before, the bulk term gets mapped into a free hopping in the bulk of the mapped chain as
\begin{align*}
\sum_{j=2}^{L-2} P_{j-1} \left( \sigma_j^- \sigma_{j+1}^+ + \sigma_j^+ \sigma_{j+1}^- \right) \, P_{j+2} \ 
& \Rightarrow \ \sum_{j=1}^{L'-1} \left( \tau_j^- \tau_{j+1}^+ + \tau_j^+ \tau_{j+1}^- \right).
\end{align*}
For the term on the left edge, $\sigma_1^- \sigma_2^+ P_3$ ($\sigma_1^+ \sigma_2^- P_3$), we note that in order to hop the first up spin unit distance to its right (left) one requires $x_2 -x_1 \geq 3\ (2)$. This translates in terms of new variables as $y_2 -y_1 \geq 2 \ (1)$ and leads to the mapped operator
$\tau_1^- \tau_2^+ \ (\tau_1^+ \tau_2^-)$. Similarly for the term on right edge, $P_{L-2}\, \sigma_{L-1}^- \sigma_L^+ \ (P_{L-2}\, \sigma_{L-1}^+ \sigma_L^-) $, hopping of $N$-th up spin unit distance to its right (left) requires the constraint $x_N - x_{N-1} \geq 2 \ (3)$. This translates to the condition $y_N - y_{N-1} \geq 1 \ (2)$ and yields the term $\tau_{L'-1}^- \tau_{L'}^+ \ (\tau_{L'-1}^+ \tau_{L'}^-)$. Hence, under OBC, one gets the mapped version of $H_{\text{od}}$ of Eq.~\eqref{5M} as 
\begin{align} \label{eqM1}
H_{\text{od}} \ & \Rightarrow \ J \sum_{j=1}^{L'-1} \left( \tau_j^- \tau_{j+1}^+ + \tau_j^+ \tau_{j+1}^- \right) = \frac{J}{2} \sum_{j=1}^{L'-1} \left( \tau_j^x \tau_{j+1}^x + \tau_j^y \tau_{j+1}^y \right).
\end{align}

\subsection{Diagonal term}
\label{dt}
Next, we consider the diagonal term $H_{\rm d}$ (\ref{eq4M}) written explicitly in OBC as 
\begin{align} \label{eq6M}
H_{\mathrm {d}} &= J \sum_{j=2}^{L-1} P_{j-1} \, \sigma_j^z \, P_{j+1} \ + \ J \sigma_1^z P_2 \ + \ J P_{L-1} \sigma_L^z.
\end{align}
Using the relations $P_k = (\mathbbm{1} -n_k), \ \sigma_k^z = (2n_k - \mathbbm{1})$ and the constraint $n_k \ n_{k+1}=0$, we find 
\begin{equation*}
\begin{aligned}
 P_{j-1} \ \sigma_j^z \ P_{j+1} &= (\mathbbm{1} - n_{j-1}) \ (2n_j-\mathbbm{1}) \ (\mathbbm{1} - n_{j+1}) \\
 &= 2n_j-\mathbbm{1} + n_{j+1} + n_{j-1} - n_{j-1}\, n_{j+1}.
\end{aligned}
\end{equation*}
The term $n_{j-1} \, n_{j+1}$ annihilates a Fock state unless it hosts up spins at both sites $j \pm 1$. It essentially means $\sum_j \ n_{j-1} \, n_{j+1}$ searches for $\uparrow \downarrow \uparrow$ on three consecutive sites. If the up spin at $(j-1)$ is the $k$'th one, under the map the distance $x_{k+1} - x_k=(j+1)-(j-1)=2$ gets reduced by unity and hence leads to an $\uparrow \uparrow$ configuration. Thus one can map 
$$ \sum_{j=2}^{L-1} \ n_{j-1} \, n_{j+1} \Rightarrow \sum_{j=1}^{L'-1} \ n^{(\tau)}_{j} \, n^{(\tau)}_{j+1},$$
which is essentially a density-density term (Hubbard-like). In terms of the spin variables $\tau$'s, apart from some constant and boundary terms, this leads to $ZZ$ interaction terms given by
\begin{equation*}
 \begin{aligned}
 -\sum_{j=1}^{L'-1} n^{(\tau)}_j \, n^{(\tau)}_{j+1} &= -\frac{1}{4} \sum_{j=1}^{L'-1} (\mathbbm{1}+ \tau_j^z) \, (\mathbbm{1}+ \tau_{j+1}^z) \\
 &= -\frac{(L'-1)}{4} -\frac{1}{4} \sum_{j=1}^{L'-1} \tau_j^z -\frac{1}{4} \sum_{j=1}^{L'-1} \tau_{j+1}^z - \frac{1}{4} \sum_{j=1}^{L'-1} \tau_j^z \, \tau_{j+1}^z \\
 &= -\frac{(L'-1)}{4} -\frac{1}{2} \sum_{j=1}^{L'} \tau_j^z - \frac{1}{4} \sum_{j=1}^{L'-1} \tau_j^z \, \tau_{j+1}^z + \frac{1}{4} (\tau_1^z + \tau_L^z) \\
 &= \frac{L' -4N+1}{4} - \frac{1}{4} \sum_{j=1}^{L'-1} \tau_j^z \, \tau_{j+1}^z + \frac{1}{4} (\tau_1^z + \tau_L^z).  
 \end{aligned}
\end{equation*}
Here we have used $$\sum_{j=1}^{L'} \tau_j^z = 2N -L'\, , ~~ \text{where} ~ \ \sum_{j=1}^{L'} n_j^{(\tau)} = N ~~ \text{is a given constant.}$$

The left over terms in the bulk are
\begin{equation*}
 \begin{aligned}
 \sum_{j=2}^{L-1} \left( 2n_j-\mathbbm{1} + n_{j+1} + n_{j-1}\right) &= 4 N - (L-2) -3(n_1 + n_L) - (n_2 + n_{L-1})
 \end{aligned}
\end{equation*}
which, together with 
\begin{align*}
 \sigma_1^z P_2 + P_{L-1} \, \sigma_L^z &= (2n_1 - \mathbbm{1})(\mathbbm{1}-n_2) \ + (\mathbbm{1} - n_{L-1}) (2n_L -\mathbbm{1}) \\
 &= 2(n_1 + n_L) + (n_2 + n_{L-1})-2,
\end{align*}
allows us to rewrite Eq.~\eqref{eq6M} as 
\begin{align*}
H_{\mathrm{d}} &= 4J N - J(n_1 + n_L) - JL - J\sum_{j=2}^{L-1} n_{j-1} n_{j+1}.
\end{align*}
This now gets mapped as follows.
\begin{equation}\label{eqM2}
\begin{aligned} 
H_{\mathrm{d}} &\Rightarrow \ 4J N - J(n^{(\tau)}_1 + n^{(\tau)}_{L'}) - JL - J\sum_{j=1}^{L'-1} n^{(\tau)}_{j} n^{(\tau)}_{j+1}\\
&= \ - \frac{J}{4} \sum_{j=1}^{L'-1} \tau_j^z \tau_{j+1}^z \ - \frac{J}{4} (\tau_1^z + \tau_{L'}^z) + \frac{J}{4} (11N -3L +2).
\end{aligned}
\end{equation}
Hence the diagonal term gets mapped to the $ZZ$ interaction term in addition to magnetic field terms at the two boundaries and a constant term. Combining Eqs.~\eqref{eqM1} and \eqref{eqM2} then gives 
\begin{equation}
\begin{aligned} \label{eqM3}
 H_F^{(2)} \ &\Rightarrow \ J' \sum_{j=1}^{L'-1} \left[\left( \tau_j^x \tau_{j+1}^x + \tau_j^y \tau_{j+1}^y \right) - \frac{1}{2} \tau_j^z \tau_{j+1}^z \right] \ - \frac{J'}{2} (\tau_1^z + \tau_{L'}^z) \ + \frac{J'}{2} (11N -3L +2) \\
 &= H_{\mathrm{XXZ}} \ + H_{ext} \ + \mathrm{const}, \\
 H_{ext} & = - \frac{J'}{2} (\tau_1^z + \tau_{L'}^z),
\end{aligned}
\end{equation}
where $J'=J/2$. Here we note that under the exact mapping in the case of PBC in $K=0$ sector where we also delete the down-spin to the left of $x_1$, the exact XXZ Hamiltonian that $H_F^{(2)}$ gets mapped into has the same anisotropic parameter $\Delta=-1/2$ as one gets under this mapping here in OBC. The extra term $H_{ext}$, which is symmetric under space reversal, gives an $\mathcal{O}(1)$ contribution in comparison to $H_{\mathrm{XXZ}}$ which is of $\mathcal{O}(L)$. Hence for large $L$, $H_{\mathrm{ext}}$ makes a negligible contribution to the bulk spectrum. This explains the presence of integrable features in the eigenspectrum of $H_F$ manifested by, for example, dips of $r$ to $0.39$, at the special drive frequencies where $H_F^{(2)}$ forms the leading term. 

Before ending this section, we note that $H_F^{(2)}$ is integrable for PBC in any $K$ sector as demonstrated in Refs.\ \onlinecite{intryd1,intryd2}. However, in these sectors, a mapping of such a constrained model to an unconstrained spin-$1/2$ XXZ 
model does not exist; such a mapping is possible only in the $K=0$ sector for PBC as shown in the main text and, with additional boundary terms, for OBC as discussed in this section.

\section{ Details of numerical procedure}
\label{numdetails} 

In this section, we provide some details of the numerical procedures used to obtain our results on the level spacing statistics, entanglement entropy and spectral form factors.

\subsection{Floquet eigenspectrum}
\label{flsp} 

To obtain the Floquet spectrum, we first diagonalize the evolution operator $U(T_1,0)$. 
For this, we use exact diagonalization (ED) technique (in particular, standard linear algebra routines from the LAPACK library) and obtain eigenvalues $\Lambda_p= e^{i \theta_p(T_1;q)T_1/\hbar }$
(which lie on a unit circle) and eigenvectors $|p\rangle$. The latter are also the eigenvectors of the Floquet Hamiltonian $H_F$; the eigenvalues of $H_F$ are obtained from $\Lambda_p$ as $E_F^p =\hbar \arccos({\rm Re}[\Lambda_p])/T_1$. 

To this end, we first note that for the square-pulse, one can denote the instantaneous Hamiltonian as 
\begin{eqnarray}
H[a,b] = \sum_j (w_0 +b w_1) \tilde \sigma_j^x +a \lambda_0 \sigma_j^z), \label{habeq} 
\end{eqnarray}
where $a,b=\pm 1$. The eigenvalues and eigenvectors of $H[a,b]$ are given by 
\begin{eqnarray} 
H[a,b] |n_p^{a,b}\rangle &=& \epsilon[n_p^{a,b}] | n_p^{a,b} \rangle. \label{eigen1}
\end{eqnarray}
The time-evolution operator corresponding to any specific $[a,b]$ can be expressed as
\begin{eqnarray}
U_{a b}(t,0) &=&\sum_{n_p^{a,b}} e^{-i \epsilon[n_p^{a,b}] t/\hbar} |n_p^{a,b}\rangle \langle n_p^{a,b} |. \label{uab1}
\end{eqnarray}
Using these decomposition one can express $U(T,0)$ as a matrix. This follows from the fact that $U(T,0)$ for $\omega_2=3 \omega_1$ can be written as 
\begin{eqnarray} 
U(T_1,0) &=& U_{-1 -1}(T_1/6,0) \, U_{-1 1}(T_1/6,0) \,U_{-1 -1}(T_1/6,0)\,U_{1 1} (T_1/6,0) \,U_{1 -1}(T_1/6,0) \, U_{1 1}(T_1/6,0) \nonumber\\
 &=& \sum_{n_{p_6}^{-1,-1} .. n_{p_1}^{1,1}} \exp \left[-i T_1 (\epsilon[n_{p_1}^{1,1}] + \epsilon[n_{p_2}^{1,-1}]\epsilon[n_{p_3}^{1,1}]\epsilon[n_{p_4}^{-1,-1}]\epsilon[n_{p_5}^{-1,1}]\epsilon[n_{p_6}^{-1,-1}])/(6\hbar) \right] \nonumber\\
 && \times c_{1,-1;1,1}^{n_{p_2}^{1,-1} n_{p_1}^{1,1} } \,c_{1,1;1,-1}^{n_{p_3}^{1,1} n_{p_1}^{1,-1} } \,c_{-1,-1;1,1}^{n_{p_4}^{-1,-1} n_{p_3}^{1,1} } \, c_{-1,1;-1,-1}^{n_{p_5}^{-1,1} n_{p_4}^{-1,-1} } \,c_{-1,-1;-1,1}^{n_{p_6}^{-1,-1} n_{p_5}^{-1,1} } \, \, |n_{p_6}^{-1,-1}\rangle\langle n_{p_1}^{1,1}|, \label{umat1}
 \end{eqnarray} 
 where $ c_{i,j; k,\ell}^{n_{p_a}^{i,j}, n_{p_b}^{k,\ell}} = \langle n_{p_a}^{i,j}|n_{p_b}^{k,\ell}\rangle $. Having obtained the matrix elements of $U$, we utilize a Schur decomposition of the unitary operator,
\begin{equation}
U' = V \ \widetilde{T} \ V^\dagger,
\end{equation}
where $\widetilde{T}$ is an upper triangular matrix whose diagonal entries are the eigenvalues of $U$. For normal matrices such as $U$, this decomposition yields a unitary matrix $V$, ensuring that the eigenvectors of $U$ form an orthonormal basis. 

\subsection{Calculation of level spacing ratio}

The computation of the level spacings ratios require a list of quasienergies $\{E^F_{p+1}\}$ arranged in ascending order. The quasienergies are first sorted in ascending order within the first Floquet Brillouin zone extending from $-\pi/T_1$ to $\pi/T_1$. We then define the $p^{\rm th}$ adjacent level spacing ratio as 
\begin{equation}
r_p := \frac{\mathrm{min}(s_p, s_{p+1})}{\mathrm{max}(s_p, s_{p+1})},
\end{equation} 
where $s_p = (E^F_{p+1}-E^F_{p})$ is the $p$-th level spacing. The mean level spacing ratio is then calculated as the arithmetic mean of $\{r_p\}$ using the distribution $P(r_p)$; for this purpose, we construct a histogram of the set ${r_p}$. Specifically, the interval $r \in [0,1]$ is divided into bins of width $\Delta r$. The bin width $\Delta r$ is chosen such that it provides a balance between statistical resolution and noise. In practice, we fix $\Delta r = \frac{1}{N_{\mathrm{bin}}}$, where $N_{\mathrm{bin}}$ is the total number of bins (typically $N_{\mathrm{bin}} \sim 50$–$100$). We have verified that our results are insensitive to moderate variations of $\Delta r$ within this range. The number of values $r_p$ that fall within each bin is counted to obtain the histogram, which is then normalized to obtain the probability distribution $P(r_p)$. The resulting $P(r_p)$ provides a detailed characterization of the spectral correlations.

\subsection{Symmetry Sector Resolved Entanglement Entropy}

We compute the half-chain von Neumann entanglement entropy of Floquet eigenstates in the constrained Hilbert space of the PXP model, restricted to the symmetry sector of zero momentum ($K=0$) and even parity ($P=+1$). For this purpose, we consider a spin chain of length $L$ with periodic boundary conditions. The spins on the sites of the chain are subjected to the Rydberg blockade constraint which forbids adjacent excited Rydberg atoms. The Hilbert space is therefore restricted to configurations ${ n_i \in {0,1} }$ that satisfy
\begin{equation}
n_i n_{i+1} = 0 \quad \forall i.
\end{equation}
The dimension of this constrained Hilbert space grows as $\varphi^L$, where $\varphi=[(\sqrt{5}+1)/2]$ is the golden ratio, for large $L$. To further reduce the Hilbert space, we exploit translation and reflection symmetries. For each configuration $|\alpha\rangle$, we construct its translation orbit
\begin{equation}
\mathcal{O}_\alpha = \{ T^r |\alpha\rangle ,| r = 0, \dots, R_\alpha-1 \},
\end{equation}
where $R_{\alpha}$ is the orbit size. Next, we identify a representative state $|\tilde{\alpha}\rangle$ as the configuration with the minimal integer value (in decimal representation of the binary Fock states) within the orbit. The reflection symmetry is implemented via the operator $\mathcal{P}$, which reverses the order of the sites. Depending on whether the reflected configuration lies within the same translation orbit, we define a parity multiplicity $p_\alpha = 1$ (self-symmetric) or $p_\alpha = 2$ (distinct partner). The symmetry-adapted basis states in the $(K=0, P=+1)$ sector are constructed as
\begin{equation}
|\phi_\alpha\rangle = \frac{1}{\sqrt{R_\alpha p_\alpha}}
\sum_{r=0}^{R_\alpha-1} \left( T^r |\tilde{\alpha}\rangle + \mathcal{P} T^r |\tilde{\alpha}\rangle \right),
\end{equation}
where $\mathcal{P}$ commutes with translation operator $T$ in $K=0$ (and $K=\pi$) momentum sector.

To compute the entanglement entropy, we partition the system into two equal halves, $A$ and $B$, each of size $L/2$. A given symmetry-adapted basis state $|\phi_\alpha\rangle$ corresponds to a superposition of $R_\alpha p_\alpha$ configurations in the full Hilbert space. For a Floquet eigenstate having quantum numbers $K=0,P=+1$
\begin{equation}
|\psi\rangle = \sum_\alpha \psi_\alpha |\phi_\alpha\rangle,
\end{equation}
the amplitude of a configuration belonging to the orbit of $\alpha$ is given by
\begin{equation}
\psi_{\text{full}} = \frac{\psi_\alpha}{\sqrt{R_\alpha p_\alpha}}.
\end{equation}
We construct the reduced density matrix $\rho_A = \mathrm{Tr}_B \ |\psi\rangle\langle\psi|$ by grouping configurations according to subsystem $B$. Denoting configurations as $|a\rangle \otimes |b\rangle$, where
\begin{equation}
|\psi\rangle = \sum_{a,b} \psi_{a,b} \ |a\rangle \otimes |b\rangle,
\end{equation}
we obtain
\begin{equation}
\rho_A(a,a') = \sum_b \psi_{a,b} \ \psi^*_{a',b}.
\end{equation}
In practice, this is implemented efficiently by organizing contributions into blocks labeled by $b$, avoiding explicit construction of the full Hilbert space. The von Neumann entanglement entropy is computed from the reduced density matrix as
\begin{equation}
S = - \mathrm{Tr}(\rho_A \log \rho_A),
\end{equation}
where the eigenvalues of $\rho_A$ are obtained via exact diagonalization. Small eigenvalues below a numerical threshold are discarded to ensure stability while taking logarithm.
This procedure yields the half-chain entanglement entropy of Floquet eigenstates directly within the $(K=0, P=+1)$ symmetry sector, while fully accounting for the underlying symmetry-induced normalization factors.

\section{Spectral form factor and transverse magnetization}
\label{sffcorr} 

\begin{figure}[t]
\centering
\includegraphics[width=0.32\linewidth]{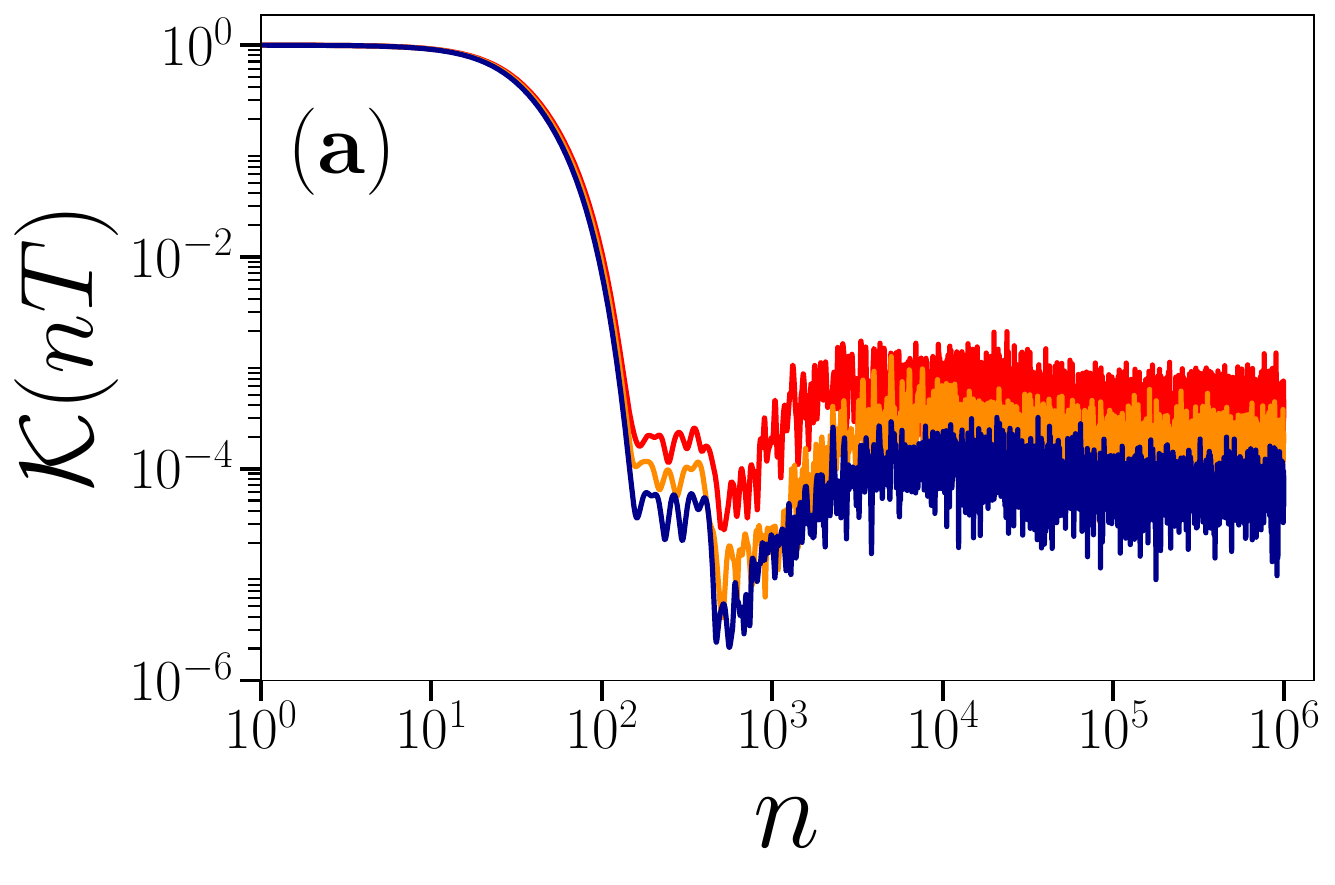}\hfill
\includegraphics[width=0.32\linewidth]{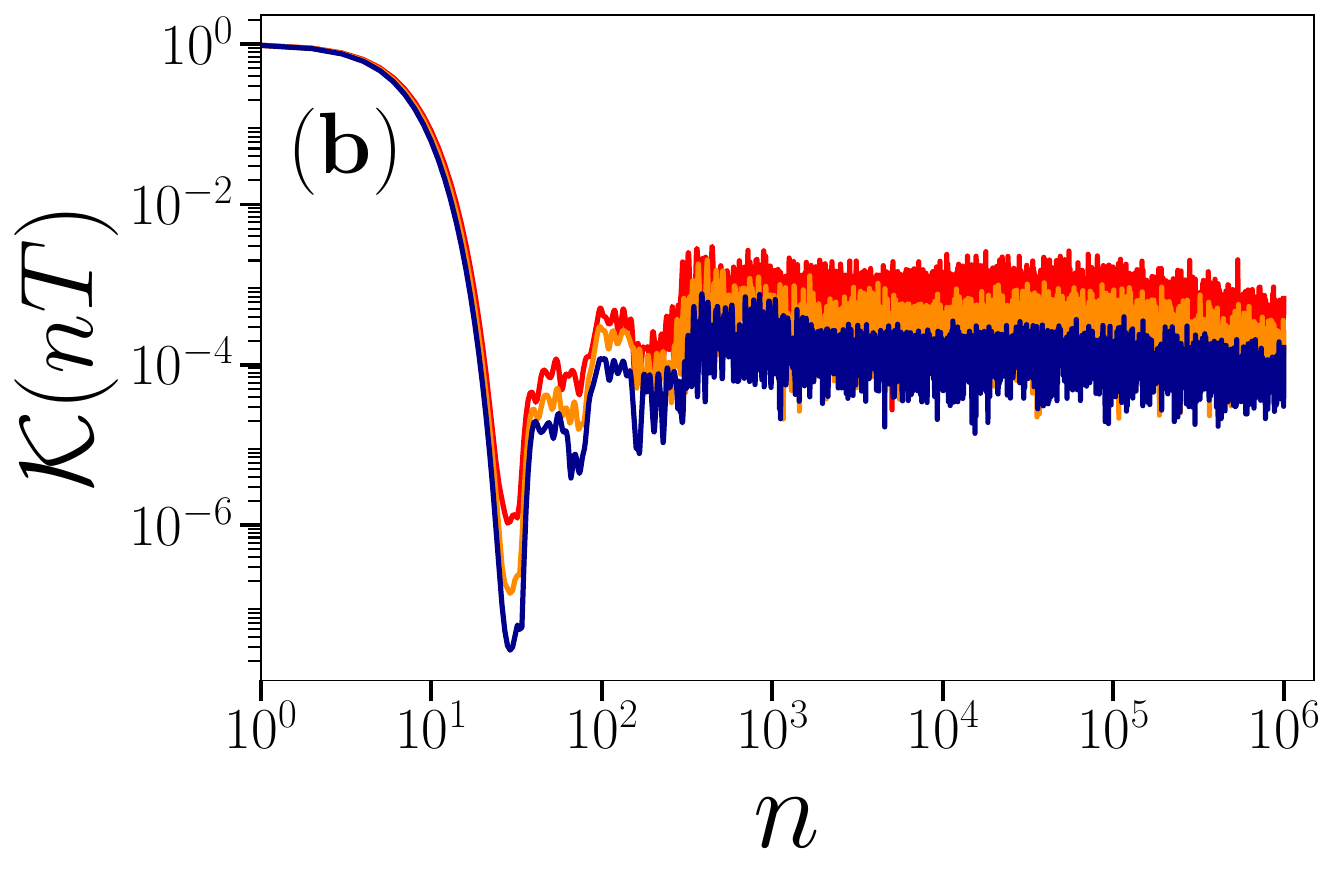}\hfill
\includegraphics[width=0.32\linewidth]{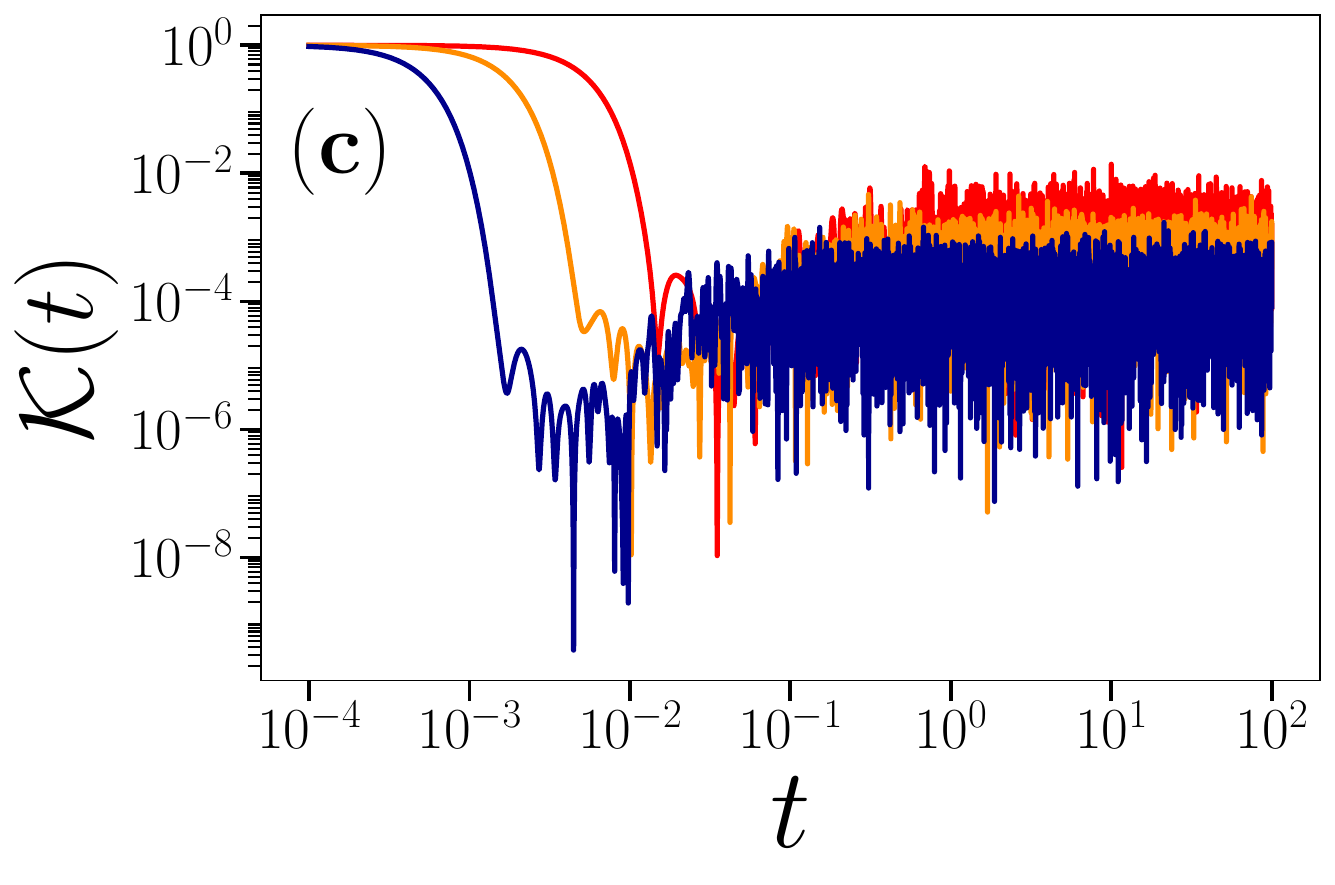}
\centering
\caption{(a) Plot of $\mathcal{K}(nT_1)$ as a function of the number of drive cycles $n$ for $L=24$ (red), $L=26$ (orange) and $L=28$ (blue); the driven chain is assumed to have PBC and we are in the $K=0,P=+1$ sector. For all plots $w_0/w_1=1$ and $\gamma/\pi=2$. (b) same as in (a) but with $\gamma/\pi=1.67$. (c) Plot of $\mathcal{K}(t) := \frac{1}{D_{\mathrm{sec}}^2} \sum_{j_1,j_2=1}^{D_{\mathrm{sec}}} e^{i (E_{j_1}-E_{j_2}) t/\hbar} $ for the eigenspectrum $\{E_j\}$ of an undriven XXZ Hamiltonian with anisotropy parameter $\Delta = -1/2$ so that $H_F^{(2)}$ can be directly mapped to such an ${\rm XXZ}$ chain at $\gamma=2 \pi$. For all plots, $D_{\rm sec}$ is the Hilbert space dimension of the chosen magnetization sector, the chain lengths are chosen to be $L=18$ (red), $L=20$ (orange) and $L=22$ (blue), and we have used PBC with $K=0$, $P=+1$, $M^z=0$. For all plots we have averaged the SFF over a small range of $w_0$ ((a) and (b)) and $J$ (c) to reduce fluctuations in ${\mathcal K}$. 
\label{figsff}}
\end{figure} 

In this section, we consider the spectral form factor and the transverse magnetization of the driven model. The spectral form factor of the model is defined as 
\begin{eqnarray}
 {\mathcal K}(nT_1) &=& \frac{1}{{\mathcal D}^2} \sum_{p ,q} e^{i (\epsilon_p^F -\epsilon_q^F) n T_1/\hbar }. \label{sff1} 
\end{eqnarray}
For the driven model, ${\mathcal K}$ is known to exhibit the usual dip-ramp-plateau structure consistent with ergodic systems \cite{tb1,sff1,sff2,sff3,sff4} with the presence of the linear ramp
regime indicating thermalization \cite{sff2}. For the present driven model, SFF indicated such a structure for generic drive frequencies; it was shown that for $w_0 \ll w_1,\lambda_0$, the dip time grows as inverse Floquet bandwidth $\Lambda_p^{-1} \sim 1/w_0$ indicating proximity to the flat band regime \cite{tb1}. 

A plot of ${\mathcal K} (nT_1)$ as a function of $n$ is shown in Fig.\ \ref{figsff}(a) for $\gamma= 2\pi$. The plot indicates drastic reduction of the linear ramp regime compared to its counterpart shown in Fig.\ \ref{figsff}(b) where $\gamma= 2 \pi/3$. These plots indicates absence of a thermal ergodic regime at the special drive frequencies. The presence of the small linear ramp regime in Fig.\ \ref{figsff}(a) seems to be an effect of finite $L$; this can be inferred from comparing it to the SFF
for the integrable ${\rm XXZ}$ spin chain in Fig.\ \ref{figsff}(c) showing similar features. The larger dip time for ${\mathcal K}$ in Fig.\ \ref{figsff}(a) is an indication of slower dynamics at special drive frequencies where $H_F^{(1)}$ vanishes; such dynamics is mediated by second-order terms in the Floquet Hamiltonian. In contrast, a much faster dynamics leading to a shorter dip time occurs away from the special frequencies where $H_F^{(1)}$ is finite; this is seen in Fig.\ \ref{figsff}(b). 

\begin{figure}[t] 
\centering
\includegraphics[width=0.47\linewidth]{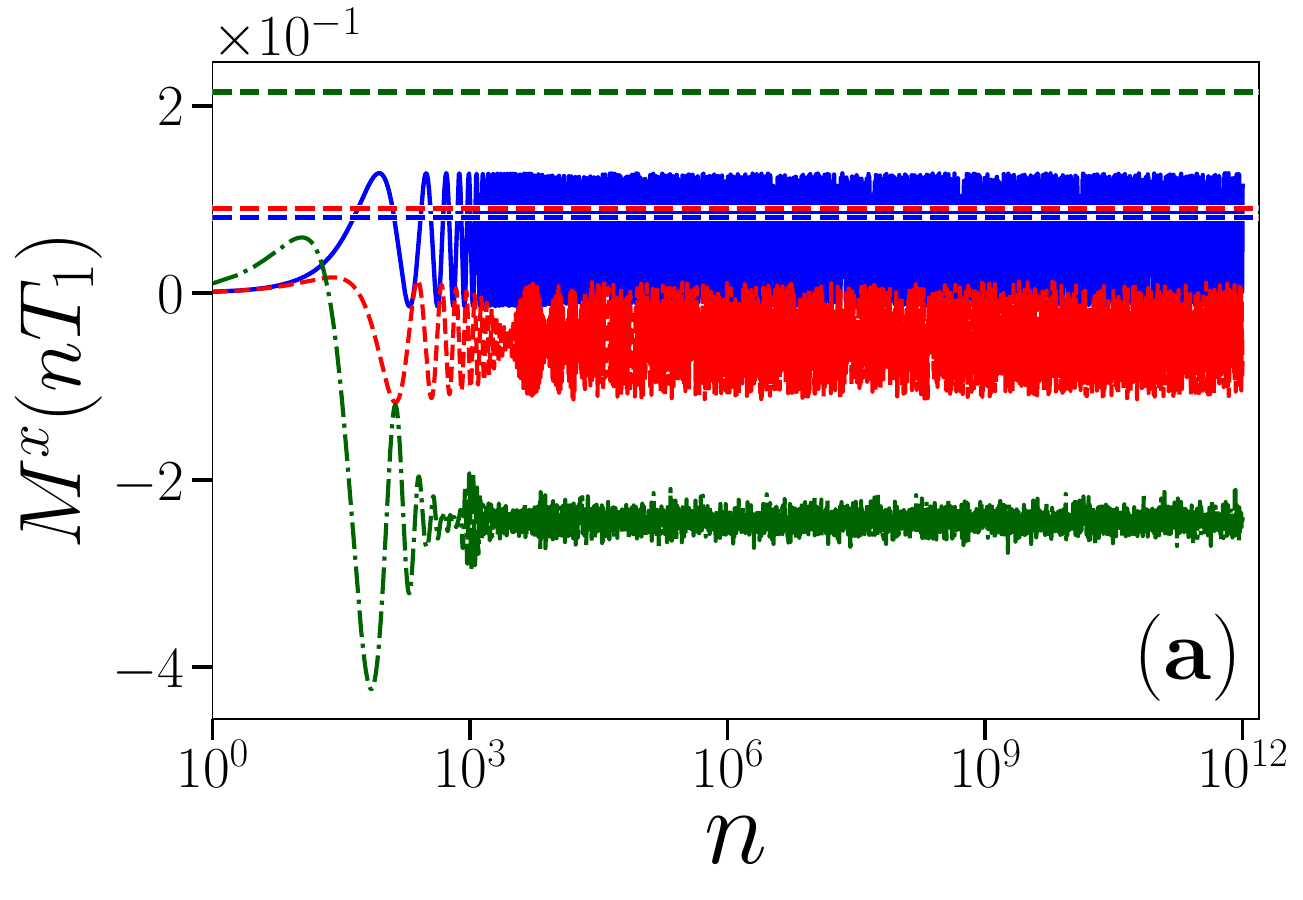}
\includegraphics[width=0.47\linewidth]{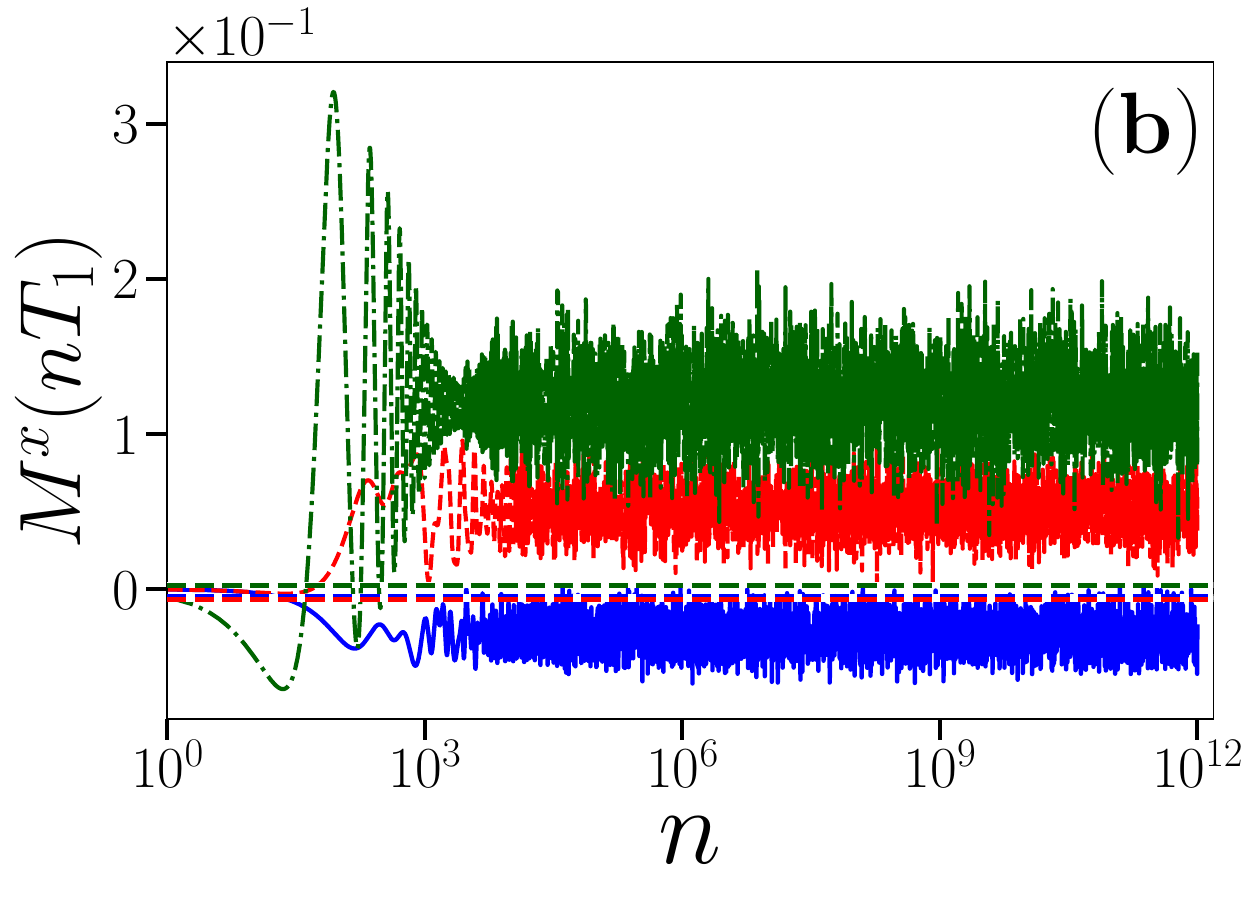}
\centering
\caption{(a) Plot of $M^x(nT_1)$ as a function of the number of drive cycles $n$ for $\gamma/\pi=2$ (blue), $1.99$ (green) and $1.9$ (red) for $|\psi(0)\rangle=|{\rm vac}\rangle$; the driven chain is assumed to have PBC and we are in the $K=0,P=+1$ sector. For all plots $w_0/w_1=1$ and $L=28$. (b) same as in (a) but for $|\psi(0)\rangle= |{\rm AFM}\rangle$. The dotted lines show the corresponding ETH predicted DE values while the black dashed line indicated the ITE value. \label{figmx}}
\end{figure} 

The transverse magnetization of the driven PXP chain is given, for a chain of length $L$, by 
\begin{eqnarray} 
M^{x}(nT_1) = \frac{1}{L} \, \langle \psi(nT_1) \vert \sum_{j} \tilde{\sigma}_j^{x} \vert \psi(nT_1) \rangle, \label{mxdef1}
\end{eqnarray} 
where $|\psi(nT_1)\rangle = U(nT_1,0)|\psi(0)\rangle$ and $|\psi(0)\rangle$ is the initial state. In what follows, we choose $|\psi(0)\rangle$ to be a Fock state.

The plot of $M^x(nT_1)$ shown for the initial $|{\rm vac}\rangle$ and $|{\rm AFM}\rangle$ states for Figs.\ \ref{figmx}(a) and (b) respectively. For both case, $M^x$ remain close to its initial value for $\gamma =2 \pi$ for a large number of drive cycles; this is a consequence of slow dynamics which occurs due to vanishing of $H_F^{(1)}$ at these points. We note that unlike its longitudinal counterpart discussed in the main text, $M^x$ is not a conserved charge and hence it eventually deviates from its initial value and fluctuates around a steady state value. The slow nature of change of $M^x$ in the transient regime ($n\le n_0 \sim 500$) serves an indication of the proximity of the driven chain to the special frequencies. Near the special points, the steady state value is close to the thermal value predicted by the diagonal ensemble (DE); however away from the special points where the dynamics is controlled by $H_F^{(1)}$, it is significantly different. The steady state reaches a superthermal value of $\langle M^x\rangle$ due to the presence of quantum scars for the AFM initial state; in contrast it has a subthermal value for $|\psi_0\rangle=|{\rm vac}\rangle$.
These properties were discussed in details, for a driven PXP chain with a single-tone drive, in Refs.\ \onlinecite{scar2,scar5}. 

\section{Equivalence of $H_F^{(2)}$ and $\mathrm{XXZ}$ models : Bethe ansatz for two particles}

In this section, we will use the Bethe ansatz to
find the spectra of two particles (i.e., two up spins)
for $H_F^{(2)}$ and the $\mathrm{XXZ}$ model for systems with PBC.
We will show that the energy levels in the two models are identical only 
in the $K=0$ sector, thereby confirming the analysis
in the main text that the
models are equivalent only in that sector.

We will begin with the Hamiltonian of an $\mathrm{XXZ}$ model 
defined on a system with $L'$ sites and PBC,
\beq H_{XXZ} ~=~ \sum_j [2 (\tau_j^+ \tau_{j+1}^- ~+~
{\rm H.c.}) ~+~ \Delta (\tau_j^z \tau_{j+1}^z ~-~ 1)]. 
\label{hamxxz} \eeq
Note that we have added a constant to ensure that
the state with all spins down is an eigenstate of
$H_{XXZ}$ with zero eigenvalue. We now consider a
state with exactly one particle (up spin)
which can be at any site $n$, and we look at a momentum
eigenstate given by the superposition
\beq | k \ra ~=~ \sum_n ~e^{ikn} ~| n \ra. \label{psi1} \eeq
The eigenvalue equation $H_{XXZ} | k \ra = E_k | k \ra$
then gives the dispersion
\beq E_k ~=~ 4 \cos k ~-~ 4 \De. \label{disp1} \eeq
The PBC implies that $| n+L' \ra = | n \ra$, and the
relation $e^{ik(n+L')} = e^{ikn}$ then implies that
$k$ must be quantized as $k = (2 \pi/L') q$, where $q$
is an integer.

Next we consider a state with exactly two particles which
are located at sites $n_1$ and $n_2$, and we will assume that $n_1 < n_2$. (Physically, these coordinates
must satisfy $1 \le n_1 < n_2 \le L'$, but, as we will
see below, it is sometimes mathematically convenient to 
ignore the limits 1 and $L'$ and allow $n_1 < n_2$
to take values outside these limits). We now consider the Bethe ansatz for a state in which the particles have momenta
$k_1, ~k_2$,
\bea | k_1, k_2 \ra &=& \sum_{n_1 < n_2} ~\psi (n_1,n_2) 
~| n_1, n_2 \ra, \non \\ 
{\rm where} ~~~\psi (n_1, n_2) &=& e^{i(k_1 n_1 + k_2 n_2)} ~+~ 
\al_{21} ~e^{i(k_1 n_2 + k_2 n_1)}. \label{bethe1} \eea
Here $\al_{21}$ can be a function of $k_1, ~k_2$ and
$\De$ and it will be determined by the eigenvalue condition
$H_{XXZ} | k_1, k_2 \ra = E_{k_1,k_2} | k_1, k_2 \ra$. Looking at the coefficients of states $|n_1 , n_2 \ra$ 
in Eq.~\eqref{bethe1} for which $n_1 < n_2 -1$,
we find from the eigenvalue condition that the 
two-particle dispersion is given by
\beq E_{k_1,k_2} ~=~ E_{k_1} ~+~ E_{k_2} ~=~ 4 (\cos k_1 + \cos k_2) ~-~ 8 \De. \label{disp2} \eeq
Next, looking at the coefficient of states with 
$n_1 = n_2 - 1$, we find from the eigenvalue equation
that $\al_{21}$ is given by the expression
\beq \al_{21} ~=~ - ~e^{i(k_2 - k_1)} ~\frac{e^{ik_1} ~+~
e^{-i k_2} ~-~ 2 \De}{e^{-ik_1} ~+~ e^{i k_2} ~-~ 2 \De}.
\label{al21} \eeq
It is clear that $|\al_{21}| =1$, and the phase of
$\al_{21}$ is called the two-particle scattering phase
shift. To find the quantization condition for $k_1, ~k_2$,
we now demand that $\psi (n_1,n_2)$ must satisfy the PBC,
namely,
\beq \psi (n_1, n_2) ~=~ \psi (n_2, n_1 + L). \eeq
Equating the coefficients of the terms $e^{i(k_1 n_1 + k_2 n_2)}$ and $e^{i(k_1 n_1 + k_2 n_2)}$ separately 
in Eq.~\eqref{bethe1}, we obtain
\beq \al_{21} ~=~ e^{-ik_1 L'} ~~~{\rm and}~~~ \al_{21}
~=~ e^{ik_2 L'}. \label{al21pbc} \eeq
These relations imply that $e^{i (k_1 + k_2) L'} = 1$, hence
the total momentum must be quantized as $k_1 + k_2 = (2\pi/L') q$. However, finding $k_1$ and $k_2$ separately
requires us to solve Eqs.~\eqref{al21} and \eqref{al21pbc}
self-consistently which is difficult to do analytically.

We now consider the model with $H_F^{(2)}$ 
defined on a system with $L$ sites and PBC,
\beq H_F^{(2)} ~=~ \sum_j [ 2 P_{j-1} (\si_j^+ 
\si_{j+1}^- ~+~ {\rm H.c.}) P_{j+2} ~+~ 2 (P_{j-1} 
\si_j^z P_{j+1} ~+1)]. \label{hampxp} \eeq
Once again we have added a constant to ensure that
the state with all spins down is an eigenstate of
the Hamiltonian with zero eigenvalue. We now consider a
one-particle state 
which can be at any site $n$, and look at the momentum
eigenstate given in Eq.~\eqref{psi1}.
The eigenvalue equation $H_F^{(2)} | k \ra = E_k | k \ra$
then gives 
\beq E_k ~=~ 4 \cos k ~+~ 8. \label{disp3} \eeq
As before, the PBC implies that
$k$ must be quantized as $k = (2 \pi/L) q$, where $q$
is an integer.

Next we consider a state with two up spins
located at sites $n_1$ and $n_2$, such that $n_1 < n_2-1$. (This condition comes from the Rydberg constraint
that two up spins cannot be at neighboring sites). We now consider the Bethe ansatz for a state in which the particles have momenta
$k_1, ~k_2$,
\bea | k_1, k_2 \ra &=& \sum_{n_1 < n_2} ~\psi (n_1,n_2) 
~| n_1, n_2 \ra, \non \\ 
{\rm where}~~~ \psi (n_1, n_2) &=& e^{i(k_1 n_1 + k_2 (n_2-1))} ~+~ 
\beta_{21} ~e^{i(k_1 (n_2-1) + k_2 n_1)}. \label{bethe2} \eea
Note that the ansatz in Eq.~\eqref{bethe2} is slightly
different from Eq.~\eqref{bethe1} in that we have replaced
$n_2$ by $n_2 - 1$.
We now determine $\beta_{21}$ using the eigenvalue condition
$H_F^{(2)} | k_1, k_2 \ra = E_{k_1,k_2} | k_1, k_2 \ra$. Looking at the coefficients of states $|n_1 , n_2 \ra$ 
in Eq.~\eqref{bethe1} in which $n_1 < n_2 -2$,
we find from the eigenvalue condition that the 
two-particle dispersion is 
\beq E_{k_1,k_2} ~=~ 4 (\cos k_1 + \cos k_2) ~+~ 16. \label{disp4} \eeq
Next, looking at the coefficient of states with 
$n_1 = n_2 - 2$, we find from the eigenvalue equation
that $\beta_{21}$ is given by the expression
\beq \beta_{21} ~=~ - ~e^{i(k_2 - k_1)} ~\frac{e^{ik_1} ~+~
e^{-i k_2} ~+~ 1}{e^{-ik_1} ~+~ e^{i k_2} ~+~ 1}.
\label{be21} \eeq
We see that the expressions for $\beta_{21}$ in Eq.~\eqref{be21} and
for $\al_{21}$ in Eq.~\eqref{al21} are identical 
if we take $\De = -1/2$. Next, we find the quantization condition for $k_1, ~k_2$,
by demanding that $\psi (n_1,n_2)$ must satisfy the PBC
\beq \psi (n_1, n_2) ~=~ \psi (n_2, n_1 + L). \eeq
Equating the coefficients of the terms $e^{i(k_1 n_1 + k_2 n_2)}$ and $e^{i(k_1 n_1 + k_2 n_2)}$ separately 
in Eq.~\eqref{bethe1}, we obtain the relations
\beq \beta_{21} ~=~ e^{-ik_1 L + i (k_1 - k_2)} ~~~{\rm and}~~~ 
\beta_{21} ~=~ e^{ik_2 L + i (k_1 - k_2)}. \label{be21pbc} \eeq
These relations imply that $e^{i (k_1 + k_2) L} = 1$, hence
the total momentum must be quantized as $k_1 + k_2 = (2\pi/L) q$. 

Eqs.~\eqref{be21pbc} differ from Eqs.~\eqref{al21pbc} in general. Hence, even though the expressions in 
Eqs.~\eqref{al21} and \eqref{be21} are identical for $\De 
= -1/2$, we find that the solutions for $k_1, ~k_2$
are generally not the same in the two cases once we put in
the information about the system sizes
$L'$ and $L$. However,
we observe that in the zero momentum sector where $k_1 + 
k_2 = 0$, Eqs.~\eqref{al21pbc} and Eqs.~\eqref{be21pbc} become identical for 
$L' = L -2$. This is precisely the relation between
the system sizes for the $H_F^{(2)}$ and $\mathrm{XXZ}$ models
for the case of two particles as discussed in the main
text.

In the zero momentum sector, we can set $k_1 = k$ and $k_2 = - k$. We then obtain a more explicit expression for the energy levels of two particles by combining Eqs.~\eqref{be21}
and \eqref{be21pbc},
\beq - ~e^{- i 2 k} ~\frac{2 e^{ik} ~+~ 1}{2 e^{-ik} ~+~ 1} ~=~ e^{-ik (L-2)}. \eeq
This equation will have several solutions for $k$; the number of solutions is equal to the number of zero momentum states
which is given by the largest integer less than or equal to
$(L-2)/2$. Eq.~\eqref{disp4} then gives the energy as
\beq E_{k,-k} ~=~ 8 \cos k ~+~ 16. \eeq

To conclude, the energy spectra for two particles 
in systems with PBC are identical for the $H_F^{(2)}$ and
$\mathrm{XXZ}$ models in the zero momentum sector but not in 
sectors with non-zero momentum. This confirms that the
two models can be mapped into each other only in the
zero momentum sector as we have demonstrated in the main
text.

The above considerations can be generalized to the case
of $N$ particles (up spins) where $N \ge 3$ for systems
with PBC. It is known that all the eigenstates of the $\mathrm{XXZ}$ can be found
using the Bethe ansatz. Hence, after mapping the
particle coordinates from $(n_1,n_2, \cdots, n_N)$ in the $\mathrm{XXZ}$ model to the coordinates $(n_1,n_2+1,n_3+2, \cdots, 
n_N + N -1)$ in the $H_F^{(2)}$ model, we can use
arguments similar to the ones
presented above to show that all the eigenstates
of $H_F^{(2)}$ can be found using the Bethe ansatz~\cite{intryd1,intryd2}. We 
then conjecture that the eigenvalues in the two models
will be identical if the system sizes are
related as $L_{XXZ} = L_{H_F^{(2)}} - N$ and if we are
in the zero momentum sector.

\bibliography{cit}